\documentclass[11pt]{article}
\pdfoutput=1
\usepackage[margin=1.0in]{geometry}
\usepackage{amsmath,amssymb,graphicx}
\usepackage{hyperref}
\usepackage{slashed}

\usepackage[utf8]{inputenc}

\usepackage[greek,english]{babel}
\newcommand{\cpi}{\text{\greektext p}}

 \usepackage[caption=false]{subfig}
\usepackage{booktabs}

\usepackage[font=small,labelfont=bf]{caption}
\usepackage{cite}
\usepackage{enumerate}
\usepackage{mathtools}

\DeclarePairedDelimiter\floor{\lfloor}{\rfloor}

\newcommand{\iso}[2]{\mathcal{I}^\text{#1}_{#2}}

\newcommand{\msi}\bf
\newcommand{\msf}\rm

\newcommand{\iu}{{\mathrm i}}
\newcommand{\E}{{\mathrm e}}

\DeclareRobustCommand{\Sec}[1]{Sec.~\ref{#1}}

\DeclareRobustCommand{\Fig}[1]{Fig.~\ref{#1}}

\DeclareRobustCommand{\Eq}[1]{Eq.~(\ref{#1})}

\newcommand{\J}[2]{J_{#1}\left( #2\right)}

\newcommand{\Hyp}{{}_2\mathrm{F}_1}

\newcommand{\FS}[1]{\textsf{S} \left(#1 \right)}
\newcommand{\FC}[1]{\textsf{C} \left(#1 \right)}

\newcommand{\be}{\begin{equation}}
\newcommand{\ee}{\end{equation}}
\newcommand{\beq}{\begin{equation}}
\newcommand{\eeq}{\end{equation}}

\usepackage[table]{xcolor}
\usepackage{bm}

\definecolor{gray}{cmyk}{0,0,0,0.05}
\newcolumntype{a}{>{\columncolor{gray}} l}

\numberwithin{equation}{section}

\title{\bf Spheres to Jets\\ {\Large Tuning Event Shapes with 5d Simplified Models}}
\author{Cari Cesarotti,  Matthew Reece, and Matthew J.~Strassler \\
{\small Department of Physics, Harvard University, Cambridge, MA, 02138}}

\begin{document}
\maketitle

\begin{abstract}
Hidden sectors could give rise to a wide variety of events at the LHC. Confining hidden sectors are known to engender events with a small number of jets when they are weakly-coupled at high energies, and  quasi-spherical  soft unclustered energy patterns (SUEPs) when they are very strongly-coupled  (large `t Hooft coupling)  at high energies.  The intermediate regime is murky,  and could give rise to signals hiding from existing search strategies.  While the intermediate coupling regime is not calculable, it is possible to pursue a phenomenological approach in which one creates signals that are intermediate between spherical and jetty.  We propose a strategy for generating events of this type using simplified models in extra dimensions. The degree to which the event looks spherical is related to the number of decays produced near kinematic threshold. We provide an analytic understanding of how this is determined by parameters of the model. To quantify the shape of events produced with this model, we use a recently proposed observable---\textit{event isotropy}---which is a better probe of the spherical regime than earlier event shape observables.
\end{abstract}

\section{Introduction}
Data taken at the Large Hadron Collider (LHC) has been essential to confirm Standard Model (SM) predictions and perform precision measurements \cite{2012PhLB..716...30C, 2012PhLB..716....1A}.
 Extensive efforts have been made to search for phenomena beyond the Standard Model (BSM) as well, such as supersymmetry, dark matter, extra dimensions, and many others. 
 Thus far, no BSM physics has been observed. 
 The lack of evidence for new physics suggests that the simplest scenarios are not good models of nature. 

 However, one should worry that more complicated signals might have evaded current LHC search strategies. 
 A large class of unusual BSM signals   arise within the ``hidden-valley" (HV) scenario,  in which the SM is weakly coupled to a hidden sector of  self-interacting  particles neutral under the SM gauge groups,  at least one of which decays visibly \cite{Strassler:2006im, Strassler:2006ri, Strassler:2006qa, Han:2007ae}.  
Similarly unusual signals arise in theories of ``quirks'' \cite{Okun:1980kw, Okun:1980mu,Strassler:2006im, Kang:2008ea}, and also in the context of unparticles \cite{Georgi:2007ek, Grinstein:2008qk, Randall:1999vf} which in the presence of a mass gap may produce hidden-valley phenomenology \cite{Strassler:2008bv}.  
 Since hidden sectors can be arbitrarily complex,   many models of this type are  poorly constrained by LHC searches, and since indirect limits on hidden sectors are often very weak, there are few other constraints.
 But it is impossible to search through the full space of HV models, or of models in other general scenarios.
 Instead, the practical approach to discovering a new signal is to carry out searches for distinct and parametrizable {\it signatures}, and to make this possible, models that produce these signatures must be developed. Among the unusual signatures identified in HV models to date are soft, unclustered energy patterns (SUEPs) \cite{Strassler:2008bv, Harnik:2008ax, Knapen:2016hky}; lepton jets \cite{ArkaniHamed:2008qp, Baumgart:2009tn}; emerging jets \cite{Strassler:2006im,Schwaller:2015gea}; semi-visible jets \cite{Strassler:2006im, Strassler:2008fv, Cohen:2015toa, Cohen:2017pzm}; and dark jets with unusual substructure \cite{Cohen:2020afv}.

General, flexible searches for new physics at  the LHC require us not only to  widen the range  of  models and signatures, but also to expand the  tools available for data analysis,  both offline and at the trigger stage (where fewer than 1\% of LHC collisions are recorded). Search strategies for general hidden sectors are confounded by the many free parameters, including the masses, couplings, and lifetimes of new particles. The overall event shape of different scenarios can also take many forms. This motivates the development of diagnostic tools to characterize anomalous event shapes which are unlikely to arise from Standard Model processes.

 In this paper we will introduce a class of simplified models that produce a range of new signatures, and our goal will be to characterize them using event shape observables.  It would be premature to consider how to search for these signals at the LHC, as we should first understand the signatures themselves.  For this reason we focus our attention on an idealized situation: a pure signal at an $e^+e^-$ collider of the future with no background.  

There exist several well-known observables that characterize the shapes of events at $e^+e^-$ colliders. A commonly used infrared and collinear (IRC) safe event shape observable  
is thrust, defined as \cite{Brandt:1964sa,Farhi:1977sg,DeRujula:1978vmq}
\begin{equation}
T = \text{max}_{\hat{n}} \frac{\sum_i|\hat{n}\cdot \vec{p}_i|}{\sum_i |\vec{p}_i|}. 
\label{eq:thrust}
\end{equation}
It has a range $T \in [0.5,1]$, where $T=1$ corresponds to two back-to-back particles, and $T=0.5$ is an isotropic radiation pattern. While thrust has provided essential insight on the perturbative nature of QCD,  it is most sensitive to event shape deviations from the two-particle dijet configuration and has less sensitivity in the quasi-spherical regime (see Fig. 8 of \cite{Cesarotti:2020hwb}). To complement such standard observables, we also make use of a recently proposed event shape observable, the event isotropy \cite{Cesarotti:2020hwb}.

Event isotropy is defined using the energy mover's distance (EMD) \cite{Komiske:2019fks,Komiske:2020qhg}.
Given two radiation patterns of massless particles $P$, $Q$, the EMD is the minimum work necessary to rearrange $P$ into $Q$. 
A radiation pattern is defined as a set of particles, each of which are specified by their position on the unit sphere and fraction of the total energy. 
To reorganize the pattern $P$ into $Q$, we construct a transport map $f_{ij}$ that tracks the total energy fraction moved from position $i$ to position $j$. 
The total work done in the rearrangement can be written as a sum over the distance $d_{ij}$  from $i$ to $j$ weighed by the fraction of energy moved $f_{ij}$.
The EMD is the minimum work for all possible rearrangements of $P$ to $Q$:
\begin{equation}
\text{EMD}\left(P,Q\right) = \min\limits_{\{f_{ij}\}} \sum_{ij} f_{ij}d_{ij}.
\end{equation}
The distance measure we use in this paper is
\begin{equation}
d_{ij} \equiv \frac{3}{2}\sqrt{1-\hat{n}_i \cdot \hat{n}_j} = \frac{3}{2}\sqrt{1-\cos\theta_{ij}}
\label{eq:dijmeasure}
\end{equation}
for $\hat{n}_i$ the unit vector proportional to the three-momentum of the element $p_i$, etc. 
Note this differs from \cite{Cesarotti:2020hwb}, where the distance measure was proportional to $d_{ij} \sim 1-\cos\theta_{ij}$.
See Appendix B of that paper for a discussion of the different distance measures.

The ideal event isotropy of an event would be its EMD to a uniform radiation pattern $\mathcal{U}$ of equal total energy:
\begin{equation}
\mathcal{I}(\mathcal{E})  = \text{EMD} \left( \mathcal{E}, \mathcal{U}\right). 
\end{equation}
which we would take as a uniform spherical distribution for $e^+ e^-$ colliders.
However 
to make computation times practical we calculate the EMD to a uniformly tiled, high multiplicity sphere generated with HEALPix \cite{2005ApJ...622..759G}. Specifically, we will use multiplicity 192 for the tiling, and we denote the event isotropy variable by $\iso{sph}{192}$.  
%

\begin{figure}[!h]
\centering
\includegraphics[width=0.55\textwidth]{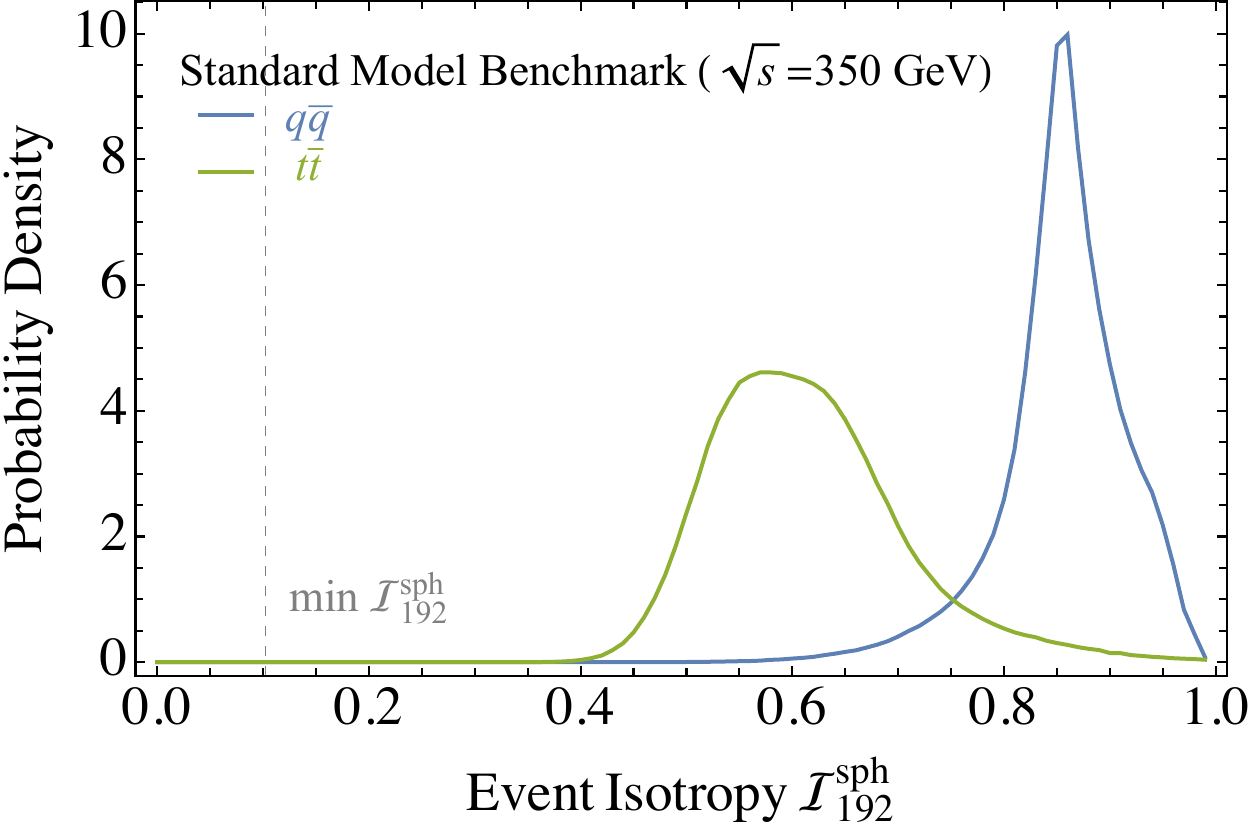}
\caption{Event isotropy distributions for  $e^+  e^- \to q {\bar q}$ and $e^+ e^- \to t {\bar t}$ (hadronic decays only) at a center-of-mass energy just above the $t\bar t$ threshold. The distance measure \Eq{eq:dijmeasure} gives higher $\iso{sph}{192}$ values than the corresponding plot in Ref.~\cite{Cesarotti:2020hwb}.}
\label{fig:SMbenchmark}
\end{figure}

Note that event isotropy is defined as a {\em distance to an isotropic event}. Thus, highly isotropic events have {\em low} values of $\iso{sph}{}$, despite the name. In particular, $\iso{sph}{}$ is 0 for a spherical event and 1 for a pencil dijet event with two back-to-back momenta.  More generally, for an event with $k$ isotropically distributed particles of equal energy,  a sphere tiled with multiplicity $N$, and the distance measure \Eq{eq:dijmeasure}, the event isotropy is\footnote{We may imagine spreading each particle's energy over a region of area $4\cpi/k$, which requires moving the energy a distance of order $1/\sqrt{k}$.  The normalization is obtained by requiring that for $k=2$, the extreme dijet limit, $\iso{sph}{192}= 1$. Any non-uniformity in the distribution generally increases the isotropy when averaged over many orientations.  For the distance measure used in \cite{Cesarotti:2020hwb}, there is no square root.
See Appendix A of \cite{Cesarotti:2020hwb} for more details.} 
\be
\iso{sph}{N}
\gtrsim \sqrt{\frac{2}{\text{min} (k,N)}}.
\label{eq:isoest}
\ee
The minimum value of $\iso{sph}{192} \approx \sqrt{\frac{2}{192}}\approx 0.1$ represents a slight loss of dynamic range, a price one pays for much faster computation. 

To illustrate the behavior of this variable, we used \texttt{Pythia 8.243} \cite{Sjostrand:2014zea} to generate, at a center-of-mass energy of 350 GeV, the processes $e^+ e^-\to q{\bar q}$ and $e^+ e^- \to t {\bar t}$.
The resulting event isotropy distributions are shown in Fig.~\ref{fig:SMbenchmark}; the values of $\iso{sph}{192}$ are higher than the corresponding plot in Ref.~\cite{Cesarotti:2020hwb} because of our choice of distance measure \Eq{eq:dijmeasure}.
QCD radiation and radiative-return to the $Z$ boson both reduce the $q\bar q$ isotropy from 1 to approximately 0.8.
Meanwhile, since the top quarks are produced near threshold, their six jets are distributed quasi-isotropically, and the distribution peaks near the value $\sqrt{1/3}\sim 0.58$ that \Eq{eq:isoest} would suggest.

\par HV extensions to  the SM often  take the form of confining hidden sectors. However, confinement is compatible with a wide range of event shapes. The 't Hooft coupling  $\lambda = \alpha_s N_c$ plays a major role in determining the shape of events. In a QCD-like, asymptotically free  theory, $\lambda \gg 1$  near the confinement scale but runs to be $\ll 1$ at energies well  above $\Lambda_\text{QCD}$.  Gluon radiation in this regime is characterized by perturbative showering, in which a hard quark or gluon is dressed with moderate amounts of collinear radiation, leading to a classic QCD jet. However, a near-conformal field theory may maintain $\lambda\gg 1$ over  a  wide  range of energies above the confinement scale. In this regime collinear radiation is extremely rapid, and all hard partons lose their energy; only soft physics survives. Such a theory will produce events which, because of greatly enhanced radiation \cite{Polchinski:2002jw}, are spherically symmetric in the extreme $\lambda\to\infty$ limit \cite{Strassler:2008bv,Hofman:2008ar,Hatta:2008tx}.

Although one can do reliable computations for $\lambda\ll 1$  (where field-theory perturbation theory is  valid)  and sometimes for $\lambda\gg 1$ (where gauge/string duality furnishes us with an alternative perturbation expansion), BSM  physics could fall in the intermediate regime.  This regime is  poorly understood, as there are no methods for detailed calculation at intermediate $\lambda$.  

Because of this obstacle, we pursue a more pragmatic approach.  In this paper we seek to develop a procedure for generating events that is physically reasonable and has parameters that allow it to interpolate between jetty and spherical.    We hope that  a flexible method  for producing  events with  intermediate values of event shape observables may allow  the  design  of  analysis and trigger  strategies  that are sensitive to a broader class of new physics models, even ones whose signals cannot currently be calculated.

One widely-adopted strategy in the search for new physics at colliders is the use of simplified models. These models can abstract away many details of a theory while preserving key elements  of the collider phenomenology, and have been designed for a broad range of signals. However, they are intrinsically `simple': the models include a small number of new light particles and interactions \cite{Meade:2006dw, ArkaniHamed:2007fw, Alwall:2008ag}. 
While it is beneficial to have fewer parameters, we would also like to consider more complex theories with many interactions and heavier particles, as in Hidden Valleys.
There have been studies of simplified models with relatively high final-state multiplicities \cite{Strassler:2008fv, Evans:2013jna, Fan:2015mxp, Cohen:2016nzv}, as well as studies of specific theories that give signals made of many soft particles \cite{Kang:2008ea, Harnik:2008ax, Knapen:2016hky}. However, there is a gap which can be filled by a straightforward approach to formulating simplified models that are flexible enough to span a wide range of event shapes.

In this paper, we show that a simplified model within an extra dimension,  with a small number of parameters, allows  for the generation of a wide range of event shapes. Specifically, we consider a warped extra dimension, in the form of a slice of a 5d AdS space \cite{RandallSundrum:1999}. This is motivated by the AdS/CFT or ``gauge/string''  correspondence, which relates gauge theories to string theory \cite{Maldacena:1997re,Gubser:1998bc,Witten:1998qj}.  The correspondence suggests that a set of interacting fields (including gravity)  on such a finite-size warped extra dimension can be interpreted as the dual of a (large $N$) confining gauge theory.  (This relation has been made precise for a few purely 4d gauge theories, as in \cite{Polchinski:2000uf, Klebanov:2000hb}.) The infinite tower of massive Kaluza-Klein modes (KK modes) in the extra dimension is equivalent to a tower of hadrons of the confining gauge theory. Rather than choosing some ansatz for the masses and couplings of hadrons in a strongly-coupled gauge theory where we cannot do calculations, we choose a small set of masses and couplings in 5d which determine the entire infinite set of masses and  couplings in 4d. This reduces the number of arbitrary choices to make, while still allowing enough flexibility to generate a wide range of collider event shapes. This use of AdS as a simplified dual to a Hidden Valley is in the spirit of previous uses of AdS to model low energy QCD \cite{Csaki:1998qr, Polchinski:2001tt, Polchinski:2001ju, Polchinski:2002jw, Csaki:2003zu, erlich2005qcd,DaRold:2005mxj}. The fact that KK-mode cascades in a warped extra dimension can produce approximately spherical events was  previously examined in \cite{csaki2009ads}. We build on this by constructing a wider range of models, which lead to a wider range of collider events. We also make use of the new tool of event isotropy to obtain an improved characterization of these events.

We emphasize that what we present here is a physical model (morally dual to a field theory at large $\lambda$) which can interpolate {\it phenomenologically} between the jetty regime (which arises at small $\lambda$ as it does in QCD) and the quasi-spherical regime (which appears at very high energy at large $\lambda$.)   
 Because multiple, qualitatively different interpolations between the two regimes likely exist, our model may have little to do with what one would observe in a real confining gauge theory whose ultraviolet value of $\lambda$ is varied from small to large.  Nevertheless, our approach widens the space of sensible targets for experimenters, and one may hope any search strategies that it inspires may be sensitive to a variety of models, not just the one proposed here, that sit between the jetty and spherical extremes.

We proceed with a brief introduction of simplified models in extra dimensions in \Sec{sec:xdimsimple}.
In \Sec{sec:simResults}, we present results from simulations of cascades generated with our models, for a variety of parameter values and interaction terms. 
We show that  these models can accommodate varied distributions of event isotropy. The features of the model are driven by basic properties of the couplings among  various KK modes. 
In simple scenarios with one self-coupled bulk field, near-threshold decays are often preferred, while decays with greater available phase space are suppressed. This leads to low-momentum daughter particles with no preferred boost axis, and thus to nearly isotropic events. The degree of isotropy depends on further details, such as the extent to which there is an approximately conserved KK number, and the number of stable KK modes at the bottom of the spectrum. In scenarios with multiple bulk fields, we find that there are ``plateaus'' in phase space with relatively high decay rates, far from  threshold. These lead to less isotropic events. Finally, cases with boundary-localized couplings can have branching ratios determined mostly by phase space, and lead to much less isotropic events. In each case, we explain how the pattern of branching ratios is reflected in properties of the event: thrust, particle multiplicity, the energy distribution of daughter particles, and the new event isotropy observable. 
The structures that we find in the patterns  of couplings among various modes are determined by overlap integrals involving products of three Bessel functions.
In \Sec{sec:analytic}, we give an analytic understanding of these integrals. In particular, we show that the overlap integrals can be separated into two terms, one of which can be computed approximately and one of which can be computed exactly. The latter term often dominates, and allows us to obtain a clear analytic understanding of both the regime in which near-threshold decays are preferred and the regime with plateaus of enhanced decays away from threshold. All of the important qualitative features determining the event spectra can thus be extracted from the analytic results.
In \Sec{sec:conclusions}, we conclude and summarize both forthcoming work and open questions for the future.

A preliminary version of some of our results was reported in \S7.3 of a recent white paper on long-lived particles at the LHC \cite{Alimena:2019zri}. This also included a comparison to a parton shower algorithm pushed to strong coupling (work of Marat Freytsis), which may be of interest to some readers.

In a companion paper \cite{paper2}, we will provide a more detailed understanding of the event shape observables, and establish that event isotropy captures features of events that are not easily extracted from traditional variables (thrust, eigenvalues of the sphericity tensor, and jet multiplicities).

\section{Extra dimensional simplified models: a brief introduction}
\label{sec:xdimsimple}

One approach to robust searches for new physics at colliders is the use of simplified models. An extensive summary can be found in \cite{Alves:2011wf}. Because these models are characterized by effective Lagrangians with only a few new particles, they are not representative of the rich spectra of new particles and decay chains that can arise within hidden sectors. It is unreasonable to select, by hand, the masses and couplings of large numbers of particles. Various approaches to this problem have been chosen in the literature, which we will briefly discuss below in \Sec{subsec:comparisons}. For our purposes, an efficient approach to generating simplified models with a small number of free parameters is to consider theories with an extra dimension containing a small number of bulk fields and interactions, which then produce many modes and couplings in the four-dimensional reduction.  This choice has the advantage that gauge/string duality  furnishes us with an interpretation of the extra-dimensional simplified model in terms of a toy model of a confining hidden sector at large 't Hooft coupling.

Readers familiar with RS models \cite{RandallSundrum:1999} can skim this section: the main message is that we  consider scalars with  trilinear bulk  interactions as a simplified model for the spectrum and  essential  
 interactions in the hidden sector.

\subsection{Spectrum of masses} 

We will imagine coupling the SM to a hidden sector which consists of states that propagate in at least five dimensions.  Such a hidden sector might in principle be dual to a gauge theory via the gauge/string  correspondence, and we will often use the language of this correspondence in describing it.  

Specifically, let us begin by considering a slice of (4+1)d AdS space (RS1). We will denote the extra spatial coordinate as $z$. This spacetime geometry is specified by the curvature radius $R$ of the 5d geometry, with metric
\be
{\rm d}s^2 = \frac{R^2}{z^2}(\eta_{\mu\nu}{\rm d}x^\mu {\rm d}x^\nu -{\rm d}z^2)
\label{eq:AdSmetric}
\ee
where $z_\text{UV}<z< z_\text{IR}$.  In this paper we take $z_\text{UV}=0$, in order to focus  purely on modeling the hidden sector; coupling the sector to the SM may require reintroducing $z_\text{UV}$ depending on the nature of the interaction between the two sectors.\footnote{SM fields do not propagate in the bulk, because they are not composite states of the hidden sector. One possibility would be to couple to them to the bulk fields by UV-brane localized interactions, but we will not pursue the details here.}   
A theory of fields propagating on AdS$_5$ for $z<z_\text{IR}$ is often called the ``hard-wall'' model and has been extensively studied as a model for QCD \cite{Polchinski:2001tt, Polchinski:2001ju, Polchinski:2002jw, erlich2005qcd,DaRold:2005mxj}. 

The dimensionful parameter $z_\text{IR}$ plays the role of the confinement length scale in pure Yang-Mills theory. Indeed, this type of model is a cartoon of sorts, representing more realistic string  constructions  that are dual to quasi-conformal field theories which vaguely resemble QCD.  More precisely, these field theories are asymptotically conformal at high energy (corresponding to small $z$) with a continuous coupling constant, and their conformal invariance is broken at a scale $\Lambda$ that corresponds to $z\sim z_\text{IR}$.  In some cases the breaking of conformal invariance is due to confinement.  Simple versions of the hard-wall model, like pure Yang-Mills theory, have a mass gap and towers of states, the details depending on the 5d fields that the model contains.  We will consider theories of this type below.

 For simplicity only, we will consider interacting scalars propagating in the bulk.  These could be a subset of fields in a more realistic theory, or could serve as warm-ups for gauge and/or gravity fields. The scalars will satisfy a 5d Klein-Gordon equation with mass $M$, and can be written as an infinite sum of scalar modes  that propagate in the 4d bulk modified by wavefunctions in the fifth dimension: 
\begin{equation}
\Phi\left(x^\mu, z \right) = \sum_{n=1}^\infty \phi_n \left( x^\mu \right) \psi_n \left(z\right). 
\end{equation}
The 5d wavefunctions have the form of Bessel functions: $\psi_n(z) \propto z^2 J_\nu(m_n z)$, where
\begin{equation}
\nu \equiv \sqrt{4+M^2R^2}.
 \label{eq:nuFromMass}
\end{equation}
The tower of massive Kaluza-Klein (KK) modes of the scalars can be interpreted, through gauge/string duality, as a tower of hadrons of the quasi-conformal confining 4d theory dual to the bulk description.  These hadrons are sourced by a field-theory operator ${\cal O}$ of scaling dimension $d_{\cal O} \equiv \nu + 2$. The smallest 5d  mass-squared  is set by the Breitenlohner-Freedman bound \cite{Breitenlohner:1982jf} 
\begin{equation}
M^2R^2 \geq -4.
\end{equation}
This corresponds to values of $\nu$ 
\begin{equation}
\nu \equiv d_{\cal O} - 2 \geq 0.
\end{equation}
The range $1 \leq d_{\cal O} < 2$ is allowed by unitarity but requires an alternative boundary condition  at $z_\text{UV}$ \cite{Klebanov:1999tb}, and will not be considered in this paper.
\begin{figure}[t]
\centering
\includegraphics[width=0.55\textwidth]{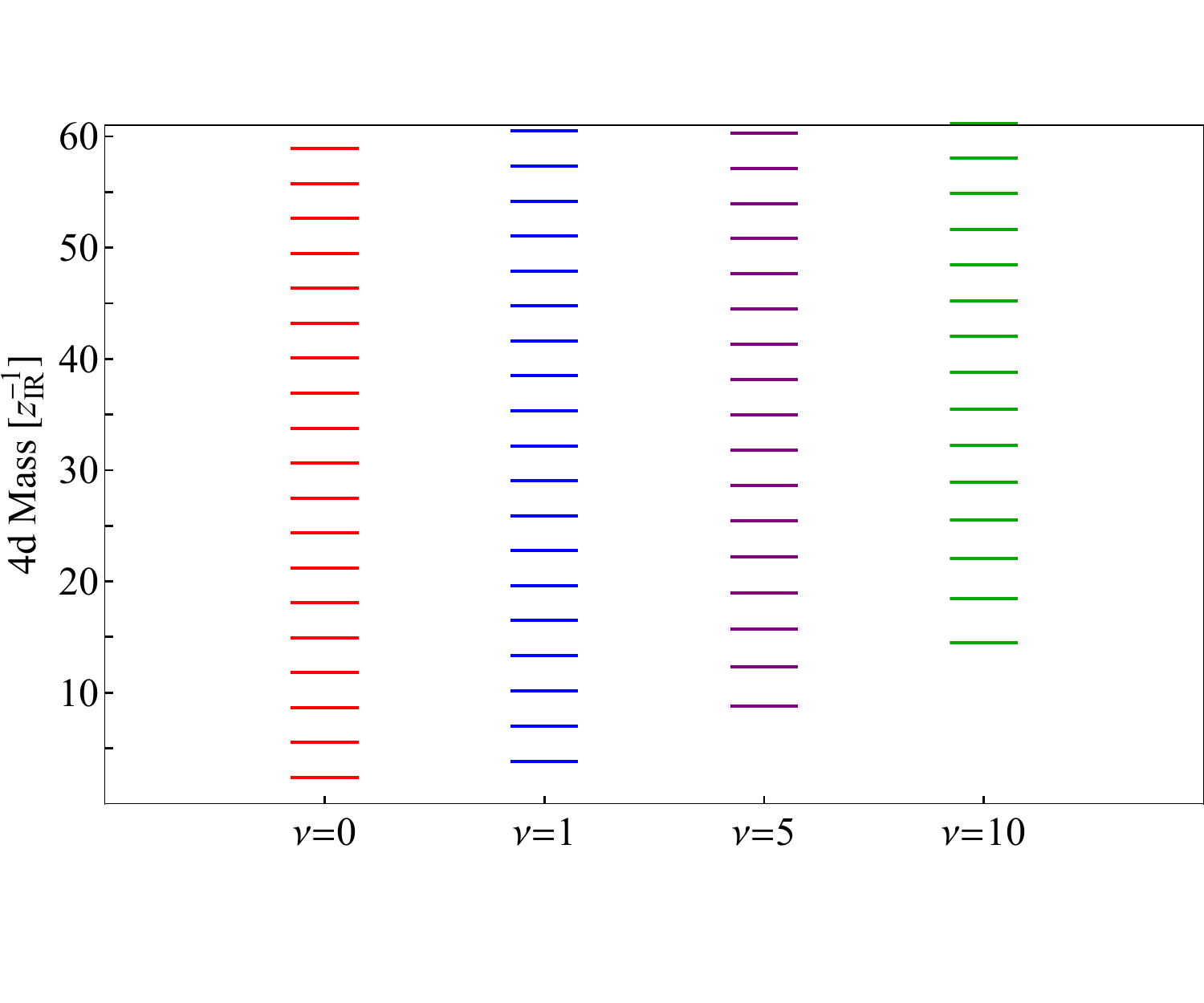}
\caption{The mass spectra of KK towers for $\nu = 0$, $1$, $5$, and $10$ respectively, in units of $1/z_\mathrm{IR}$ starting at the lowest mode. We assume Dirichlet boundary conditions on the IR brane. We highlight two important trends. First, the lowest mass in the tower increases as $\nu$ increases. Second, the mass splittings sufficiently high in each tower are approximately equal and independent of $\nu$.}
\label{fig:towers}
\end{figure}

One can estimate the masses of the Kaluza-Klein modes by using the asymptotic expansion of the Bessel function for large (positive real) arguments,
\begin{equation}
\J{\nu}{x} \approx \sqrt{\frac{2}{\cpi x}} \left[\cos\left(x - \frac{\cpi}{2} \nu - \frac{\cpi}{4}\right) - \frac{\nu^2 - \frac{1}{4}}{2 x} \sin\left(x - \frac{\cpi}{2} \nu - \frac{\cpi}{4}\right) + {\cal O}(1/x^2)\right].
\label{eq:besselasymptotic}
\end{equation}
In particular, the $n^\text{th}$ KK mode mass is approximately given by
\begin{equation}
m_n^{(\nu)} \approx \frac{\cpi}{2} (\nu + C + 2n) z_\mathrm{IR}^{-1},
\label{eq:KKmassestimate}
\end{equation}
where $C$ is an ${\cal O}(1)$ number that depends on the choice of boundary condition on the IR brane. For  Dirichlet  boundary conditions  ($\psi_n|_{z_\text{IR}} =0$), we have $C = -1/2$, while for Neumann  boundary conditions  ($\partial_z\psi_n|_{z_\text{IR}}  =0$), we have $C = -3/2$. 

The details of the mass spectrum impose constraints on particle decays; examples of spectra for different $\nu$ are given in \Fig{fig:towers}.  For the Dirichlet case, $C=-1/2$ implies that for $\nu<1/2$, a decay of KK mode $n_1$ to KK modes $n_2$ and $n_3$ of the same field, with $n_1=n_2+n_3$, is always kinematically allowed.  In particular, barring some additional constraints, the only stable mode is $n=1$.  Conversely, for $\nu>1/2$, some of these decays are always disallowed, and (to a very good approximation) $1+\floor*{(\nu+\frac{3}{2})/2}$ modes are stable against decay.  In particular, the mass spectrum for $\nu=\frac{1}{2}$  is exactly that of a 5d-massless field in a flat extra dimension, $m_n=\cpi n$; as $\nu\to\frac{1}{2}$ from below, the phase space for the decay $2\to 1+1$ closes off and the second KK mode becomes kinematically stable.  Note also that for all such spectra, a decay $n_1\to n_2+n_3$ is always forbidden for modes of the same field if $n_1<n_2+n_3$.

The wavefunction of the $n$th mode is
\begin{equation}
\psi^{(\nu)}_n(z) = N^{(\nu)}_n \,z^2 J_{\nu}(m_n^{(\nu)} z), \quad
N^{(\nu)}_n \equiv \left({\int_0^{z_\mathrm{IR}} {\rm d}z\,(R/z)^3\, \left[z^2 J_{\nu}(m^{(\nu)}_n z)\right]^2}\right)^{-\frac12} \ .
   \label{eq:wavefunctionestimate}
\end{equation}
The coefficient $N^{(\nu)}_n$ is determined by requiring that  the 4d field $\phi_n(x)$ be canonically normalized at tree level.
The Dirichlet case has a simple normalization:
\begin{equation}
N_n^{(\nu)} = \frac{1}{z_\text{IR}R^{3/2}}\frac{\sqrt{2}}{|J_{\nu+1}(m_n^{(\nu)} z_\text{IR})|}\approx 
\sqrt{\frac{\cpi m_n^{(\nu)}}{R^3 z_\text{IR}}}.\qquad \text{ (Dirichlet)}
\label{eq:dirNorm}
\end{equation}
For the Neumann case the closed form is more complicated,  but the final approximate expression in \Eq{eq:dirNorm} remains true  at large $n$. 

In the case of Neumann boundary conditions, the value of the wavefunction on the IR boundary can be approximated by:
\begin{equation}
\psi^{(\nu)}_n(z_\mathrm{IR}) \approx \frac{z_\mathrm{IR}}{R^{3/2}} \left[(-1)^{n+1} \sqrt{2} + O(1/n)\right].
 \label{eq:neumannIRvalue}
\end{equation}
This implies that a $\phi^3$ interaction localized on the IR boundary will give rise to couplings of approximately equal magnitude between any three KK modes, a  fact that we will make use of in \Sec{subsec:boundary}.

\subsection{Interaction terms}

Each scalar field in 5d provides a tower of massive particles.  To induce the decays among these particles that will create a range of signals, we next turn on interactions among the scalars.  In many models, the dominant decays are all two-body, and in such cases, the only important interactions are cubic.  We therefore consider cubic couplings of scalar fields in the 5d bulk,
\begin{equation}
\int \sqrt{g} \ {\rm d}^4x \ {\rm d}z \ \mathcal{L}_\text{int} = - \int \sqrt{g}\  {\rm d}^4x \ {\rm d}z \ c\, \Phi_{1}\Phi_{2}\Phi_{3}
\label{eq:interactions}
\end{equation}
where $\Phi_{1,2,3}$ are potentially different fields with corresponding bulk mass parameters $\nu_{1,2,3}$, and $c$ is a coupling constant.\footnote{Throughout the paper, we only use $c$ to denote this coupling constant; it should not be confused with the central charge $c$ that is often discussed in the context of AdS/CFT.}  We implicitly assume that the unboundedness of this Lagrangian is corrected by higher-order terms which make the theory well-behaved but do not affect decays.

In our studies below, we will focus only on the ``single field case'', where all three fields are the same, and the ``two field case'' where $\Phi_2=\Phi_3$.  The single-field case captures some features of self-interacting bulk fields such as a dilaton, non-abelian vectors or scalars, or the gravitational field.  The two-field case captures features of situations in which a scalar, gauge field, or gravity couples to a second field that carries charge under either a ${\bf Z}_2$ symmetry, or perhaps $\mathsf{CP}$, forbidding modes of $\Phi_2$ from being created singly.  It also is similar to cases in which the second field is complex and carries a $U(1)$ charge, since the spectrum and decay modes of $\Phi_2$ and $\Phi_2^*$ are the same in such a case.  The field $\Phi_1$ may have its own cubic interaction, but we will assume here for simplicity that its coupling is relatively small compared to the $\Phi_1\Phi_2^2$ coupling, and so plays an insignificant role in decay chains.  The cubic interaction could also be zero, as for an abelian gauge field.

We will denote the $i^\mathrm{th}$ KK mode of the field $\Phi_n$ by $\phi_{n,i}(x)$. In the 4d effective theory, the 5d interaction translates into an infinite set of couplings among the 4d modes:
\begin{equation}
\mathcal{L}_\mathrm{4d} \supset \sum_{i,j,k}^\infty c_{ijk} \phi_{1,i}(x) \phi_{2,j}(x) \phi_{3,k}(x),
\end{equation}
where the effective couplings of the 4d scalars are determined by the overlaps of the wavefunctions in the extra dimension,
\begin{equation}
c_{ijk} = c N^{(\nu_1)}_i N^{(\nu_2)}_j N^{(\nu_3)}_k \int_{z_\text{UV}}^{z_\text{IR}} \left(\frac{R}{z}\right)^5  {\rm d}z \  \big[z^2 J_{\nu_1}(m^{(\nu_1)}_i z)\big] \big[z^2 J_{\nu_2}(m^{(\nu_2)}_j z)\big]\big[z^2 J_{\nu_3}(m^{(\nu_3)}_k z)\big] \ . 
\label{eq:4dcoup} 
\end{equation}
Here we substituted $\det g = (R/z)^{10}$ for the metric \Eq{eq:AdSmetric} into \Eq{eq:interactions}.

These 4d coupling constants depend on various dimensionful 5d quantities: $c$, $R$, $M_n$, and $z_\mathrm{IR}$. However, it turns out that if we write $c_{ijk}$ in terms of the $\nu_n$ (which depend only on the product $M_n R$) and the dimensionless coupling $c_0 \equiv c \sqrt{R}$, then all remaining dependence on $R$ drops out of the equation. Hence, we never need to specify the 5d length scale $R$ to do a calculation. Furthermore, we are primarily interested in branching ratios (rather than total widths), for which the value of $c_0$ cancels out as well. Consequently, we can set both $R$ and $c_0$ to 1 for convenience. Meanwhile the scale $z_\mathrm{IR}$, corresponding to the confinement scale of a dual field theory,  is the only dimensionful quantity that appears in physical measurements. We may express masses and widths of the KK modes, and other dimensionful measurements, in units of $z_\mathrm{IR}$, and so we can also set this quantity to 1 if we choose. The only non-trivial parameters left, then, are the 5 dimensional masses, which  we express using the dimensionless parameters $\nu_i$, which in the context of a gauge/string duality are related to 4d operator dimensions.

As we will see, depending on the choices of $\nu_i$ and the boundary conditions, the $c_{ijk}$ will often (but not always) respect an approximate KK-number symmetry.  Were the extra dimension flat, it would have a conserved KK-number with Neumann boundary conditions (or with periodic boundary conditions) and an approximately conserved KK-number with Dirichlet boundary conditions.  More precisely, in the latter case, $c_{ijk}$ vanishes if $i+j+k$ is even, and falls off as $\sim 1/\Delta_\text{KK}$ when odd, where (if $i>j+k$) the violation of KK-number is $\Delta_\text{KK}= i-j-k$.  Once we replace the flat space with a slice of AdS, however, additional effects from the bulk will break the KK symmetry, sometimes leaving it approximately conserved as in the flat Dirichlet example, but  sometimes not.  

The degree and pattern of KK-number violation has an intricate structure.  Much of it emerges from the $c_{ijk}$, through the very interesting properties of the triple Bessel function integrals in \Eq{eq:4dcoup}.  Specifically, it is convenient to rewrite the integral of interest as a difference of two easier integrals:
\begin{align}
I(\nu_i, m_i) \equiv \int_0^1 {\rm d}z\, z\, \prod_{k=1}^3 \J{\nu_k}{m_k z} 
&= \left(\int_0^\infty - \int_1^\infty\right) {\rm d}z\, z\, \prod_{k=1}^3 \J{\nu_k}{m_k z}
\nonumber \\
&\equiv   I_+(\nu_i, m_i) - I_-(\nu_i, m_i)
\label{def:IplusIminus}
\end{align}
 Except right at threshold, detailed understanding of the $I_+$ and $I_-$ integrals can be obtained using approximation methods described in \Sec{sec:analytic}.  We will use specific cases in the studies of our model presented in \Sec{sec:simResults}.
 
 Additional KK-number violation can arise from kinematic constraints.  As already noted  (see \Eq{eq:KKmassestimate} and following), certain decays are kinematically forbidden in the single field case for $\nu>1/2$.  These constraints can be more complex in a two field case.

In addition, KK-number might be further violated by interactions at the  IR boundary of the space at $z = z_\mathrm{IR}$.  Specifically, in addition to or as an alternative to the bulk coupling \eqref{eq:interactions}, we could add an interaction term:
\begin{equation}
\int \sqrt{g} \ {\rm d}^4x \ {\rm d}z \ \mathcal{L}_\text{bdry} = - \int \sqrt{g}\  {\rm d}^4x \ {\rm d}z \  z_\mathrm{IR} \delta(z -  z_\mathrm{IR}) \ {\tilde c}\, \Phi_{1}\Phi_{2}\Phi_{3}.
\label{eq:bdryinteractions}
\end{equation}
This leads to nontrivial interactions if the fields satisfy Neumann boundary conditions at $z = z_\mathrm{IR}$. The boundary-localized interaction leads to approximately equal couplings among all the modes, due to \eqref{eq:neumannIRvalue}.  This contrasts with bulk couplings for small $\nu_i$, where the $c_{ijk}$ generally have more structure.

A mode's decay width is determined by the available phase space for its potential decays.
Recall that the decay width of a scalar $\phi_i \to \phi_j + \phi_k$ through a constant matrix element $c_{ijk}$ is
\begin{equation}
\Gamma = \frac{c_{ijk}^2}{16\cpi m_i^3} \lambda_{\rm PS}^{1/2}(m_i^2, m_j^2, m_k^2),
\label{eq:width}
\end{equation}
where the phase-space function $\lambda_{\rm PS}$ is defined as 
\begin{equation}
\lambda_{\rm PS}(m_1^2, m_2^2, m_3^2) \equiv (m_1 + m_2 + m_3)(m_1 - m_2 + m_3)(m_1 + m_2 - m_3)(m_1 - m_2 - m_3).
  \label{eq:lambdaPS}
\end{equation}
We will see cases where near-threshold decays are favored, because approximate KK-number conservation in the $c_{ijk}$ overcompensates the phase space suppression.  This situation leads to near-spherical distributions in decay chains.  In other cases this is not so, and decay chains lead to much less spherical events.

\subsection{Comparison to other high-multiplicity models in the literature}
\label{subsec:comparisons}

The use of a warped extra dimension to provide a model for a dark or hidden sector is natural following \cite{Randall:1999vf,Verlinde:1999fy}, and is well-established in the literature \cite{Stephanov:2007ry, Strassler:2008bv, Falkowski:2008fz, Gherghetta:2010cq, Bunk:2010gb,  McDonald:2010iq, McDonald:2010fe}. In this subsection, we will briefly comment on some related or alternative approaches to modeling high-multiplicity hidden sectors. 

Recently, cascade decays in warped dark sectors have been discussed in a series of papers \cite{Fichet:2019hkg, Brax:2019koq, Costantino:2020msc}. While the basic RS framework of these papers is similar to ours, they have focused especially on the regime in which individual KK modes become sufficiently broad that they should be described as a continuum, rather than as narrow resonances. On the other hand, all of our calculations will be done in the regime in which each KK mode is narrow and we can model a cascade decay as a sequence of $1 \to 2$ decays. This can be done consistently provided that our bulk couplings are sufficiently small, while at the same time not {\em so} small that resonances acquire a long lifetime and alter the collider phenomenology. (The long-lived regime may be of independent interest, but its additional complications are beyond the scope of this paper.) In every case that we consider, the couplings can be chosen so that this consistency condition is met. In particular, in every case the width-to-mass ratio of the $n^\text{th}$ KK mode grows more slowly than it would in a model with decays determined by pure phase space. With pure phase space decays, the $n^\text{th}$ KK mode has of order $n^2$ decay modes available to it, with comparable widths. On the other hand, the partial width of a given decay mode scales as in \eqref{eq:width}, i.e., roughly as $1/m_n$. As a result, the {\em total} width of the KK mode scales linearly with the mode number, and the width-to-mass ratio is constant. If we take the bulk couplings to be small but not too small, the width-to-mass ratio of each mode will be small but every unstable KK mode will still decay promptly on collider time scales. The regime of small bulk couplings is where gauge/string duality is well-understood. If the bulk couplings are too small, there is a potential ``empty universe problem'' in cosmology related to the slow first-order confining phase transition in the dark sector \cite{Creminelli:2001th,Randall:2006py,Kaplan:2006yi}. However, the details of such a phase transition can be model-dependent (see, e.g., \cite{Agashe:2019lhy}). We assume that potential cosmological problems can be addressed without qualitatively altering the collider phenomenology.

Other recent work \cite{Dienes:2019krh, Dienes:2020bmn} has studied dark sector particles with simple ans\"atze for the masses and couplings, e.g., $m_n = m_0 + n^\delta \Delta m$ for some positive exponent $\delta$, and couplings depending on factors $(m_\ell - m_i -m_j)^r$ and $(1 + |m_i - m_j|/(\Delta m))^{-s}$ favoring decays to lighter daughter particles or nearby daughter particles, respectively. The model of \cite{Dienes:2019krh} involves neutral particles $\chi_n$ decaying to ${\bar q} q' \chi_l$, and leads to high-multiplicity event shapes with missing momentum, which are qualitatively similar to events we will discuss.  The fully dark decay chains discussed in \cite{Dienes:2020bmn} are even more similar to ours, but are discussed in  the context of cosmology rather than collider physics. It is no accident that these decay chains have similar properties, as the ans\"atze used have been motivated, in part, by models of extra dimensions \cite{Dienes:2011ja,Dienes:2011sa,Dienes:2012jb,Buyukdag:2019lhh}. 

Further models for high-multiplicity events have been proposed based on a variety of ideas. SUEP events have been modeled using thermal spectra \cite{Knapen:2016hky}. Models of black hole production  at colliders predict similar spectra \cite{Giddings:2001bu, Dimopoulos:2001hw}, as do ``string balls,'' lower-mass precursors of black holes \cite{Dimopoulos:2001qe, Gingrich:2008di}.  Another model with interesting phenomenology, albeit without a well-motivated UV completion, achieves a wide range of couplings within a large ensemble of particles through a random mass matrix \cite{DAgnolo:2019cio} (see also \cite{Dienes:2016kgc,  Craig:2017ppp, Tropper:2020yew}). It would be interesting, in the future, to apply the event isotropy variable to more of these models.

\section{Simulation Results}
\label{sec:simResults}

In this section we study event shapes of our toy model for different parameters, by simulating decay cascades of a heavy KK mode with $n=n_p\gg 1$, in its rest frame, to light and stable KK modes.
(We will often refer to the KK modes as ``hadrons,'' using the dual viewpoint, but one should keep in mind that these are {\it hidden-sector} hadrons, not SM hadrons.) 
We further decay the hidden-sector stable hadrons (HSH) into two massless particles each, to mimic decays back into the SM.  Then we calculate observables from the collection of massless, final-state momenta. 

In this simplified model, mass and KK-number are closely related, 
as we have seen in \Eq{eq:KKmassestimate}.  It follows that the degree of violation of KK-number is directly tied to kinematics. Roughly, if KK-number is conserved or lightly violated in a two-body decay, the final state particles tend to be slow in the initial particle's rest frame, while if  KK-number is strongly violated, the final state particles are produced with a substantial boost.  It is not surprising then that  KK-number violation correlates closely with event-shape variables, as we will show in this section.  

In extreme limits, it is clear how this should work.  
Were KK-number precisely conserved in all decays, then every decay would occur at or near threshold, and the final state of the hidden sector cascade would be a collection of slow HSHs.  When the HSHs themselves decay to the visible sector, they would produce an array of roughly back-to-back massless particles, produced at random angles.  The expected distribution of such particles is roughly spherical.  Note, however, that even with dozens of particles, random fluctuations are large and observed events are far from spherical, both to the eye and to event shape variables. 

Conversely, in cases with large KK-number violation, the first decay in the cascade produces two relatively light KK states at high boost.  Once this occurs, all ensuing decays of the lighter daughters will be highly collimated, and so, independent of the details, two hard jets result.\footnote{These jets, neither pencil-like  nor QCD-like, will have opening angles and subjets that depend on the kinematics of the initial steps in the cascade, especially on the boost of the initial decay's daughters.} Thus KK-number violating decays early in the cascade leads to a highly non-spherical pattern. 

The simulations that we describe in this chapter interpolate between these extremes.  We will demonstrate this using thrust and event isotropy, both of which are sensitive to the features of the events.  

These variables are somewhat correlated with a third, namely particle multiplicity.  The possible maximum particle multiplicity in our simulations is 2$n_p$.  This occurs when KK-number is conserved in every decay, and all hidden hadrons can decay except the $n=1$ state, the unique HSH.  Then the decay cascade leads to $n_p$ HSH's, and  to $2n_p$ massless particles once these decay.  Violations of KK-number in the cascade, and the existence of multiple HSH's with $n>1$, will decrease this number.  This tends to increase the event isotropy, since, as noted in \Eq{eq:isoest}, for a multiplicity $k<192$, $\iso{sph}{192} \gtrsim \sqrt{2/k}$.  (The average isotropy tends to be higher than this estimate, which holds for maximally symmetric events, because of random fluctuations in the angles.) Despite this we will see that isotropy and particle multiplicity are not redundant.

We now present our results, progressing from the most isotropic scenario to the least.   
For each choice of bulk masses and couplings, we generate $10^4$ events starting at KK mode $n_p = 100$ and allow it to cascade into stable hadrons, each of which then decays to a pair of massless particles. In each case, we will see that the degree of KK-number conservation, as reflected in the couplings $c_{ijk}$ and the resulting branching fractions, determines the qualitative properties of the event shapes.  

Note that we mainly limit ourselves to small values of $\nu$.  This is because scalars with large 5d mass correspond to 4d operators with large scaling dimension ($d_{\cal O}=\nu+2$), and it is difficult to imagine coupling the SM to them.

\subsection{Spherical and Near-Spherical Cases}
\label{sec:sphereNearsphere}

To set a baseline, we begin with the most spherical case in our AdS-based simplified model, a single-field model with $\nu=0$. This case corresponds to a five-dimensional scalar with mass-squared $-2$, at the Breitenlohner-Freedman bound, and thus to a dual CFT operator of dimension 2 with a non-zero three-point function.  We will compare it a pair of toy models in which KK-number is exactly conserved and spherical events are to be expected.  The event shapes for $\nu=0$ are virtually the same as for the toy models, despite the former's mild KK-number violation and semi-relativistic velocities.

The $\nu=0$ single field model has a spectrum approximately given by $m_n \approx \cpi \left(n-\frac{1}{4}\right)$. The only HSH is the mode with $n=1$; all decays $\phi_i \to \phi_j + \phi_k$ are open if $i\geq j + k$.

 For any single field with a cubic self-interaction and even integer $\nu$, the  couplings of its modes satisfy a simple approximate formula.  
As noted in \Eq{def:IplusIminus}, the triple Bessel integral \Eq{eq:4dcoup} can be conveniently written as a difference of integrals.   For even $\nu\geq 0$, $I_+$ vanishes due to a factor of $1/\Gamma(-\nu/2)$ which can be seen in \Eq{eq:GandR}.  Meanwhile $I_-$ is approximately given by \Eq{eq:IminusDirichletSmallPS}. Using \Eq{eq:dirNorm}, we find
\be
c_{ijk} \approx   \frac{(-1)^{n_1+n_2+n_3+1}}{\sqrt{2}} \frac{8 m_i m_j m_k}{\lambda_\text{PS}(m_i^2,m_j^2,m_k^2)}
\label{eq:cijk_nu_even}
\ee
While this  is accurate only away from threshold, for $\nu=0$ the approximation works to within 2\% percent except for $n_2+n_3=n_1-1$, where the real coupling is smaller in magnitude by up to $5\%$, and $n_2+n_3=n_1$, where the difference reaches nearly $30\%$.

\begin{figure}[t!]
\centering
\subfloat[]{
 \includegraphics[width=0.45\textwidth]{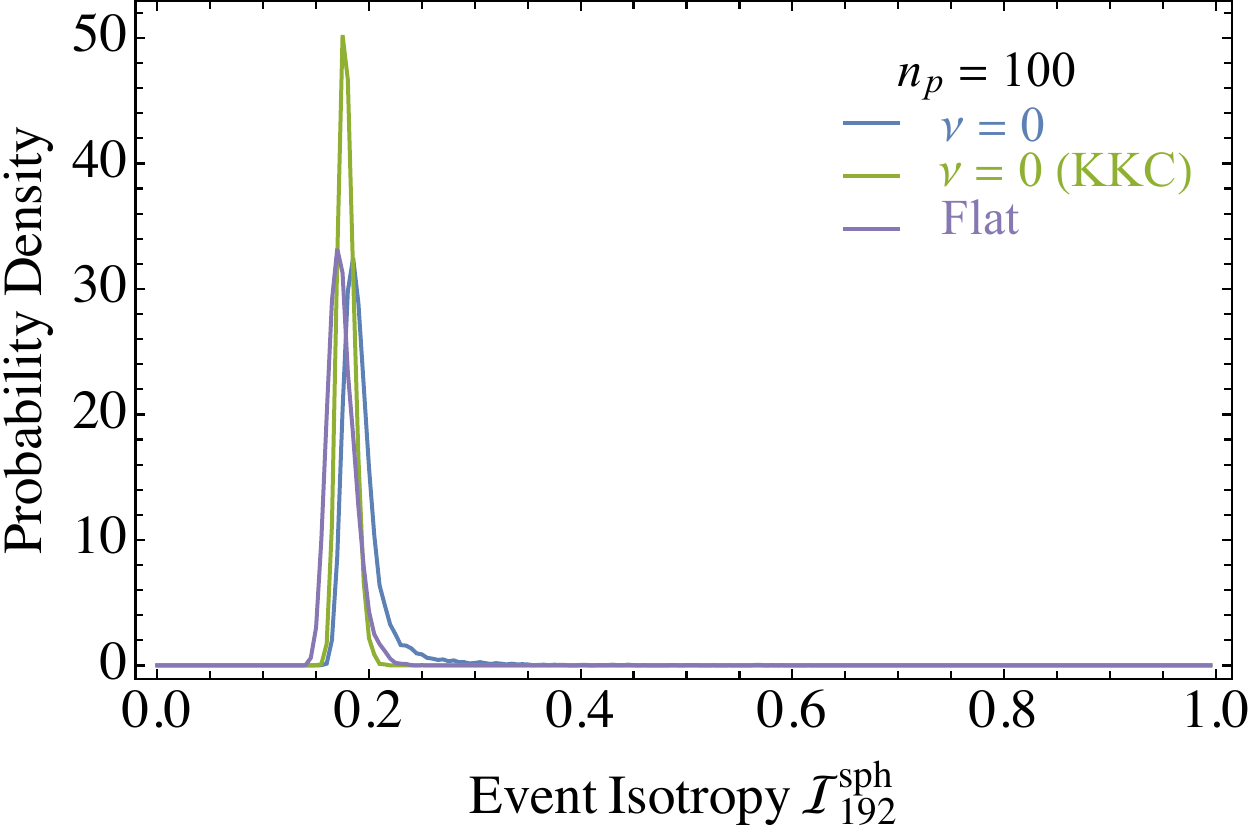}
}
\hfill
\subfloat[]{
 \includegraphics[width=0.45\textwidth]{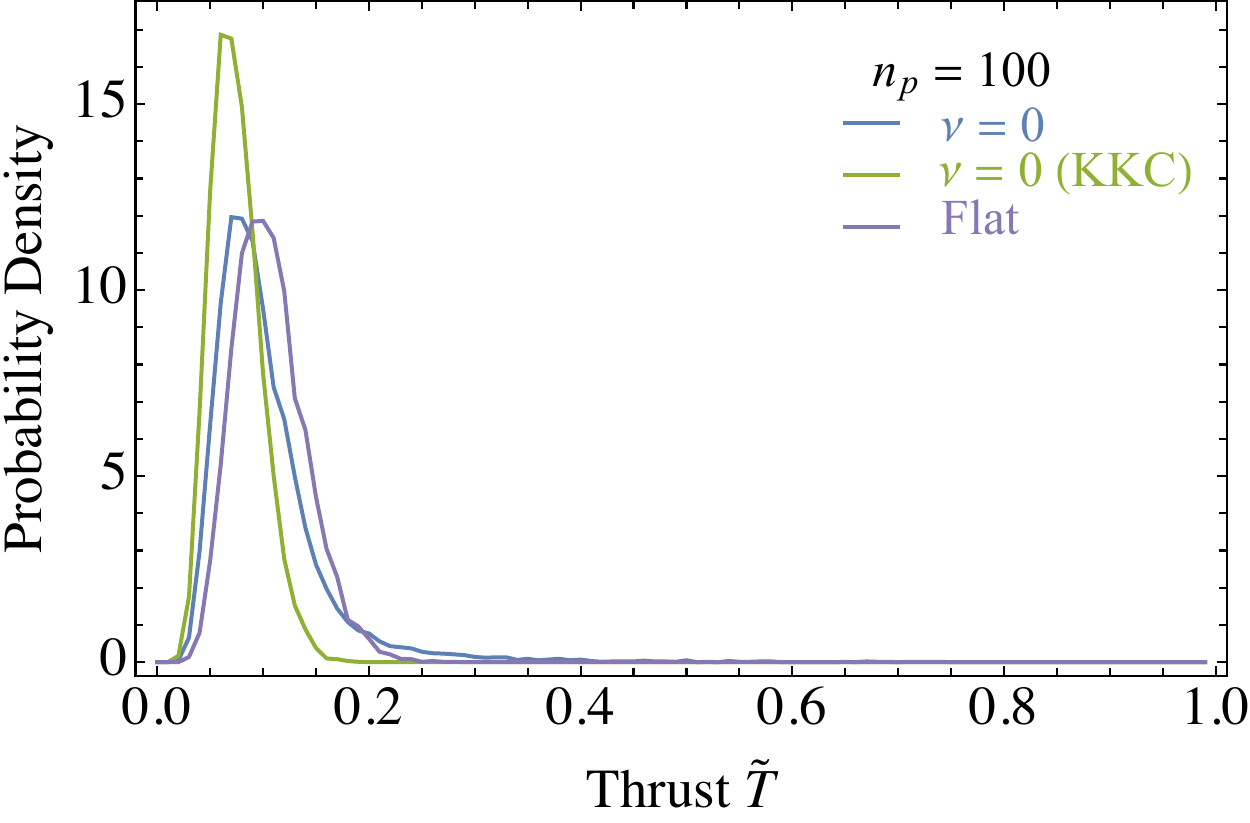}
}
\hfill
\subfloat[]{
 \includegraphics[width=0.45\textwidth]{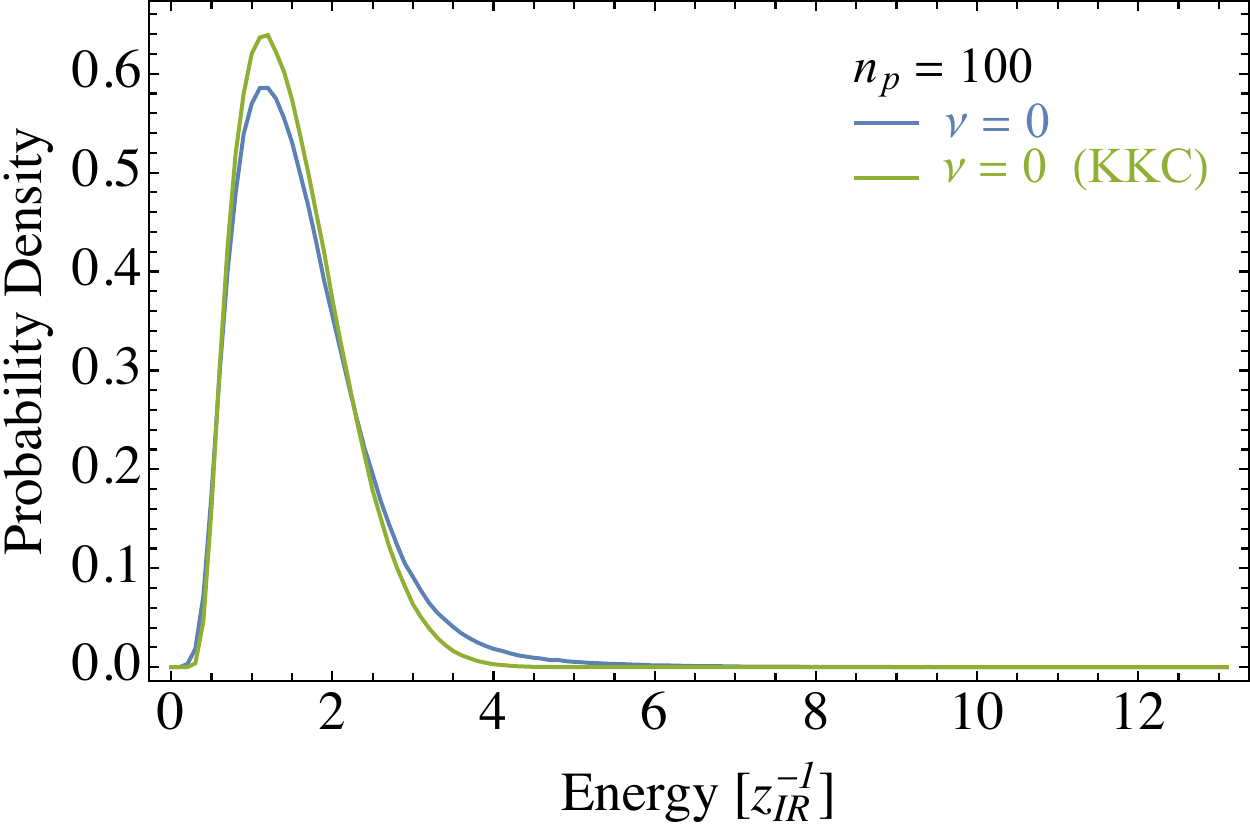}
}
\caption{The (a) event isotropy, (b) thrust, and (c) energy spectra of the final state for cascades generated with $\nu=0$ with and without KK-number conservation, and the flat case. 
The multiplicity is not shown as both the KK-number conserving $\nu=0$ and the flat case distributions are delta functions at 200 particles.
The energy spectrum of the flat case is not plotted as it is a delta function at $\cpi/2$; the other two spectra peak at approximately half the mass of the HSH.}
\label{fig:toyModels}
\end{figure}

From this formula it follows that partial widths behave as $\lambda_\text{PS}^{-3/2}$, and so decays tend to occur at or very near threshold.   This implies that  the leading decays for each hadron conserve KK-number.  (For instance, the particle with $n=n_p=100$ has a 77\% branching fraction to conserve KK-number, and this varies slowly with $n_p$.)  Even those decays that violate KK-number do so by small amounts, and in the end the average HSH multiplicity at the end of the cascade is reduced only to 93  from its maximum of 100; this is shown later in \Fig{fig:singleFieldDist}.  The decays of the HSHs produce nearly 200 massless particles with an energy distribution, shown in \Fig{fig:toyModels}(c); note that $m_1\approx 2.40\approx m_{100}/130$, and the distribution peaks at about $m_1/2$, with a tail up to $\sim 2m_1$.   Thus velocities of the HSHs tend to be only semi-relativistic, and the angular distribution of their massless daughters is largely random.

Our first toy model has the same masses and KK-number-conserving couplings as the $\nu=0$ case just described, but we set all KK-number violating couplings to zero by hand:
\be
c_{ijk} = 0 \ \ (i\neq j+k) \ .
\ee
We will call this the KK-conserving (KKC) $\nu=0$ model.  Its final state consists of exactly 200 massless particles, with an energy distribution slightly narrower than the full $\nu=0$ model, as the latter has KK-number-violating decays with more kinetic energy.  We expect it to have slightly more spherical events.

The second toy model (the ``flat case'') is a single field on Minkowski space times an interval, $M_4\times S^1/Z_2$, with Neumann boundary conditions on the field.  As noted earlier, the spectrum  has $m_n\propto n$ exactly, and KK-number is conserved.  Strictly speaking all particles are marginally stable, but we imagine deforming the model infinitesimally so that all decays can occur.   The final state from an initial heavy hadron with quantum number $n_p$ consists of $n_p$ HSHs with $n=1$, all at rest.  The decay to SM massless particles then produces events with exactly 200 massless particles in back-to-back pairs, distributed randomly in angle.  Each particle has energy exactly $m_1/2 = \frac{\cpi}{2}$.

Now we compare the event shapes for these three cases. 
In \Fig{fig:toyModels} we show the distributions in energy, event isotropy, and thrust for the particles in the final state. 
For consistency in range, we plot the scaled thrust $\tilde{T}$ 
\begin{equation}
\tilde{T} \equiv 2 (T-1)
\label{eq:ScaleThrust}
\end{equation}
such that all variables have a range of $[0,1]$ with $0$ being the most isotropic and $1$ being the least.

All three of these examples are very similar as seen by event-shape variables.  The most notable differences are percent-level shifts in event shapes, in tails that arise from small numbers of somewhat less isotropic events.   We may therefore treat any one of them as a benchmark against which to compare other cases. 

All three cases have event isotropy that peaks in the range 0.15-0.20.  
From \Eq{eq:isoest}, maximally isotropic events with $192$ particles $\mathcal{U}^\text{sph}_{192}$ would have
\be
\iso{sph}{}\left(\mathcal{U}^\text{sph}_{192}\right) \approx \sqrt{\frac{2}{192}} \approx 0.10
\ee
Naively we might have expected the flat case, with 200 particles of equal energy and random angles, to approximate this value.
However, the random fluctuations in angle (but not in energy, which remains $m_1/2$ for each particle) lead to a significantly higher event isotropy, closer to $0.16$; we do not know a method to compute this number without simulation.\footnote{For KK-number conserving scenarios, events become more spherical at high $n_p$, as shown in App.~\ref{app:npdependence}, and the theoretical limit is reached at large $n_p$.}  Despite the wider energy distribution of the $\nu=0$ and KKC cases, their event isotropies are quite similar to the flat case.  We will explore the causes of this in \cite{paper2}.

\subsection{Single field, general $\nu$}
\label{subsec:singlefield}

Next, still studying a single scalar field with Dirichlet boundary conditions, we consider other values of $\nu$.  Each has a different degree of KK-number violation.
Although the amount of KK-number violation is still relatively small, and the decays are still mostly close to threshold, the effects are large enough to observably shift event shapes relative to the $\nu=0$ benchmark.

As noted in \Eq{def:IplusIminus}, the integral \Eq{eq:4dcoup} can be written in terms of a difference of two integrals $I_+$ and $I_-$.  Substituting $\nu_i=\nu$ into \Eq{eq:Iplus2F1} and using the Euler reflection formula, one finds  
\be
I_+ = \sin\left(\frac{\nu}{2}\cpi\right) \left(\frac{m_2m_3}{m_1^2}\right)^\nu
{\lambda_\text{PS}^{-1/2}} H(\nu,m_i)
\label{eq:Iplussingle}
\ee
where $H$ is a function whose dependence on $\nu$ and the $m_i$ is subleading compared to the terms shown explicitly.\footnote{The full integral $I$ in \Eq{def:IplusIminus} is elementary at $\nu=\frac{1}{2}$, and the ${\lambda_\text{PS}^{-1/2}}$ factor is easily seen there.}

When $\nu$ is an even integer, $I_+$ vanishes, so the couplings are determined entirely by $I_-$.  Since, from \Eq{eq:IminusDirichletSmallPS}, $I_-\sim \lambda_\text{PS}^{-1}$ near but below threshold, the branching fractions decrease as $\sim {\lambda_\text{PS}^{-3/2}}$, strongly suppressing KK-number violation.  For $\nu=0$, where there are no other sources of KK-number violation, this gives quasi-spherical events.    For general $\nu$, however, $I_+\sim {\lambda_\text{PS}^{-1/2}}$ falls off more slowly away from threshold than $I_-$.  Although for $\nu$ very large this does not matter, because the power of the mass ratio in \Eq{eq:Iplussingle} decreases rapidly with $\nu$ and tends to disfavor decays to light hadrons, for $\nu\sim 1$ one finds $|I_+|\sim |I_-|$, and so KK-number violation in the couplings is considerably larger than for $\nu\approx 0,2,4,\dots$.   

The dominant decays remain KK-number conserving for $\nu\lesssim 0.1$.   For the $\nu=0.15$ case below, KK-number violation in the couplings is already significant, and reduces the multiplicity of HSHs well below the $\nu=0$ benchmark, as we will see shortly.

For $\nu\geq 0.5$, a new effect reduces multiplicity further: $m_n < m_{n'} + m_{n-n'}$, so KK-number conserving decays all become kinematically forbidden, and the minimal amount of KK-number-violation per decay is $> 0$, even for even integer $\nu$.  This reduces the number of typical decays in the cascade and the number of hadrons at the end of the cascade. However, even though KK-number conservation is forbidden, decays with large KK-number violation are still somewhat suppressed, and so the leading decays are {\it minimally KK-number-violating} --- that is, they have the smallest amount of violation consistent with kinematic constraints.  These decays are generally the ones closest to threshold.

When decays  can typically only violate KK-number by a small amount (per decay), they remain near kinematic threshold, so boosted hadrons and ensuing jetty structures in the events do not arise.  Nevertheless, $\iso{sph}{192}$ increases.  In part, this is due to a decrease in particle multiplicity. The total amount of KK-number $\Delta_{\text{KK},\text{tot}}$ lost in the cascade (equal to the sum over decays of the KK violation $\Delta_{\text{KK}, s}$ in each decay $s$) is given by the parent mode number $n_p$ at the beginning of the cascade minus the sum over KK-numbers $n_i$ of the $N_{\text{HSH}}$ HSHs in the final state, 
\begin{equation}
\Delta_{\text{KK},\text{tot}}=
 \sum\limits_s^\text{S} \Delta_{\text{KK}, s}=
n_p - \sum\limits_{i}^{N_\text{HSH}} n_i \ ,
\end{equation}
where 
$S$ is the total number of decays in the cascade.  The reduced final-state hadron multiplicity $N_{\text{HSH}}$ increases the minimum value of $\iso{sph}{192}$ by a factor of order $\sqrt{100/N_{\text{HSH}}}$, from \Eq{eq:isoest}, even if the HSHs are rarely boosted and their final decay products are quasi-isotropically distributed.

If KK-number is exactly conserved, $S=n_p-1$ and $N_{\text{HSH}}=n_p$.  More generally $N_{\text{HSH}}\leq n_p-
\Delta_{\text{KK},\text{tot}}
$, with the equality holding only if the only HSH has $n=1$. This is true for $\nu<\frac{1}{2}$, and in particular for the case $\nu=0.15$ that we show below.  For larger $\nu$, there are HSH's with $n>1$, so $N_{\text{HSH}}$ is even smaller; and on top of this,   decays with $\Delta_{\text{KK}, s}=0$ are forbidden, so 
$\Delta_{\text{KK},\text{tot}}$ is of the same order as $S$ and $n_p$, leading to a substantial reduction in $N_{\text{HSH}}$.  For the $\nu=0.75$ case we show below, $S\sim n_p/2$ and $N_{\text{HSH}}<n_p/2$, leading to a reduction of the multiplicity by more than half compared to the $\nu=0$ case, and a corresponding substantial increase in $\iso{sph}{192}$ .

\begin{figure}[t!]
\centering
\subfloat[]{
       \includegraphics[width=0.45\textwidth]{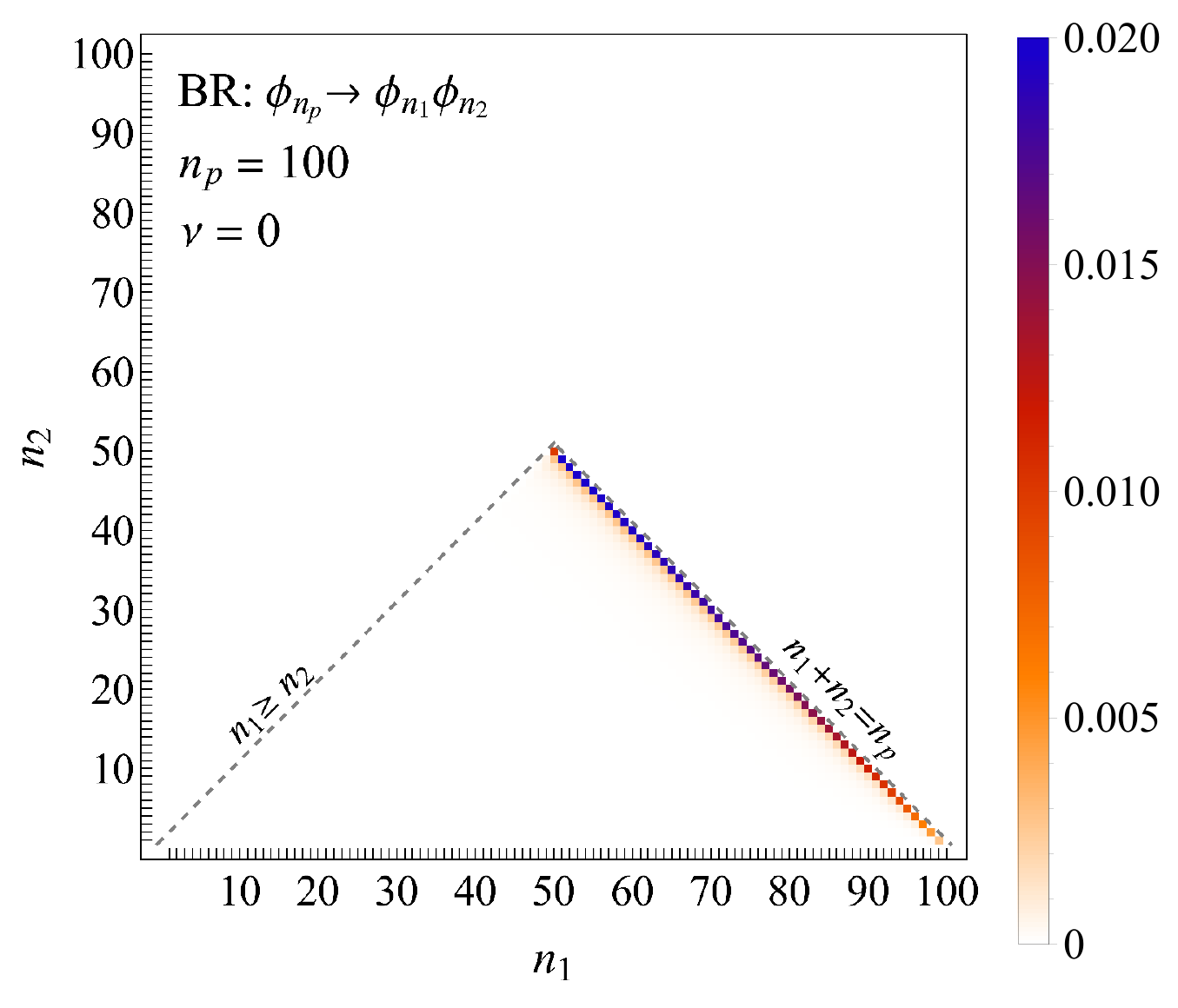}
     }
     \hfill
\subfloat[]{
       \includegraphics[width=0.45\textwidth]{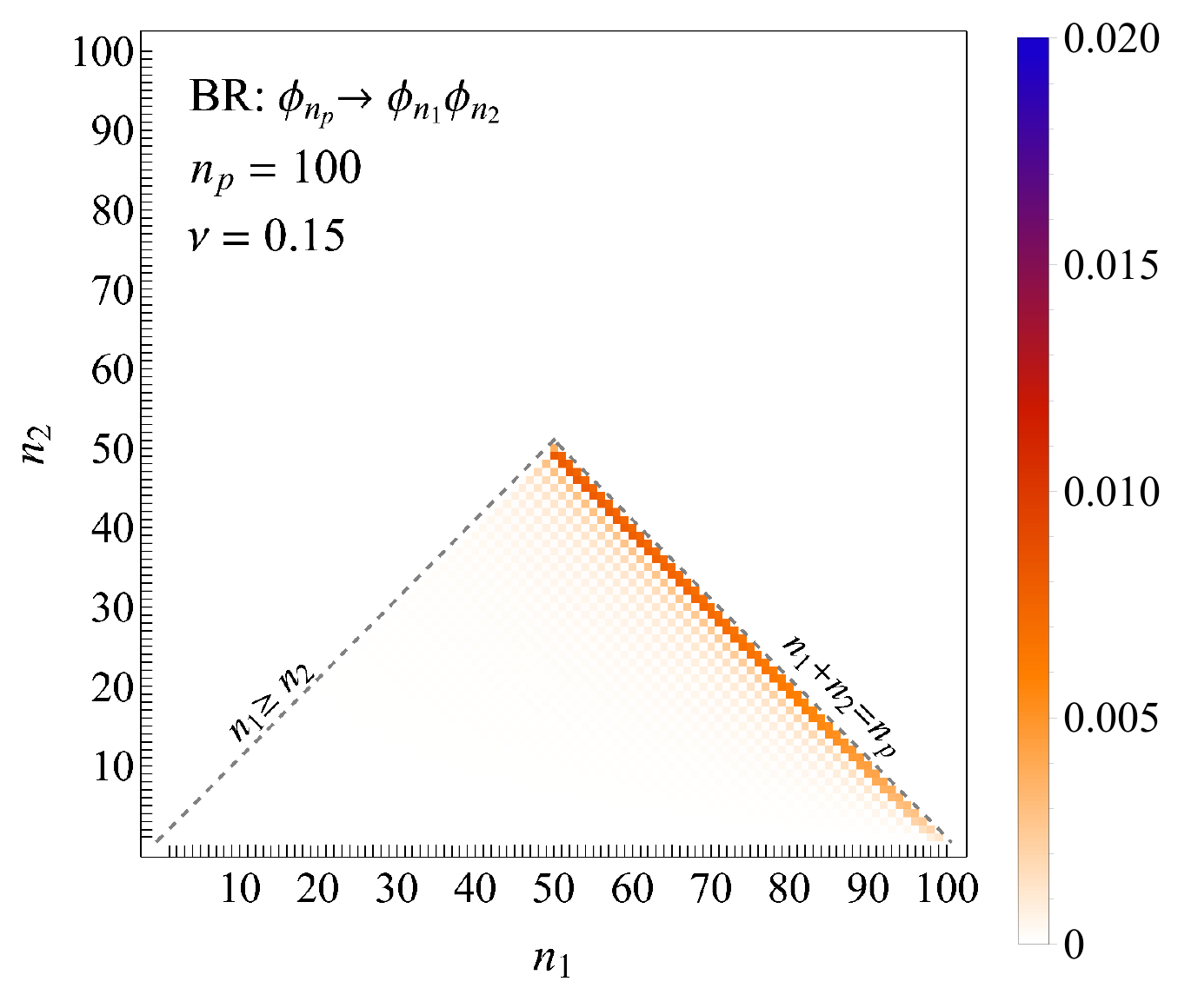}
     }
    \hfill
\subfloat[]{
       \includegraphics[width=0.45\textwidth]{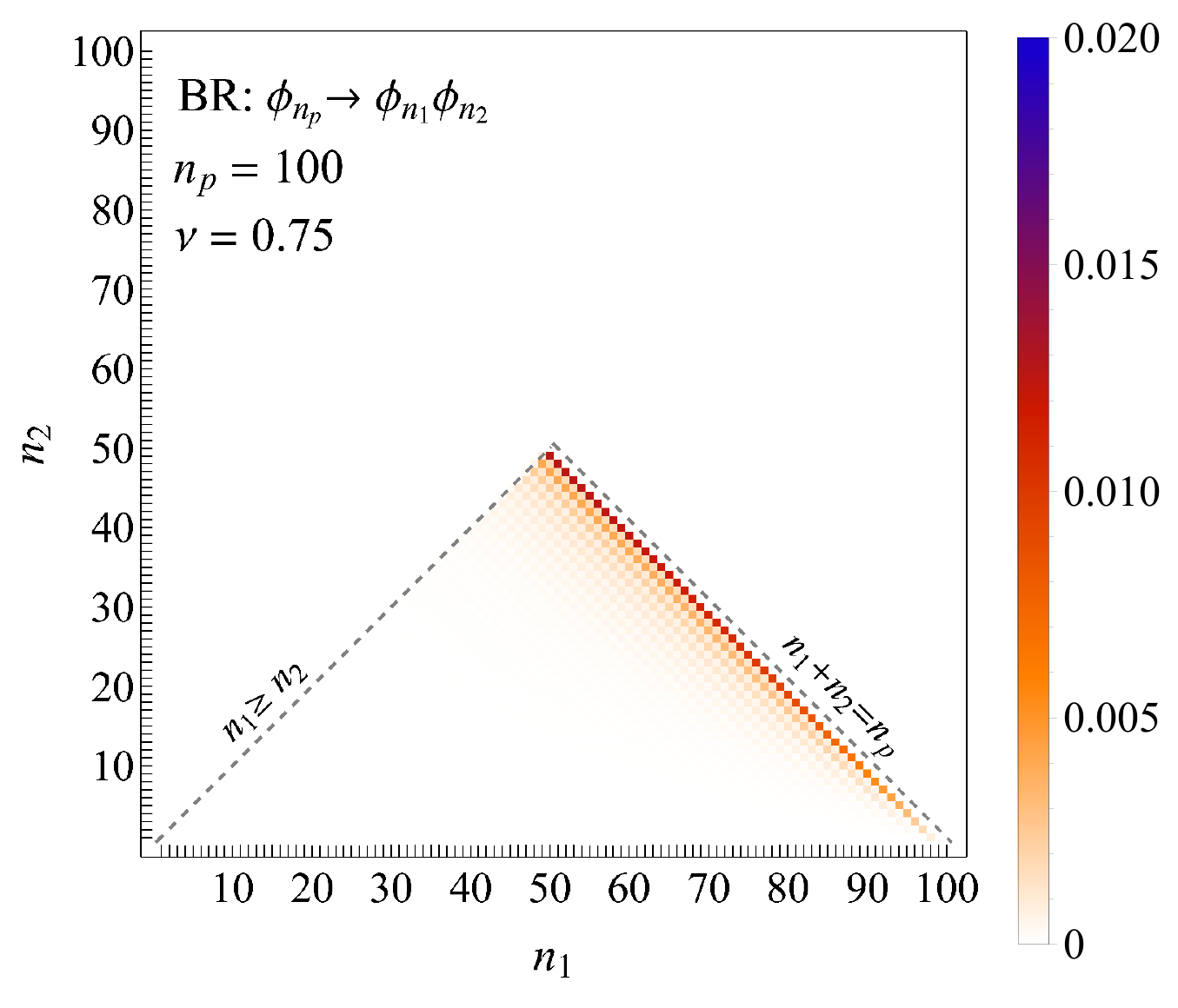}
     }
     \hfill
\subfloat[]{
\includegraphics[width=0.45\textwidth]{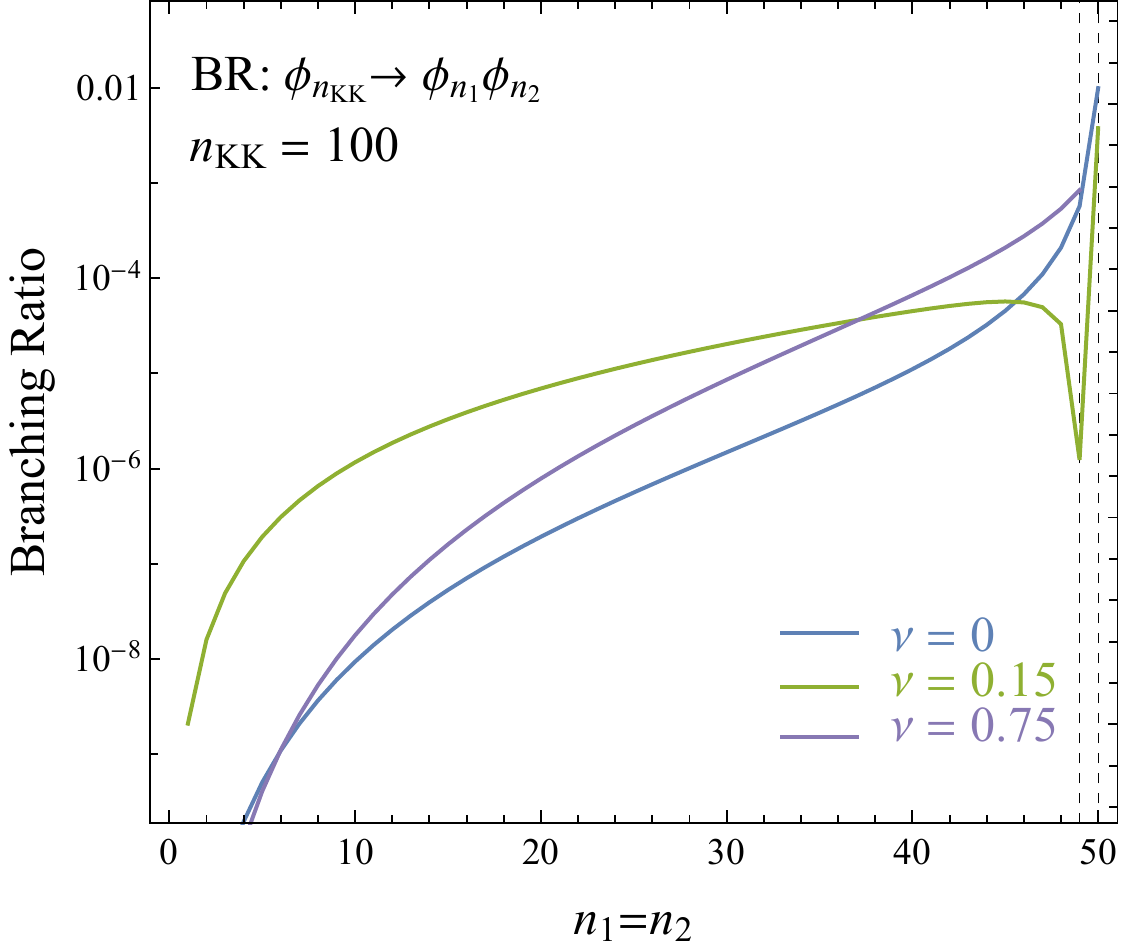}
}    
\caption{(a--c) Branching ratios, for $\nu=\{0, 0.15, 0.75\}$, of the $100$th KK mode into all kinematically allowed two-body final states, as a function of the daughter KK-numbers $n_1 \geq n_2$. 
Dominant decays occur at or near kinematic threshold, along the line of minimum KK violation. 
The projection of the branching ratios for all values of $\nu$ on the $n_1=n_2$ line is shown in (d), where the relative suppression of KK violation is made obvious especially for the $\nu=0$ case.}
\label{fig:brratios}
\end{figure}

The branching ratios for the $n_p=100$ mode into daughter modes $n_1$, $n_2$ with $\nu = \{0, 0.15, 0.75\}$ are shown in \Fig{fig:brratios}. For convenience we will refer to these plots throughout the paper as ``branching fraction triangles." For $\nu=0$, the only HSH is the $n=1$ state at the bottom of the tower, and conservation of KK-number is both kinematically allowed for all decays in the cascade and dominant; as noted earlier, about 77\% of decays are KK-number conserving. 
For $\nu=0.15$, KK-number conservation is kinematically allowed, but the probability of conservation in the decay of heavy modes is only $\sim 30\%$. 
Decays with KK-number violation $\Delta_\text{KK} = 1$ occur with comparable probability ($\sim 30\%$).  Technically, this is due to a cancellation in the integral for $c_{ijk}$;  the integrals $I_+$ and $I_-$ are similar in magnitude for $\nu \sim 0.1-1.9$ and interfere destructively (constructively) when $i+j+k$ is even (odd).  
The checkerboard pattern in \Fig{fig:brratios} (b,c) arises from this effect.

The pattern of branching ratios for $\nu = 0.75$ is similar to that of $\nu=0.15$.  However, $\Delta_\text{KK}=0$ decays are kinematically forbidden, as can be seen by careful examination of the upper right edge of the triangle.

While \Fig{fig:brratios} applies for $n_p=100$, it illustrates qualitative features that apply, at fixed $\nu$, for smaller values of $n$, and thus for the whole cascade.  This is because the integrals $I_+$ and $I_-$ have relatively simple behavior under changes of $n$; see App.~\ref{app:npdependence} for some discussion.  


\begin{figure}[t!]
\centering
\subfloat[]{
\includegraphics[width=0.45\textwidth]{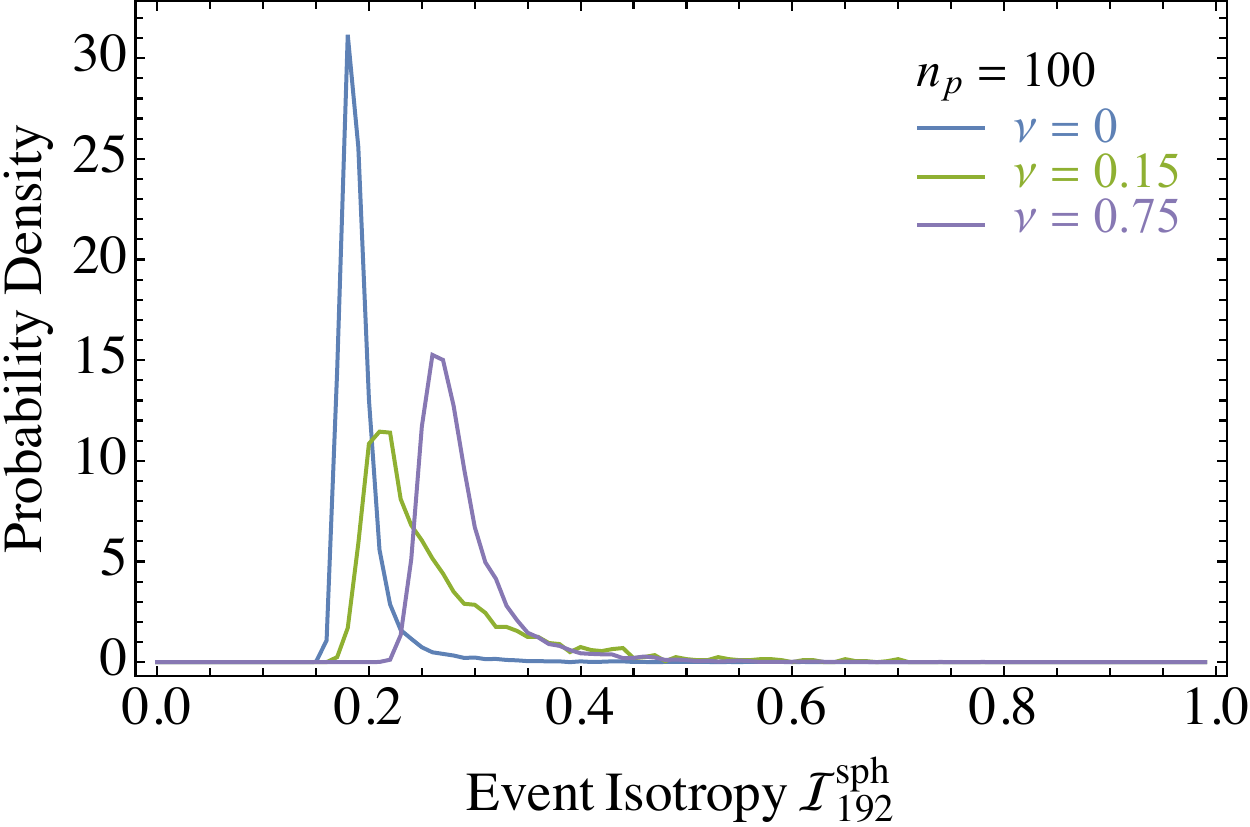}
}
\hfill
\subfloat[]{
\includegraphics[width=0.45\textwidth]{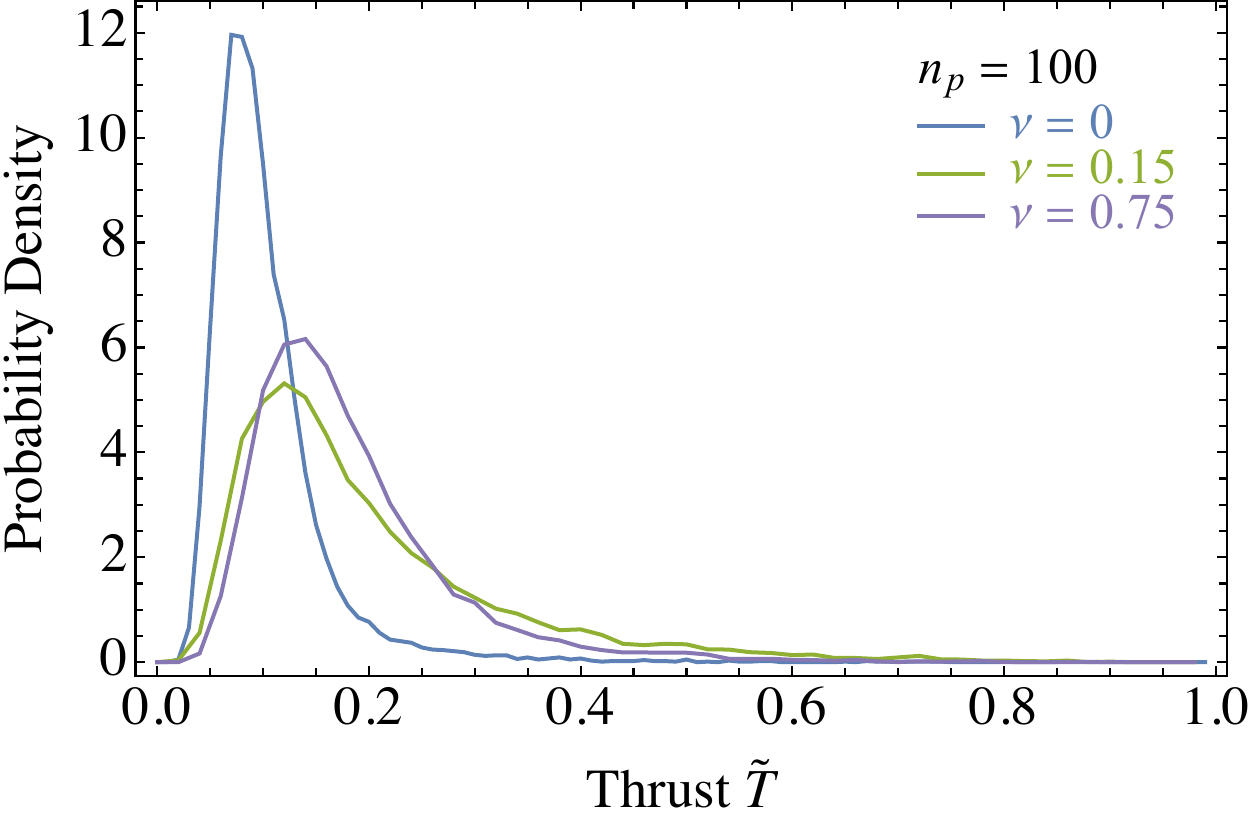}
}
\hfill
\subfloat[]{
\includegraphics[width=0.45\textwidth]{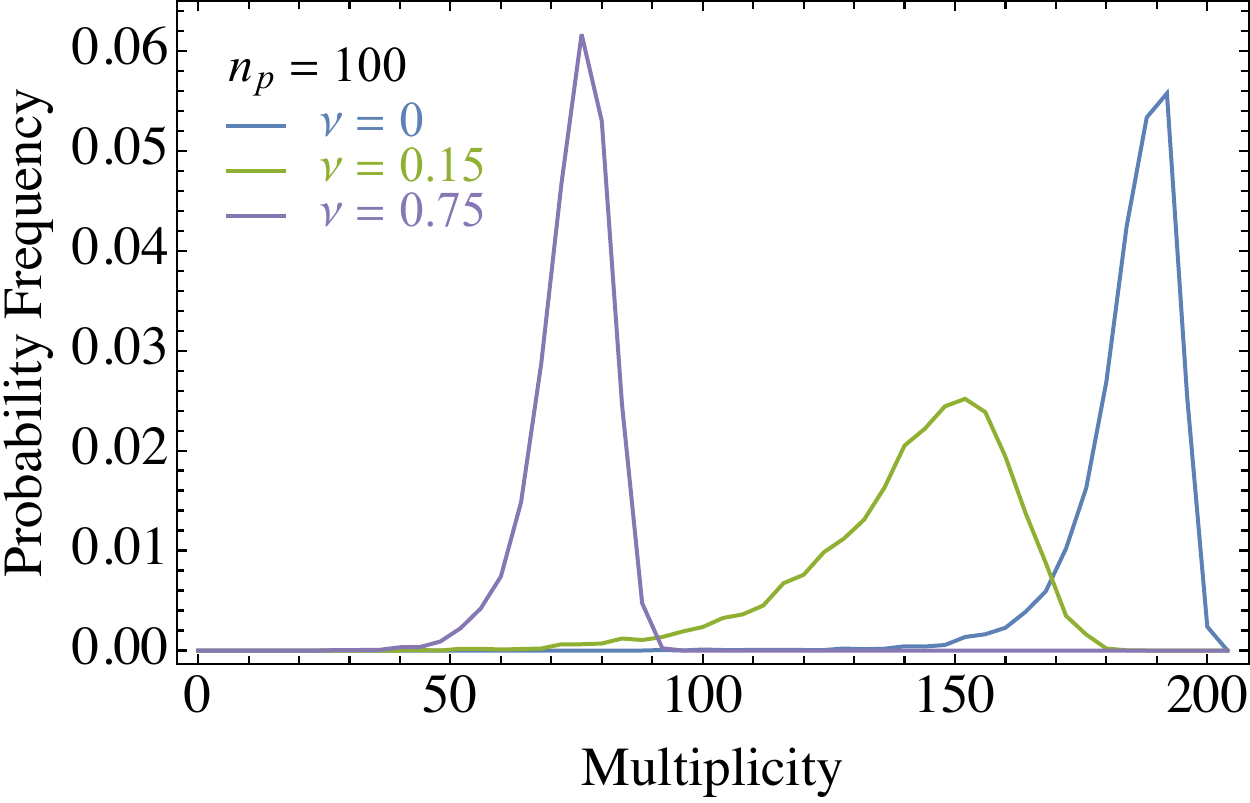}
}
\hfill
\subfloat[]{
\includegraphics[width=0.45\textwidth]{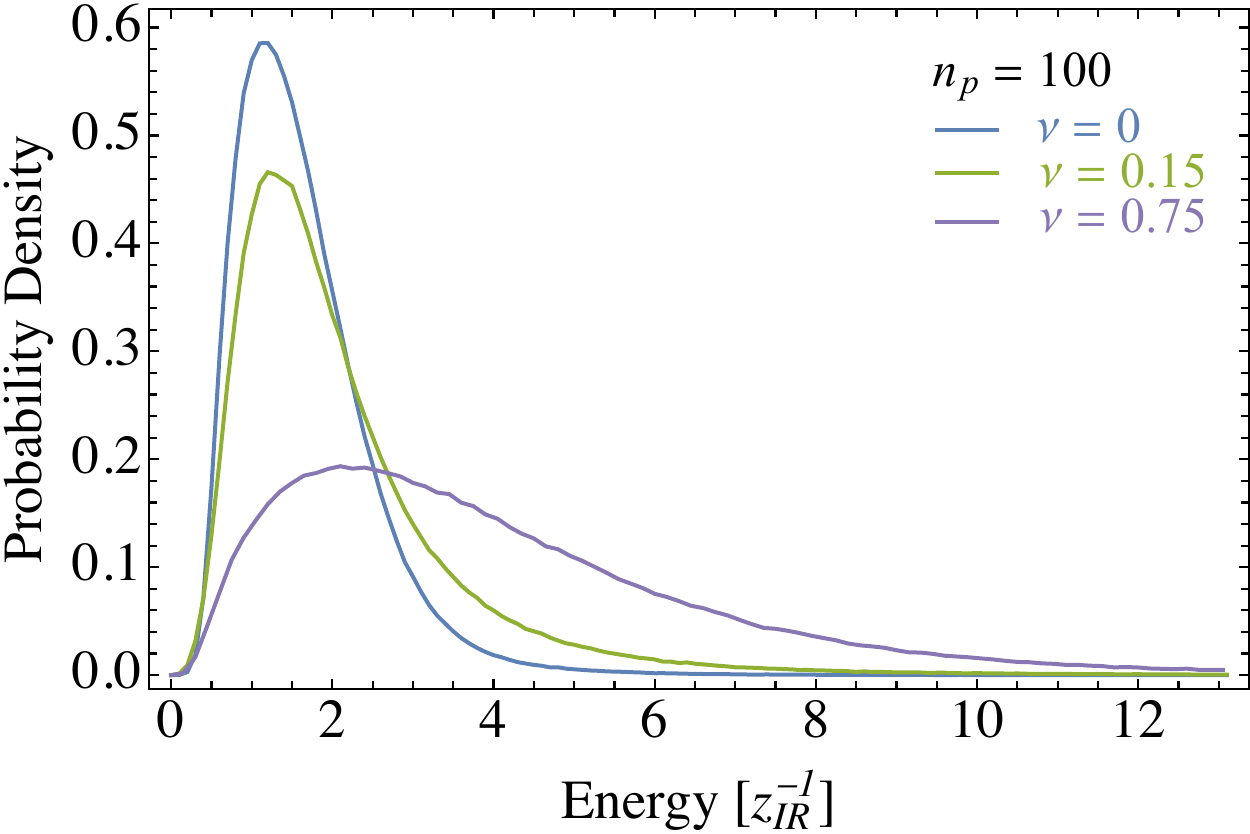}
}
\caption{The (a) event isotropy, (b) thrust, (c) multiplicity, and (d) energy spectra of the final state of $\nu=\{0, 0.15, 0.75\}$ cascades. 
The samples comprise $10^4$ events with $n_p = 100$, and all final state observables are computed after splitting the HSHs into two massless particles.  
}
\label{fig:singleFieldDist}
\end{figure}
We present results for the event shapes of the cascades in \Fig{fig:singleFieldDist}, where  we show distributions of the multiplicity and energy of the massless particles in the sample, along with the sample's event isotropy and thrust distributions.
KK-number violation reduces the multiplicity for $\nu=0.15$ relative to the $\nu=0$ benchmark.  The $\nu=0.75$ case has even smaller multiplicity and a wider energy distribution, due to additional sources of KK-number violation and the fact that the $n=2$ state is also an HSH.
Meanwhile the event isotropy increases (the events become less spherical) for $\nu=0.15$ and even more so for $\nu=0.75$.  Interestingly, this pattern does not apply for thrust; the thrust distributions for $\nu=0.15$ and $0.75$ are very similar.  
This suggests that event isotropy and thrust are sensitive to different event shape characteristics and are not redundant variables. 
Further study of this issue will be presented in \cite{paper2}.

Comparing \Fig{fig:singleFieldDist} (a) and (c), one might wonder if event isotropy and particle multiplicity are redundant variables, especially considering the natural correlation between them that we described earlier.  Although the correlation seems particularly strong in these examples, other simulations shown below clearly demonstrate these two variables are independent. 

\begin{figure}[!h]
\centering
\subfloat[]{
\includegraphics[width=0.3\textwidth]{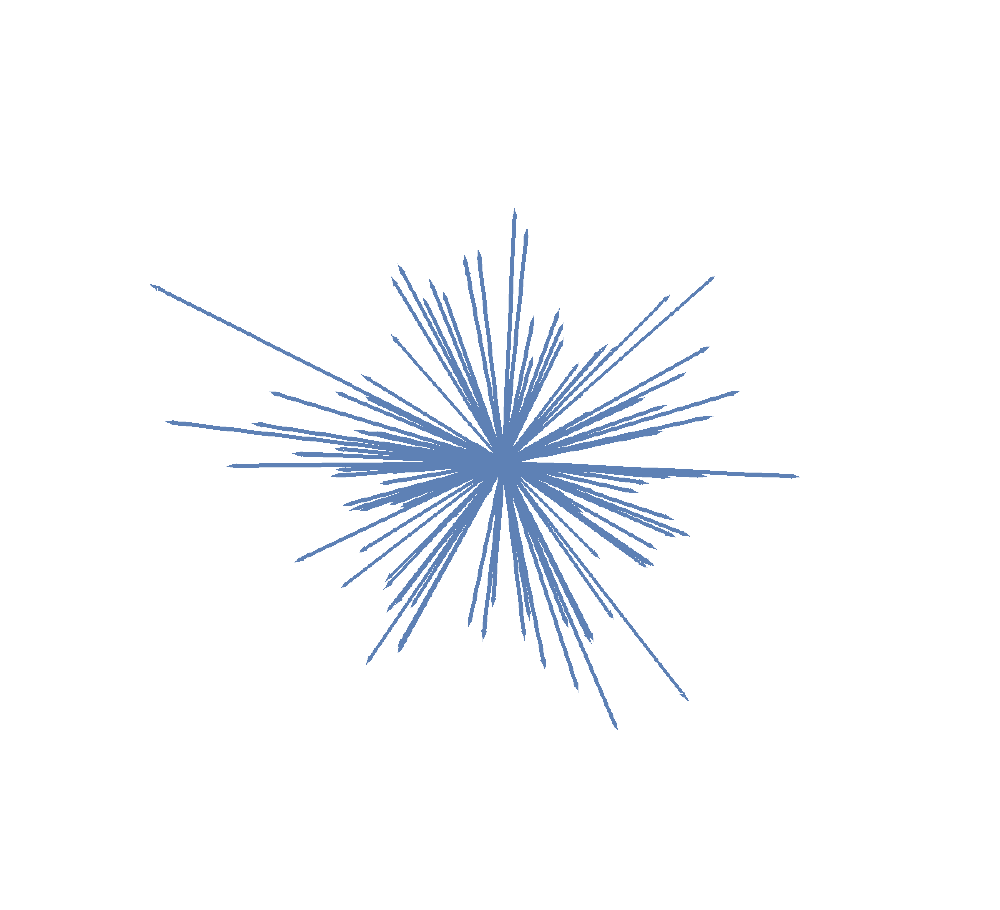}
}
\hfill
\subfloat[]{
\includegraphics[width=0.3\textwidth]{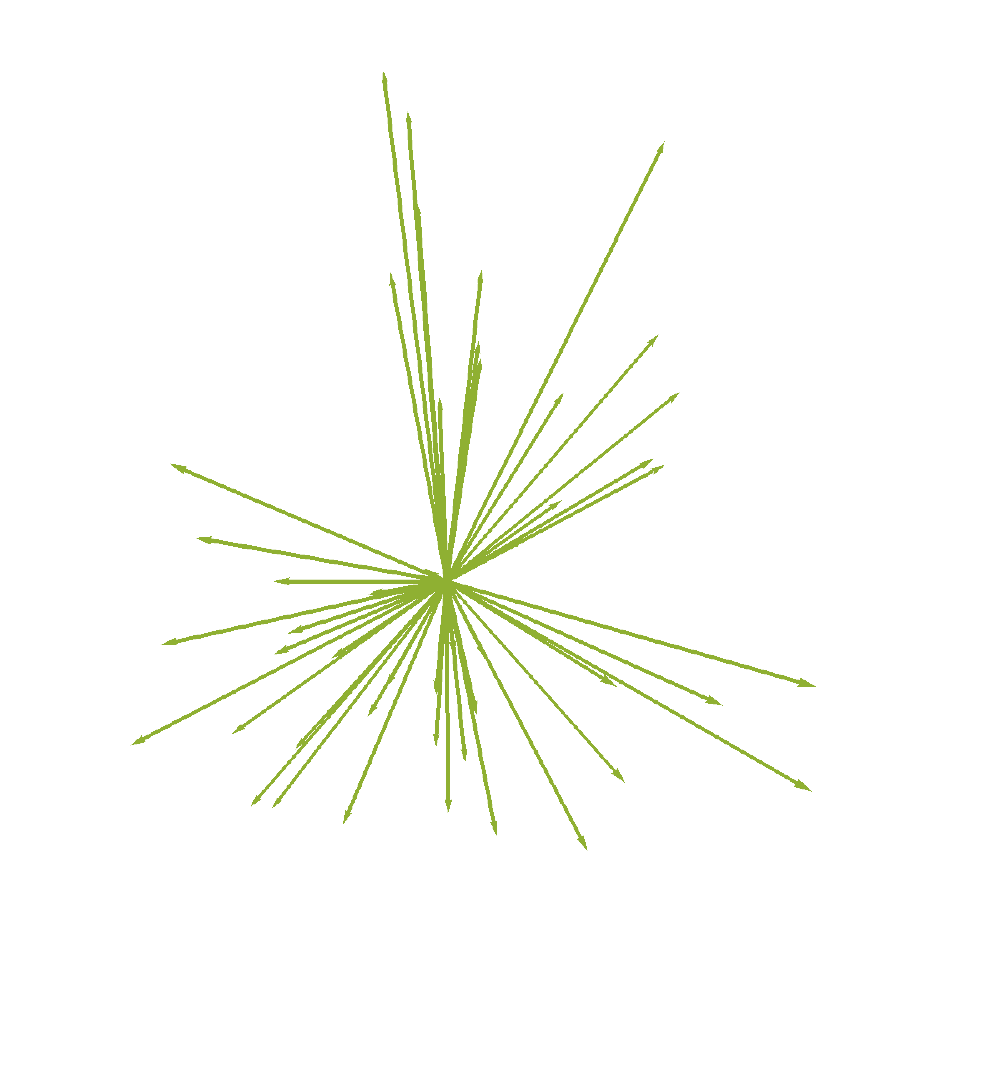}
}
\hfill
\subfloat[]{
\includegraphics[width=0.3\textwidth]{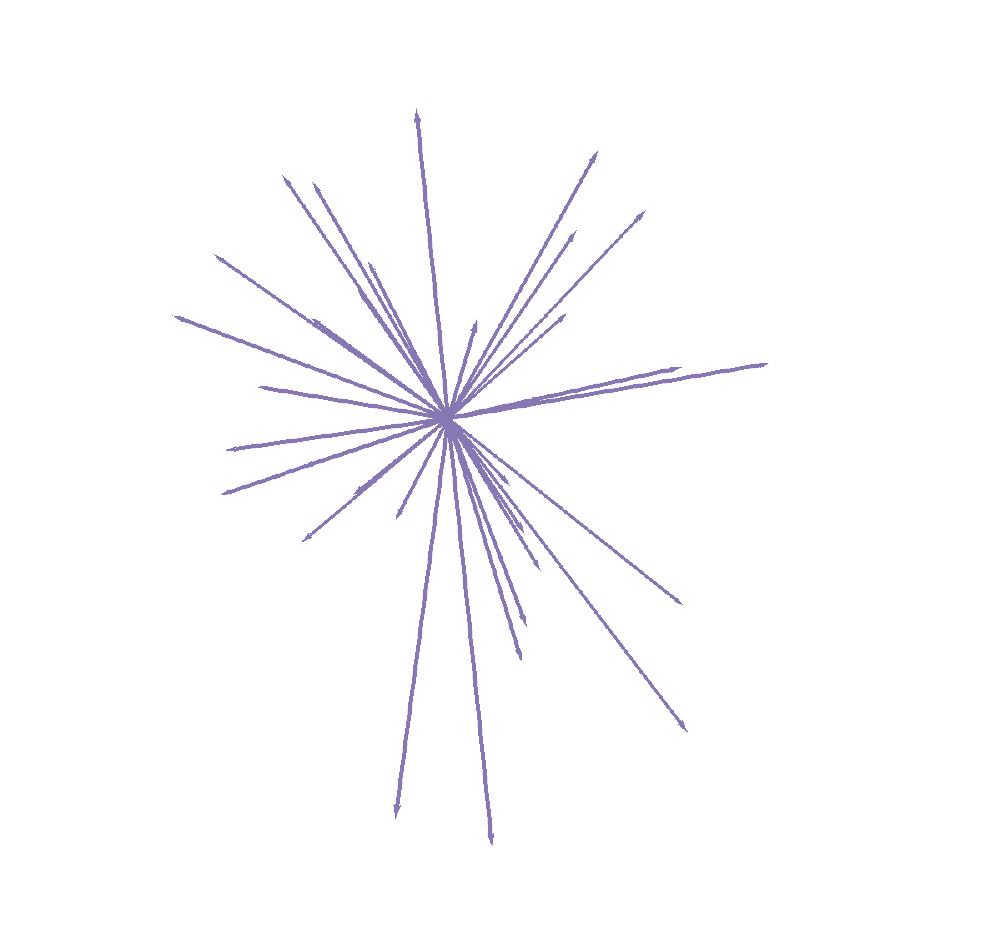}
}
\caption{Visualizations of a characteristic final state for (a) $\nu = 0$,  (b) $\nu = 0.15$,  and (c) $\nu = 0.75$. 
The chosen events have $\iso{sph}{192}$ equal to the mean in each distribution: (a) $\langle \iso{sph}{192} \rangle = 0.20$, (b) $\langle \iso{sph}{192} \rangle = 0.26$, (c) $\langle \iso{sph}{192} \rangle = 0.29$.
For ease of viewing, all momenta with magnitude above the average in the event are plotted, with the length proportional to the magnitude. 
}
\label{fig:singleFieldVis}
\end{figure}

We illustrate some characteristic events for each of these samples in \Fig{fig:singleFieldVis}.
An event of average isotropy in the most isotropic sample ($\nu=0$) does not show any clear boost axis, whereas the samples with greater amounts of KK violation ($\nu =\{0.15, 0.75\}$) begin to show collimated prongs of energy. 

We end our discussion of the single field case with a few comments on large values of $\nu$.  With increasing $\nu$ but fixed $n_p$, $I_+$ is suppressed by a factor of  $\left(m_2m_3/m_1^{2}\right)^\nu$, so $I_-$ dominates in most decay channels.  When this happens, the couplings conserve KK-number similarly to \Eq{eq:cijk_nu_even} even when $\nu$ is not an even integer.  However, this fact is irrelevant for $n_p=100$ because the kinematic constraints imposed by the mass spectrum require large violation of KK-number in each decay, which grows with $\nu$.  We noted earlier that KK-number conservation becomes kinematically disallowed for $\nu>1/2$.  More generally, decays with KK-number violation $\Delta_\text{KK}$ begin to be kinematically constrained when $\nu\gtrsim 2\Delta_\text{KK}+1/2$, starting with the most symmetric decays, and once $\nu$ reaches the values $0.5, 2.92, 5.56, 8.30, 11.1, \dots$, all decays with $\Delta_\text{KK} =0,1,2,3,4,\dots$ are forbidden. Moreover, there are roughly $1+\floor*{\frac{1}{2}(\nu+\frac{3}{2})}$ HSHs in the spectrum.  All of these effects drastically reduce multiplicity and widen energy distributions, further increasing the event isotropy and thrust, so we do not expect highly spherical distributions to be common in this regime.

%
%
\subsection{Two field ($\nu_1 \neq \nu_2 = \nu_3$)}
\label{subsec:twofield}

Now we consider the two field scenario, where \Eq{eq:interactions} includes two distinct 5d scalar fields.
Here the cascade will be populated with $\phi_{1,i} \rightarrow \phi_{2,j} \phi_{2,k}$ and $\phi_{2,i} \rightarrow \phi_{1,j} \phi_{2,k}$ decays. 

Because there are two towers of hadrons with different mass spectra, the correlation between KK-number and mass is not as simple as in the single-field case.  
However, these complications are relatively unimportant compared to the dramatic change in the pattern of couplings $c_{ijk}$.  
In the single field case, we have seen that  couplings near threshold, with zero or minimal KK-number violation, are strongly enhanced, because  $I_-\sim \lambda_\text{PS}^{-1}$ (away from threshold) and $I_+\sim \lambda_\text{PS}^{-1/2}$.
This leads to events that, to a greater or lesser extent depending on $\nu$, tend to be quasi-spherical;  jetty events are rare.
But this behavior does not extend to general $\nu_i$.

For the two field case, once $\nu$ for a decaying particle is significantly larger than the sum of the $\nu_i$ for its daughters,  KK-number violation becomes large.  
We can see this by examining $I_+$ (which typically is much greater than $I_-$ in this regime) using \Eq{eq:GandR} and the paragraph following it; see also \Sec{sec:Cheby}.
If, without loss of generality, we  take the decaying particle to be from field $\Phi_1$ and its daughters from fields $\Phi_2,\Phi_3$, and set $\Delta\nu\equiv \nu_1-\nu_2-\nu_3$, then when  $\Delta\nu=2k$, where $k$ is any positive integer, $I_+$ goes to a constant at threshold. 
Consequently branching fractions are {\it suppressed} near threshold by  $\lambda_\text{PS}^{+1/2}$, and instead peak elsewhere in the branching fraction triangle.  
Moreover, $I_+$ has $k-1$ lines of zeroes, and so the branching fraction triangle has $k$ plateaus separated by valleys.  
Some of these plateaus have large KK-number violation.  
When $\Delta\nu\neq 2k$ these plateaus survive, but are supplemented by a return of the $\lambda_\text{PS}^{-1/2}$ behavior near threshold.  
Although this near-threshold enhancement favors KK-number-conserving decays, the large KK-number violating decays in the more distant plateaus often remain dominant, making jetty events common.\footnote{In this discussion we have neglected $I_-$.  It can be seen from the formulas of \Sec{sec:analytic} that  $|I_+|\gg |I_-|$ for most decays, but there is a subtlety near threshold, where our approximation $I_-\sim \lambda_{\text{PS}}^{-1}$ seems to blow up faster than $I_+$ does.  This effect is merely due to our approximation \Eq{eq:IminusDirichletSmallPS}, however, which is not valid at threshold.  Instead, when $m_1-m_2-m_3\to 0$ for some decay, which requires tuning of the $\nu_i$, both $I_+$ and $I_-\sim \lambda_{\text{PS}}^{-1/2}$ extremely close to threshold, as noted in \Eq{eq:IminusDirichletVerySmallPS}, and in fact the original integral in \Eq{eq:4dcoup} is always finite there.   (See also App.~\ref{app:npdependence}.) In practice, then, our discussion here of $I_+$ captures all the important features of the branching fraction triangles, except for the precise details in the near-threshold region.}

\begin{figure}[t!]
\centering
\subfloat[]{
\includegraphics[width=0.45\textwidth]{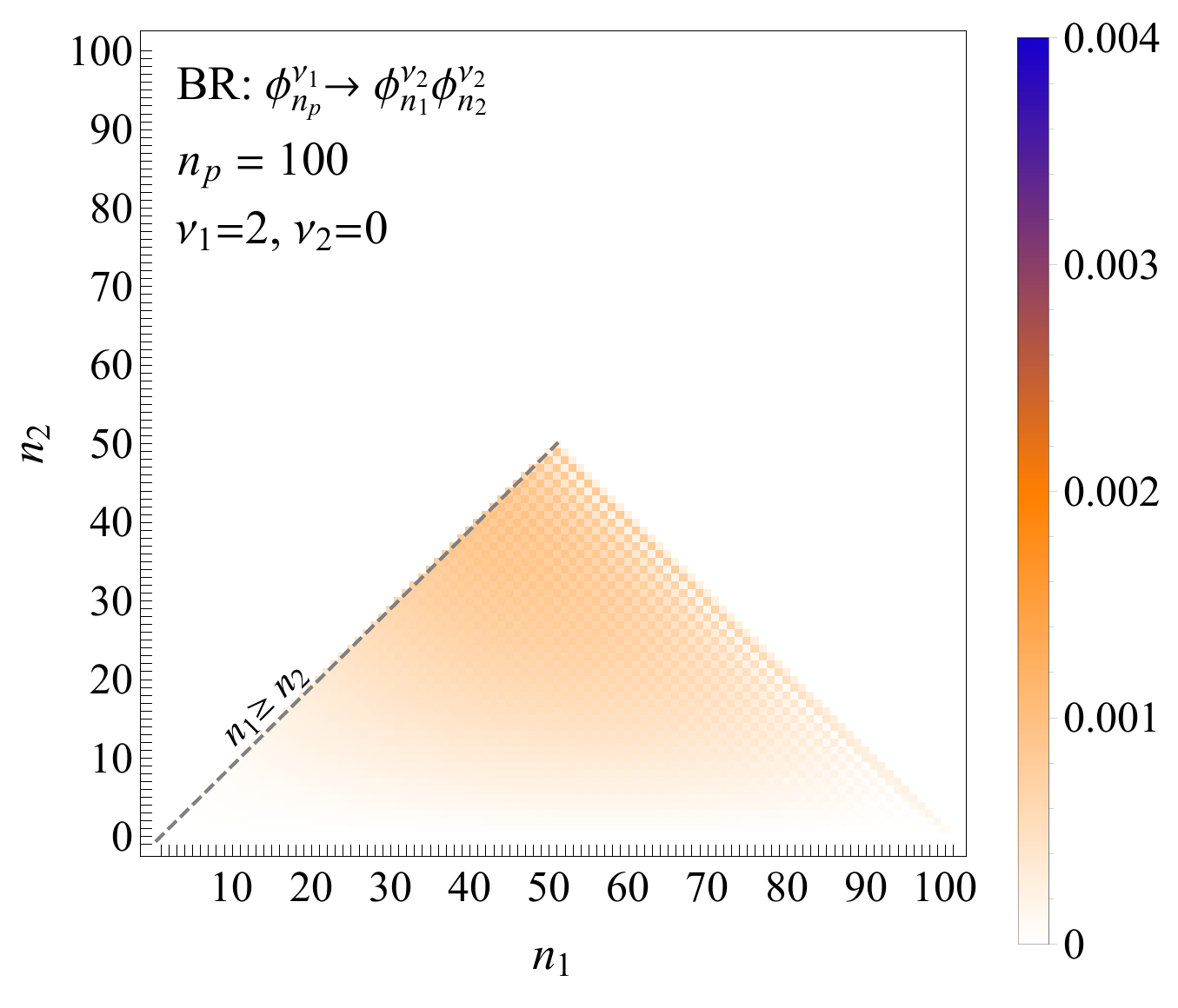}
}
\hfill
\subfloat[]{
\includegraphics[width=0.45\textwidth]{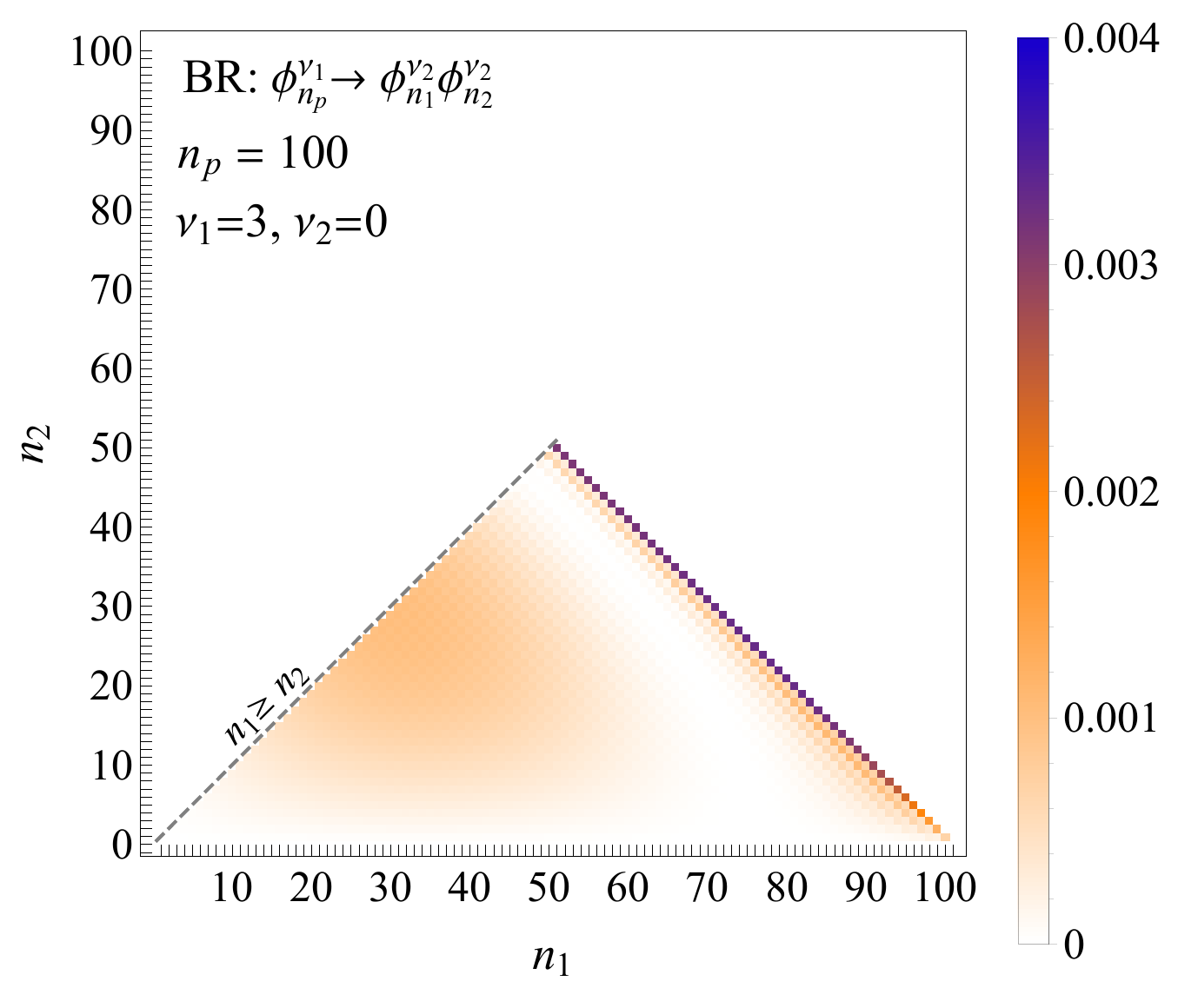}
}
\hfill
\subfloat[]{
\includegraphics[width=0.45\textwidth]{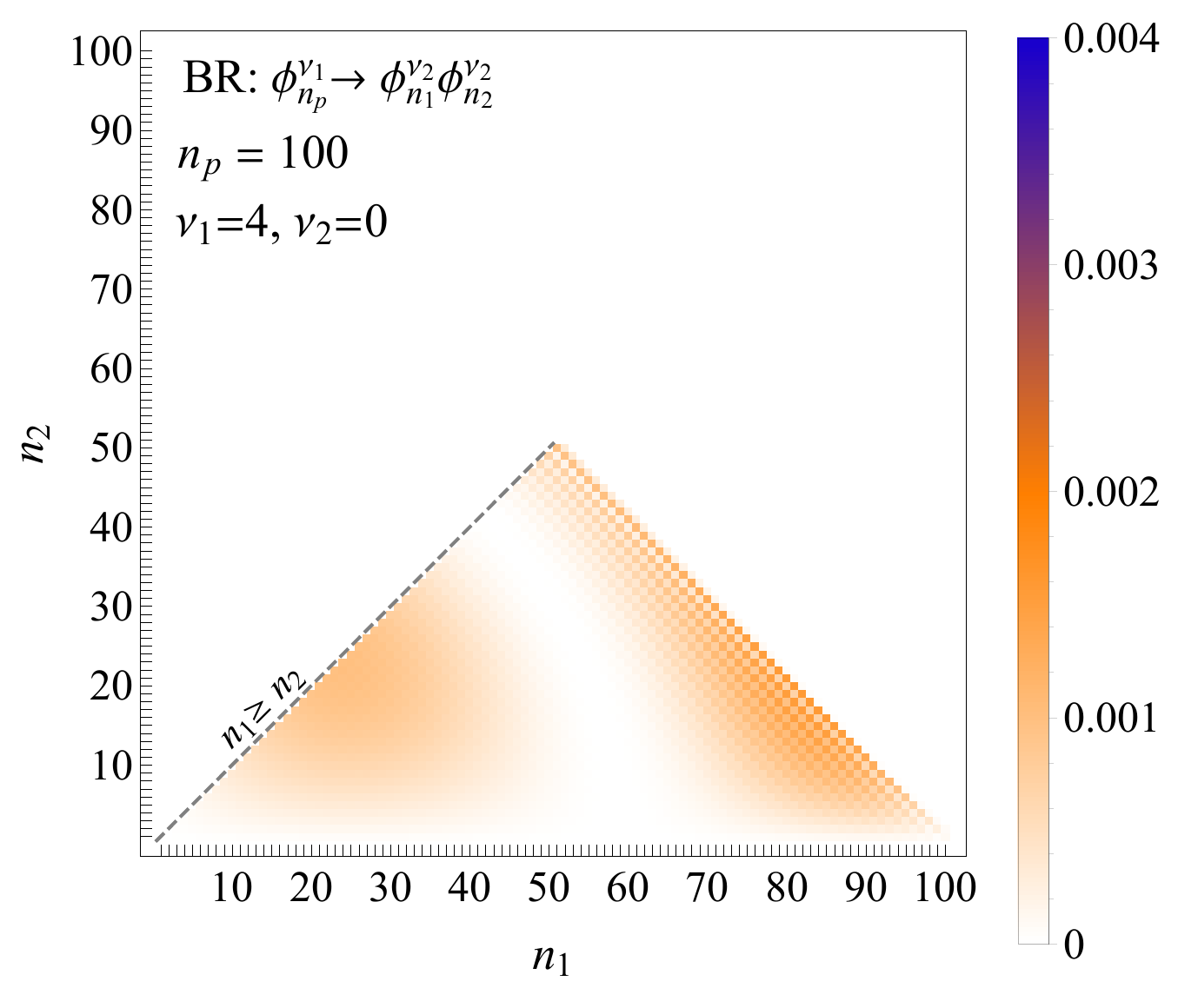}
}
\hfill
\subfloat[]{
\includegraphics[width=0.45\textwidth]{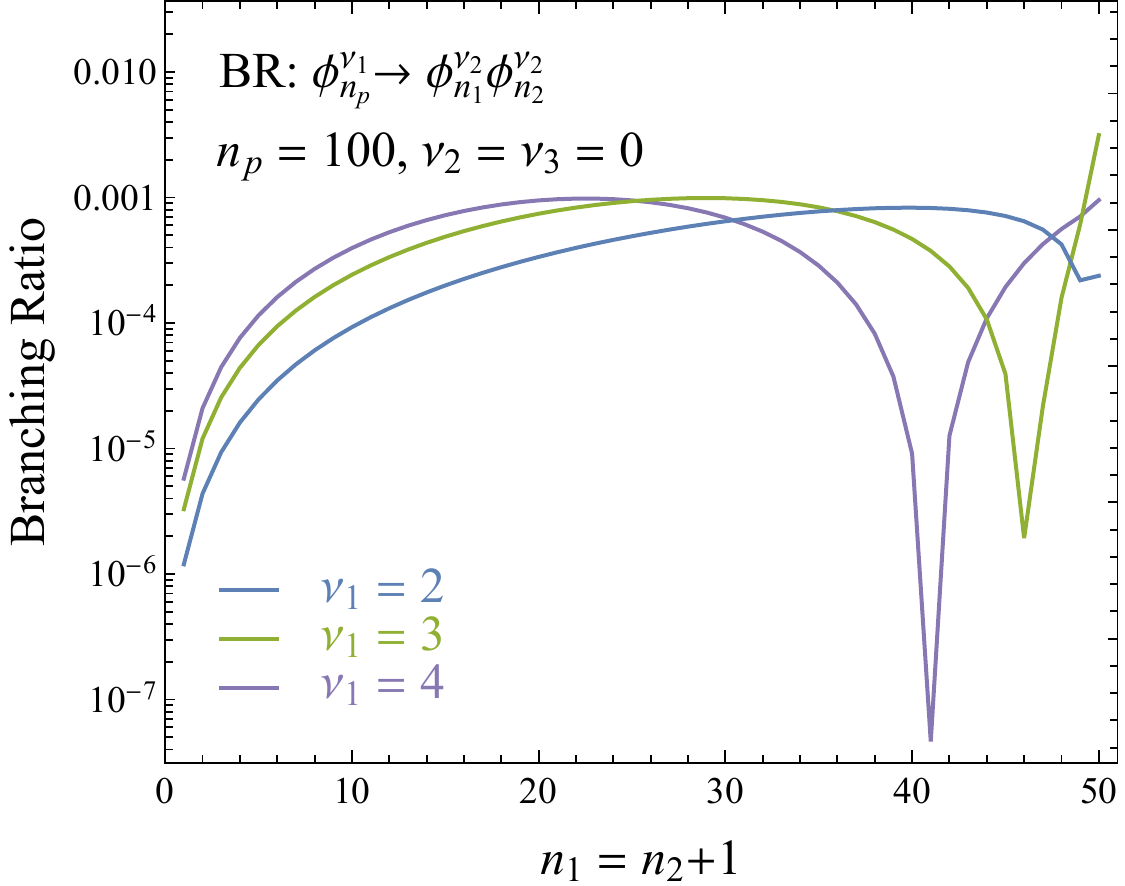}
}
\caption{The branching ratios of the $n_p=100$ mode for decays of the field corresponding to (a) $\nu_1=2$, (b) $\nu_1=3$, and (c) $\nu_1=4$ into two $\nu_2 = 0$ fields with KK modes $n_1$, $n_2$. %
The centralization of the plateau is a function of all $\nu_i$, but generally the plateaus are more strongly centralized for larger values of $\Delta \nu \equiv \nu_1 -\nu_2-\nu_3$. 
The 1d projection of the branching ratios for all values of $\nu_1$ along the $n_1 = n_2 + 1$ line is given in (d). 
}
\label{fig:brTwoField}
\end{figure}

Focusing now on $\nu_2=\nu_3=0$, we will illustrate the behavior just described in the cases $\nu_1=2,3,4$, whose branching fractions for $n_p=100$ are shown in \Fig{fig:brTwoField}.  
 For  $\nu_1=2$, using \Eq{eq:dirNorm}, \Eq{eq:4dcoup}, \Eq{eq:GandR}, and the remark below \Eq{eq:GandR} that $\mathrm{F}_4 \to 1$ in this case, we find (to a very good approximation)
\be
c_{ijk} \propto \sqrt{m_2 m_3} \ .
  \label{eq:cijknu2}
\ee 
The branching fractions are then proportional to $m_2m_3\lambda_\text{PS}^{1/2}$, so they vanish at all boundaries of the branching fraction triangle and are broadly distributed, as seen in  \Fig{fig:brTwoField}(a), peaking $\sim10\%$ below threshold. 
 For $\nu_1=4$, \Eq{eq:F4k2} gives
\be
c_{ijk} \propto \sqrt{m_2 m_3} \left(1-3\frac{m_2^2 + m_3^2}{m_1^2}\right),
  \label{eq:cijknu4}
\ee
which creates a zero between the threshold region at upper right and the $m_2,m_3\to 0$ corner at lower left.
The branching fraction triangle then has two plateaus, one far from threshold and one nearby, as seen in  \Fig{fig:brTwoField}(c).
  The probability for a particle to decay via either plateau is comparable. 
  Finally, for $\nu_1=3$, the zero seen for $\nu_1=4$ is still present, closer to threshold, but in addition there is enhancement right near threshold.  
  Despite this enhancement, there are so many decay paths in the plateau that the total probability to decay at or near threshold is only $\sim 1/4$, and so large KK-number violation is the norm.

\begin{figure}[t!]
\centering
\subfloat[]{
\includegraphics[width=0.45\textwidth]{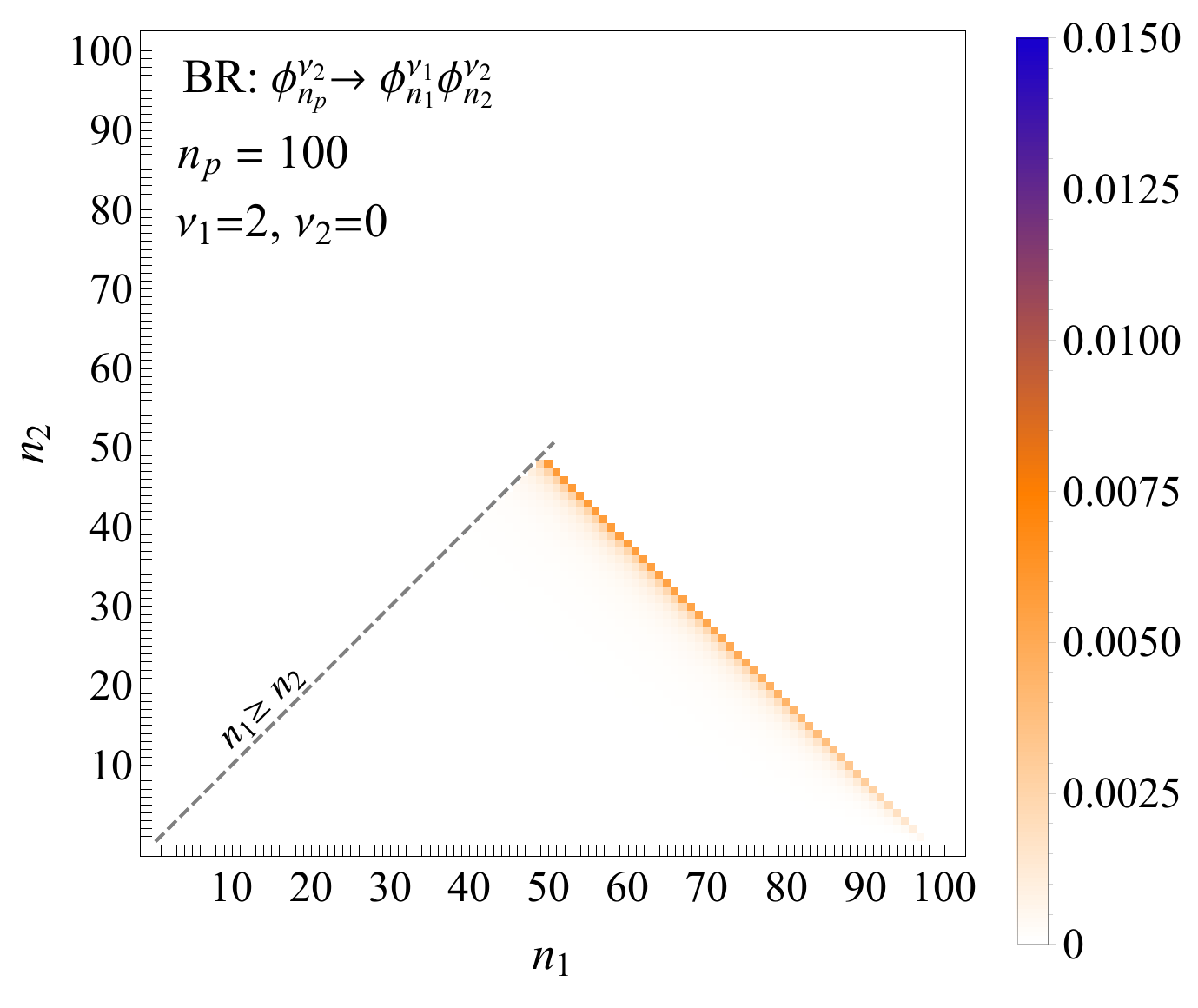}
}
\hfill
\subfloat[]{
\includegraphics[width=0.45\textwidth]{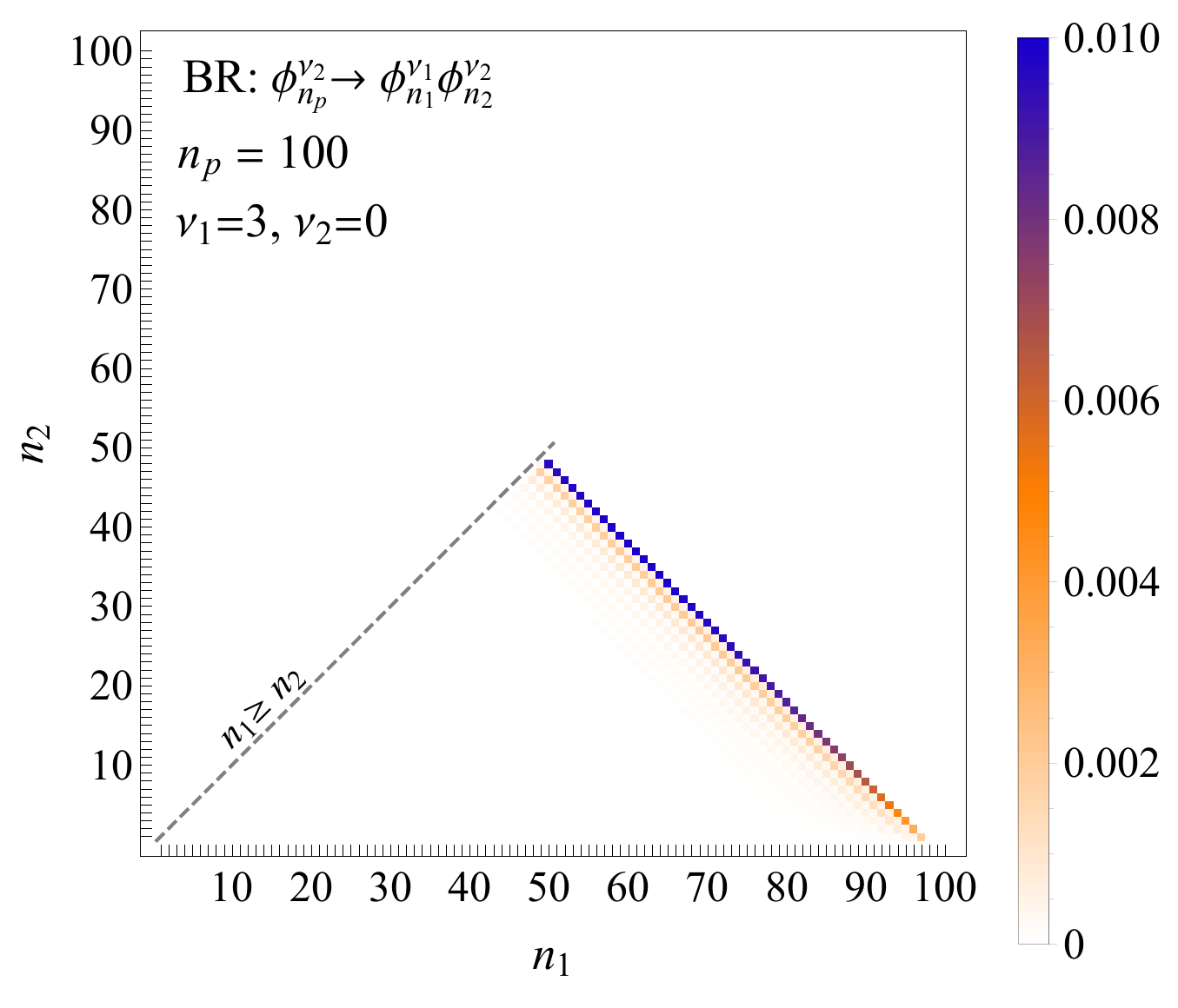}
}
\hfill
\subfloat[]{
\includegraphics[width=0.45\textwidth]{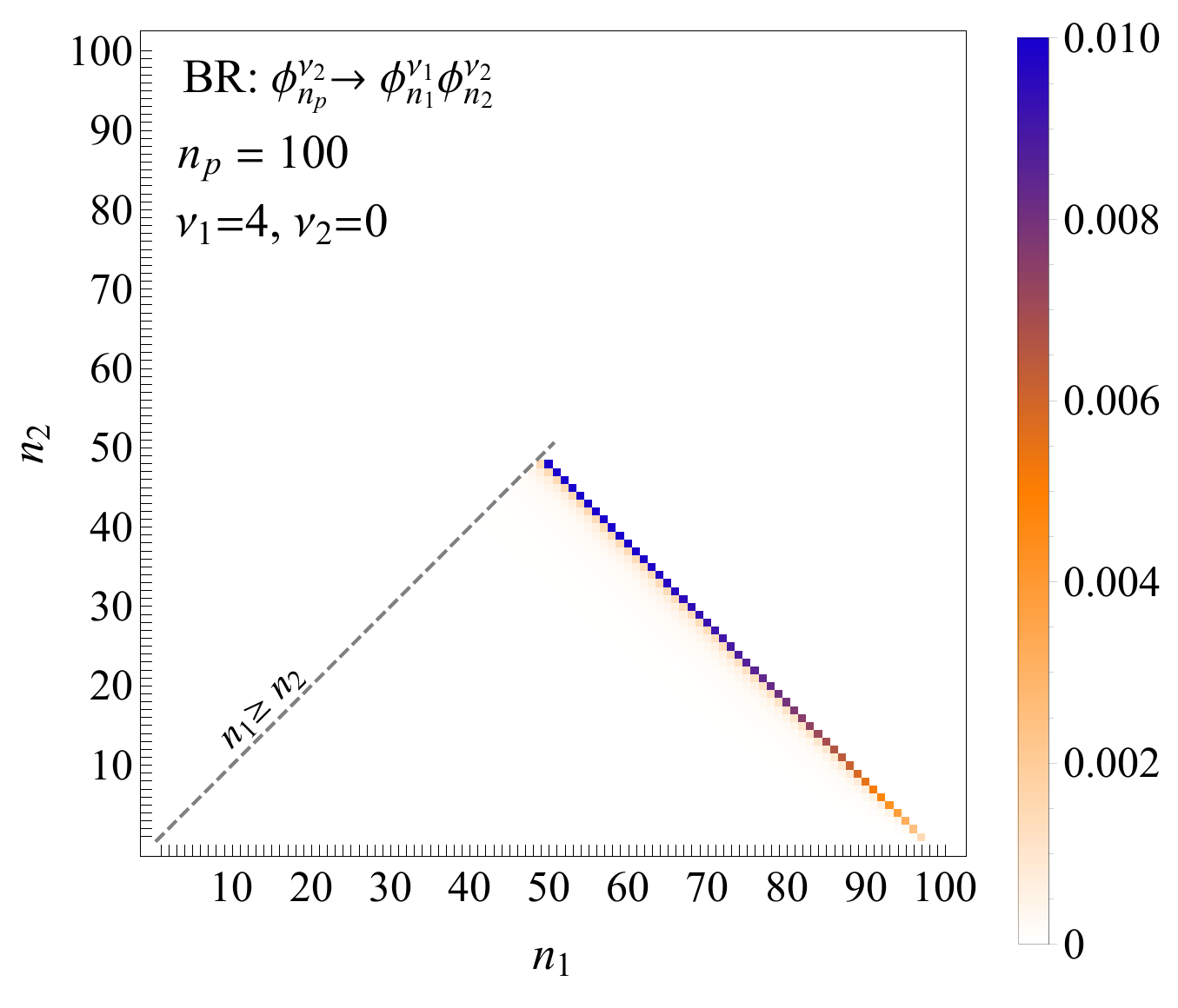}
}
\hfill
\subfloat[]{
\includegraphics[width=0.45\textwidth]{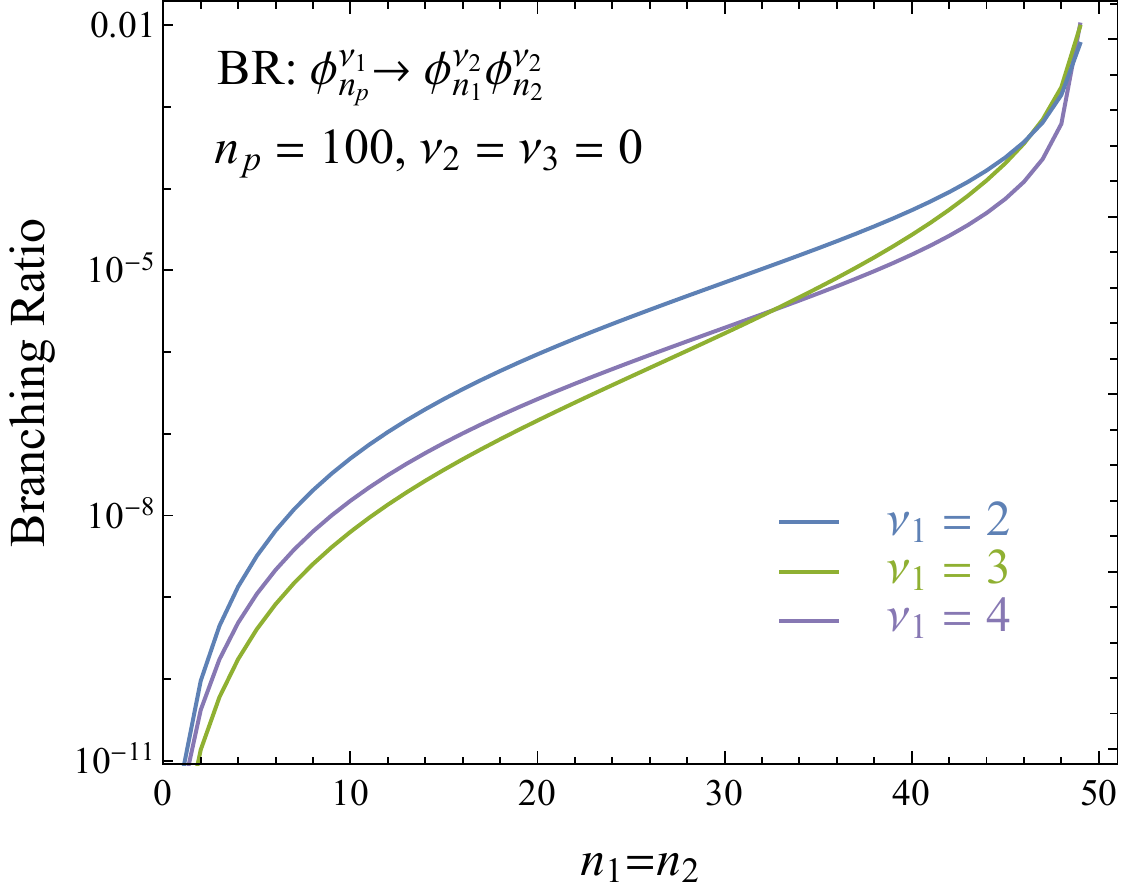}
}
\caption{The branching fraction triangle of the $n_p=100$ mode for the field corresponding to $\nu_2 =0$ into another $\nu_2$ field and a (a) $\nu_1=2$, (b) $\nu_1=3$, or (c) $\nu_1=4$ field. %
These decays do not exhibit the same plateau structure as in \Fig{fig:brTwoField}.
The 1d projection of the branching ratios for all values of $\nu_1$ along the $n_1 = n_2$ line is given in (d). 
}
\label{fig:brTwoField2}
\end{figure}

A cascade in the two field case involves both $\phi_{1,i} \rightarrow \phi_{2,j} \phi_{2,k}$ and $\phi_{2,i} \rightarrow \phi_{1,j} \phi_{2,k}$ decays.
To compute the branching fractions for the latter, we need to exchange $\nu_1$ and $\nu_2$ (but not $\nu_3$) in our analytic formulas. 
For reasons similar to those leading to \Eq{eq:Iplussingle}, $I_+$ vanishes for $\nu_1=2,4$, giving the nearly KK-number-conserving result \Eq{eq:cijk_nu_even}, but is important for $\nu=3$, leading to slightly higher KK-number violation. 
Thus it is no accident that the branching fraction triangles for these decays, shown in \Fig{fig:brTwoField2}, resemble those of the single field case, \Fig{fig:brratios}.
However, the KK-number violation in $\phi_{2,i}$ decays is subleading compared to the much larger KK-number violation that can occur in $\phi_{1,i}$ decays, and its details do not much impact the results.

As was true also for \Fig{fig:brratios}, the qualitative features of the branching fraction triangles are present also for smaller values of $n_p$, and thus apply for the whole cascade.  See App.~\ref{app:npdependence} for further discussion of the $n_p$ dependence.

Particle multiplicities are affected not only by these KK-number-violating processes but also by the increasing number of HSHs.  For $\nu_2=\nu_3=0$, and $\nu_1\gtrsim 1.75$, the decay $\phi_{1,1}\to \phi_{2,1}+\phi_{2,1}$ is always open, so only states of $\Phi_2$ are stable against decays within the hidden sector.  The mode $\phi_{2,1}$, as the lightest mode in the hidden sector, is of course an HSH, while $\phi_{2,2}$ is stable for $\nu_1\geq 0.5$,  $\phi_{2,3}$ is stable for $\nu_1\gtrsim2.90$, and $\phi_{2,4}$ is stable for $\nu_1\gtrsim5.53$.  But the effect on event shapes of reduced multiplicity is subleading compared to jet creation through boosts of daughter particles, as we will now see.


\begin{figure}[t!]
\centering
\subfloat[]{
\includegraphics[width=0.45\textwidth]{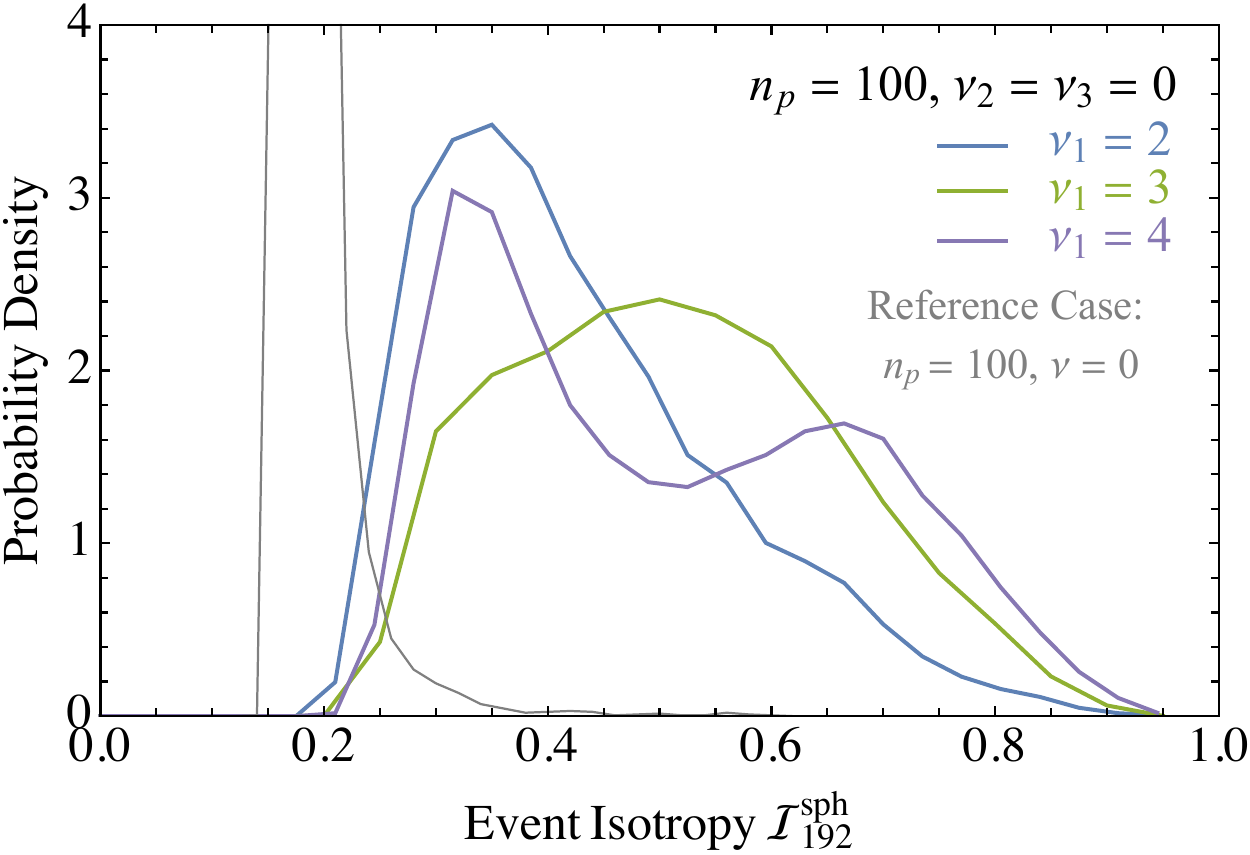}
}
\hfill
\subfloat[]{
\includegraphics[width=0.45\textwidth]{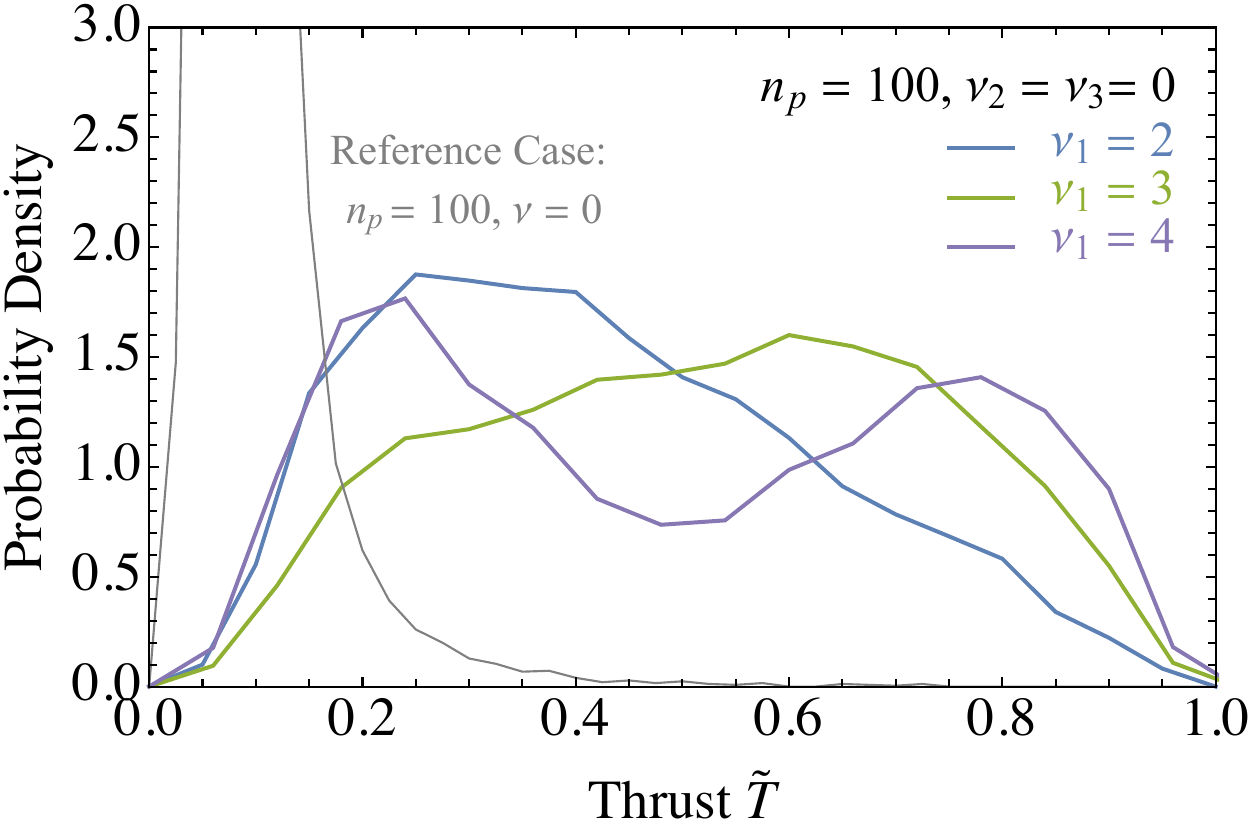}
}
\hfill
\subfloat[]{
\includegraphics[width=0.45\textwidth]{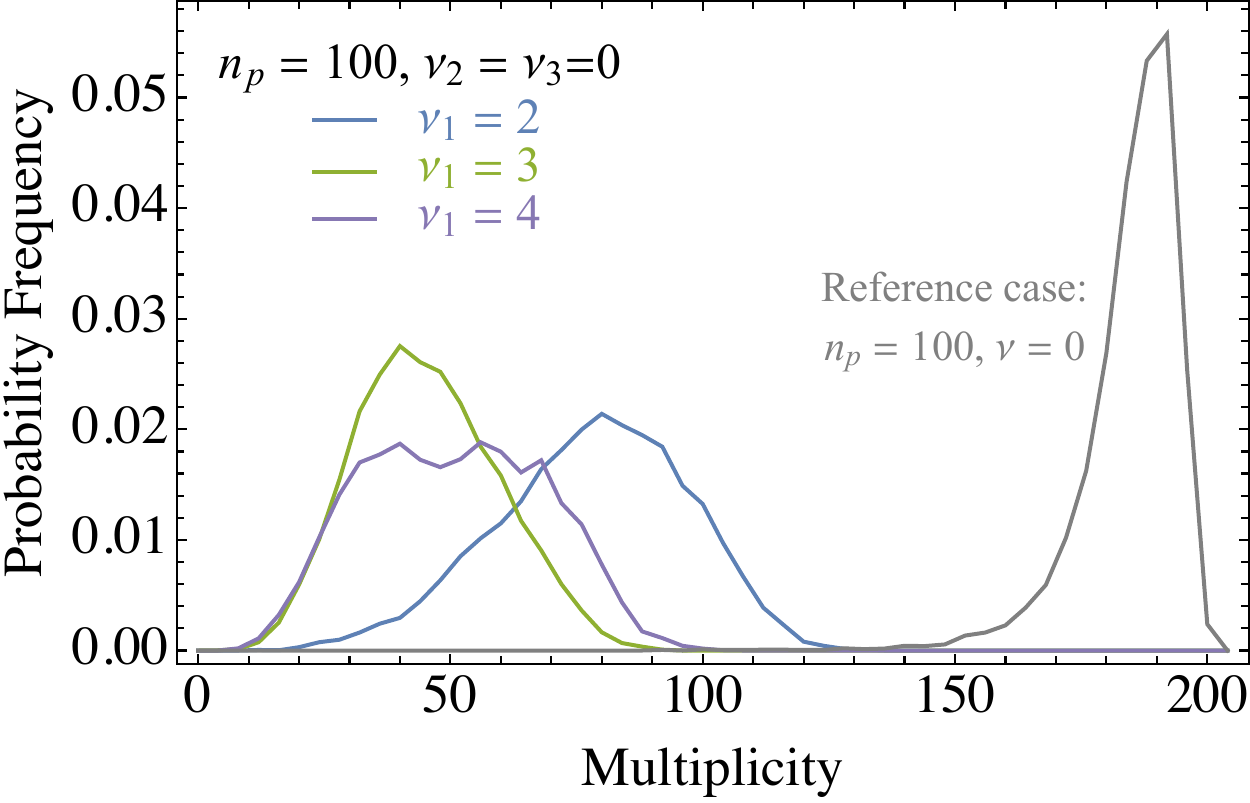}
}
\hfill
\subfloat[]{
\includegraphics[width=0.45\textwidth]{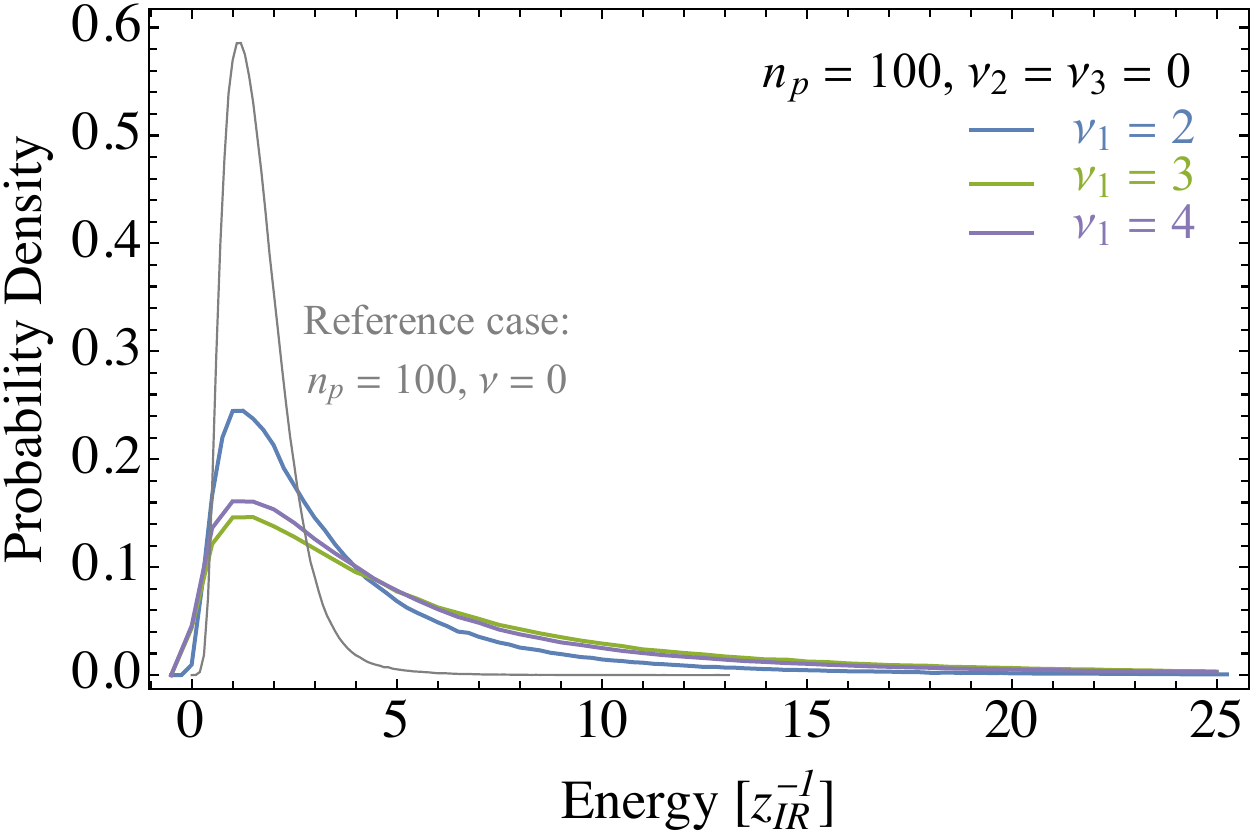}
}
\caption{Same as \Fig{fig:singleFieldDist} but for the two field scenario with $\nu_2 = \nu_3 = 0$, and $\nu_1 =2$ (blue), $\nu_1=3$ (green), and $\nu_1=4$ (red). 
Note that average multiplicity is lower and event isotropy is higher than the single field cases in \Fig{fig:singleFieldDist}.
The appearance of plateaus in the branching ratios can greatly increase the amount of KK violation per event. 
}
\label{fig:emdMultTwoField}
\end{figure}

Our results are shown in \Fig{fig:emdMultTwoField}, based again on cascades starting at the 100$^\text{th}$ KK mode of the field $\Phi_1$, with $10^4$ events per sample.  
Just as the branching fractions are much more widely distributed in \Fig{fig:brTwoField} than in \Fig{fig:brratios}, all the distributions in event isotropy and thrust are much wider than for the single field case in \Fig{fig:singleFieldDist}.
As is evident by comparing these figures with \Fig{fig:brTwoField},  the plateaus and valleys in the branching fraction triangles of the $\nu_1$ field lead directly to structure in the event shape variables.

It is easy to see why this is so.
The early stages in the cascade are most important, because decays far from threshold early in the cascade create boosted particles with a substantial fraction of the event energy, and their collimated decay products lead to hard jets, while reducing the overall multiplicity of hadrons.
Such events will differ strongly from the near-spherical events we saw in the single-field examples. 
If instead the initial decays nearly conserve KK-number and create several slow heavy hadrons, KK-number violation in their decays will still lead to jets, but these will have a much smaller fraction of the event's energy.
Thus, depending on the details of a particular cascade decay, an event may present a small number of hard jets at one extreme, or a small number of soft jets on top of a quasi-spherical background at the other.
As the probability for far-from-threshold decays increases, so will the fraction of events with larger thrust and event isotropy.

This is seen most dramatically for $\nu_1=4$, where the bimodal feature in the branching fractions, \Fig{fig:brTwoField}(c), with one plateau that weakly violates KK-number and another that violates KK-number significantly, leads to bimodal event shapes at small and large event isotropy and thrust respectively.
These correspond to two classes of events, one relatively spherical and the other relatively jetty.  Note that the lobe at larger event isotropy is comparable to the event isotropy distribution for threshold $t\bar t$ events, shown in \Fig{fig:SMbenchmark}, and actually peaks at higher $\iso{sph}{192}$, though it is not as jetty as a $q\bar q$ sample.

Comparing $\nu_1=2$ and $\nu_1=3$, one sees the latter's ensemble of events is less spherical on average.  
Despite the near-threshold enhancement for $\nu_1=3$, the majority of decays occur in the plateau far from threshold.  
This plateau is located further from threshold than that of $\nu_1=2$, so the most common decays for $\nu_1=3$ have larger boost on average than those for $\nu_1=2$, and this leads to harder jets, moving the isotropy and thrust distributions to larger values.  
The near-threshold decays partly compensate for this effect, making the event-shape variable distributions particularly broad.

The multiplicities of massless particles are of order 80 for $\nu_1=2$ and (partly because of the additional HSH) only 50 for $\nu_1=3,4$.  
Although this was not evident in the single-field cases, here one can see clearly that the event shape variables are not perfectly correlated with multiplicity.
For instance  the multiplicity distribution for $\nu_1=4$ does not show the obvious two-lobe structure seen in the event shape variables.  
We will return to this issue briefly in \Sec{subsec:evolmult}. 
It is also noteworthy that the isotropy distribution is narrower than the thrust distribution, a hint of imperfect correlation which we will study further in \cite{paper2}.

Turning our attention to the energy distributions of \Fig{fig:emdMultTwoField}, it is interesting to note that all the spectra (including the reference case $\nu=0$) are peaked at the same value. 
The peak is at half the lightest mass in the cascade: $m^{\nu=0}_1/2 = 1.2$.
This tells us that for all cascades there are many soft particles produced in the decay of nonrelativistic $n=1$, $\nu=0$ particles.
The width of the distribution is much wider for $\nu_1 = 2$, $\nu_2 =0$ than for $\nu=0$, and even wider for $\nu_2 = 3, 4$. 
This tail is due to the fact that modes heavier than the lightest mode are stable in the two field cases, whereas for $\nu=0$, only the lightest mode is stable. 
Note that these observations also apply to all the single field cases shown in \Fig{fig:singleFieldDist}.

\begin{figure}[!h]
\centering
\subfloat[]{
\includegraphics[width=0.23\textwidth]{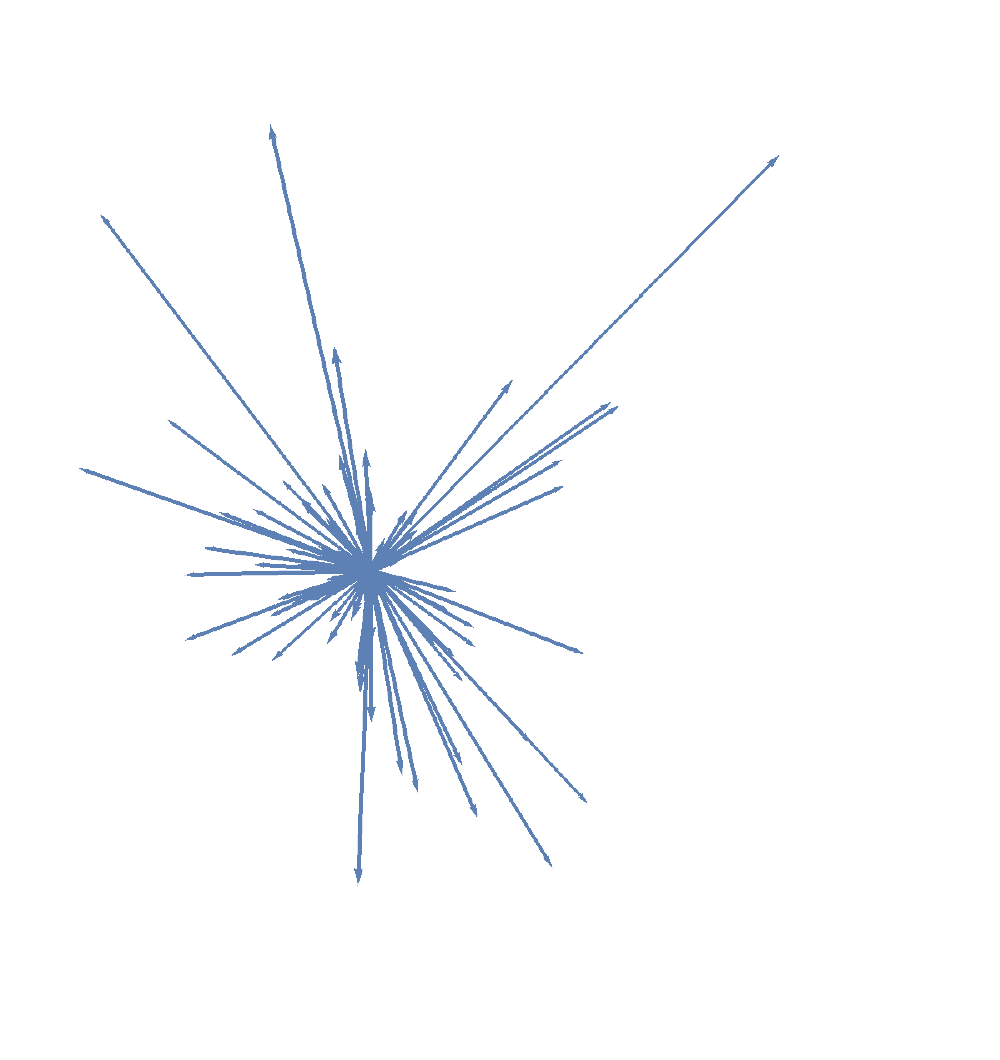}
}
\hfill
\subfloat[]{
\includegraphics[width=0.23\textwidth]{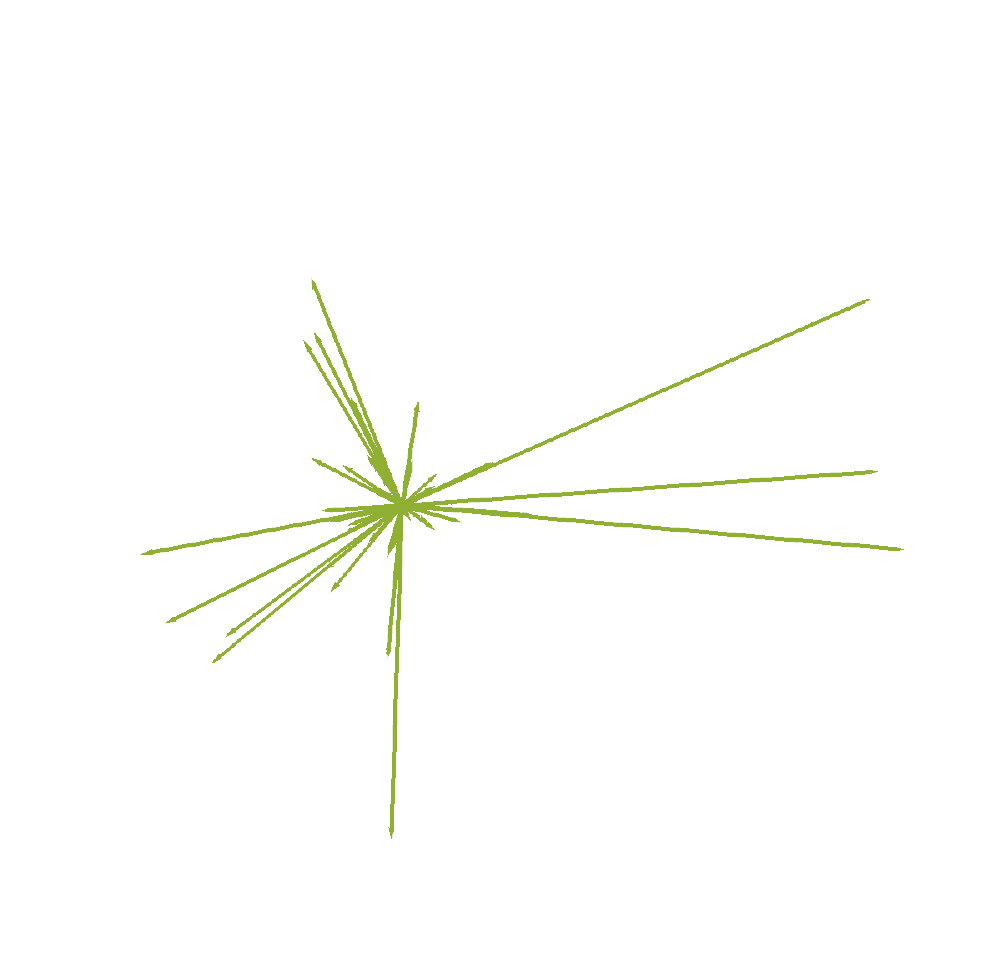}
}
\hfill
\subfloat[]{
\includegraphics[width=0.23\textwidth]{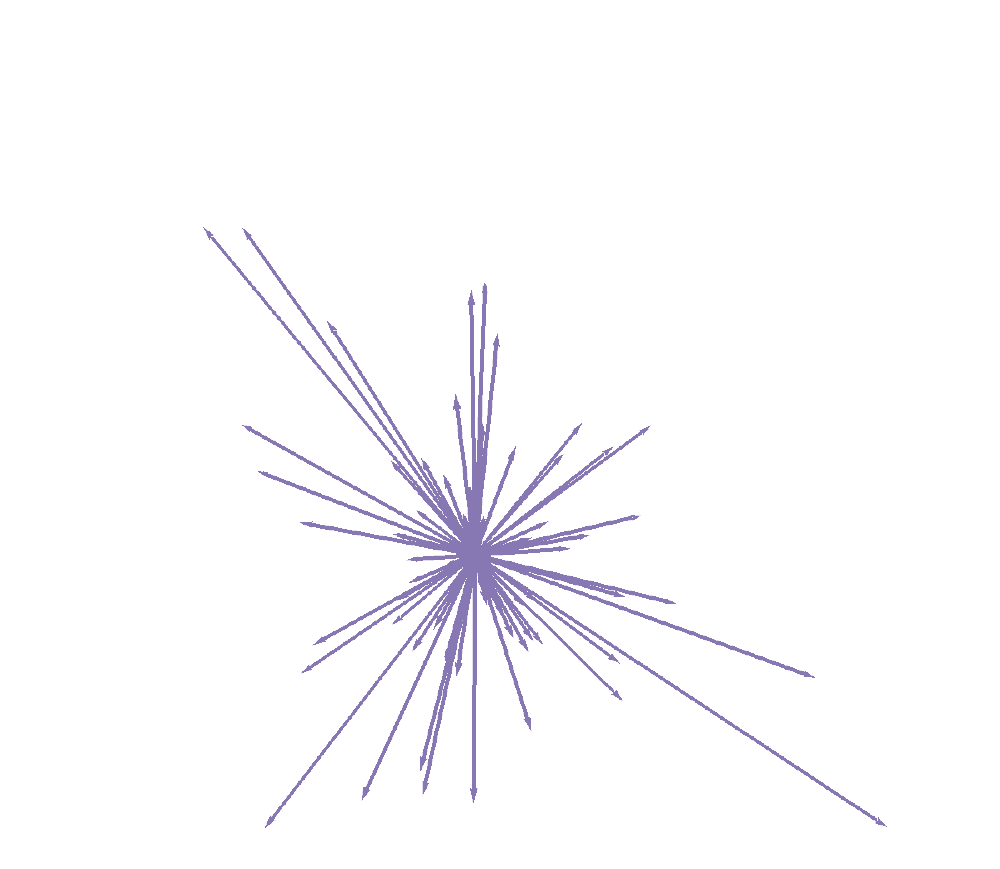}
}
\hfill
\subfloat[]{
\includegraphics[width=0.23\textwidth]{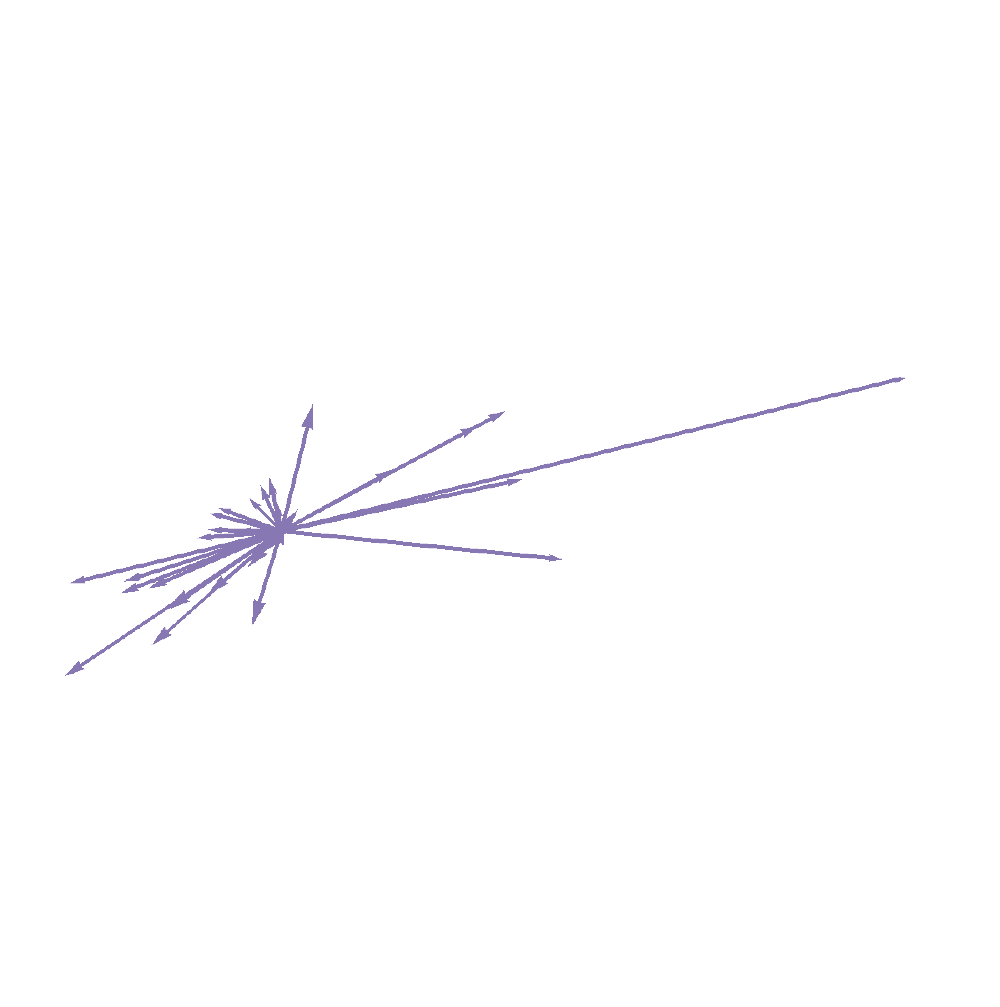}
}
\caption{ Visualizations of a characteristic final state radiation pattern for the (a) $\nu_1 = 2$,  (b) $\nu_1 = 3$,  (c,d) $\nu_1 = 4$ and $\nu_2  = \nu_3 = 0$ samples. 
The chosen events for $\nu_1 = \{2,3\}$ have event isotropy that are equal to the mean value of the distributions: (a) $\langle \iso{sph}{192} \rangle = 0.44$ and (b) $\langle \iso{sph}{192} \rangle = 0.54$.
We show two events for the $\nu_1=4$ sample, selected from the peaks of the bimodal distribution in event isotropy:
(c) $\langle \iso{sph}{192} \rangle = 0.31$  and (d)$\langle \iso{sph}{192} \rangle = 0.69$.
All momenta in the event are plotted, with the length proportional to the magnitude. 
}
\label{fig:twoFieldVis}
\end{figure}

Visualizations of characteristic events from each sample are shown in \Fig{fig:twoFieldVis}. 
The effects of the unsuppressed KK violating decays are evident.

\subsection{Cascades with boundary couplings}
\label{subsec:boundary}

\begin{figure}[!h]
\centering
\subfloat[]{\includegraphics[width=0.47\textwidth]{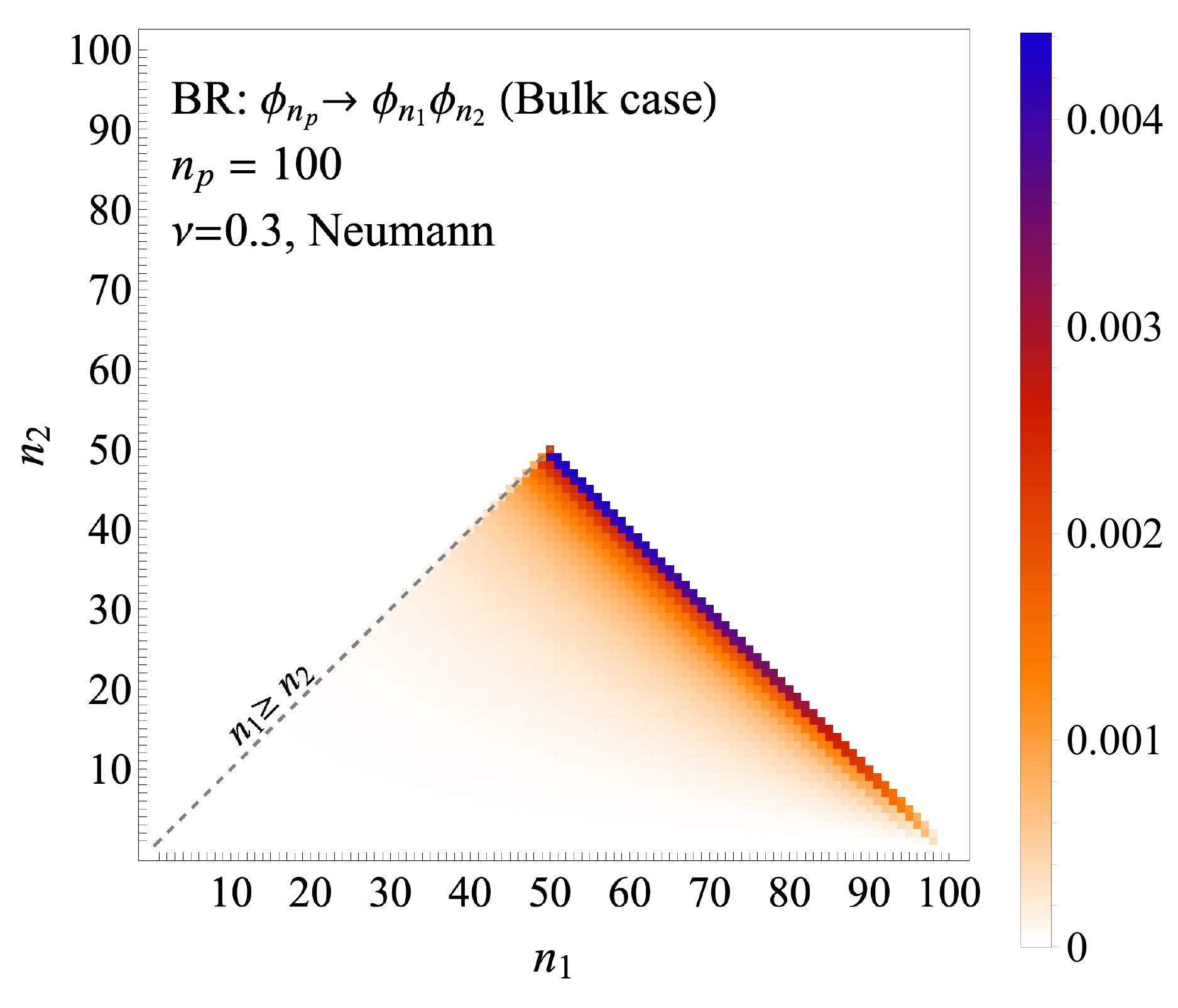}}\qquad
\subfloat[]{\includegraphics[width=0.47\textwidth]{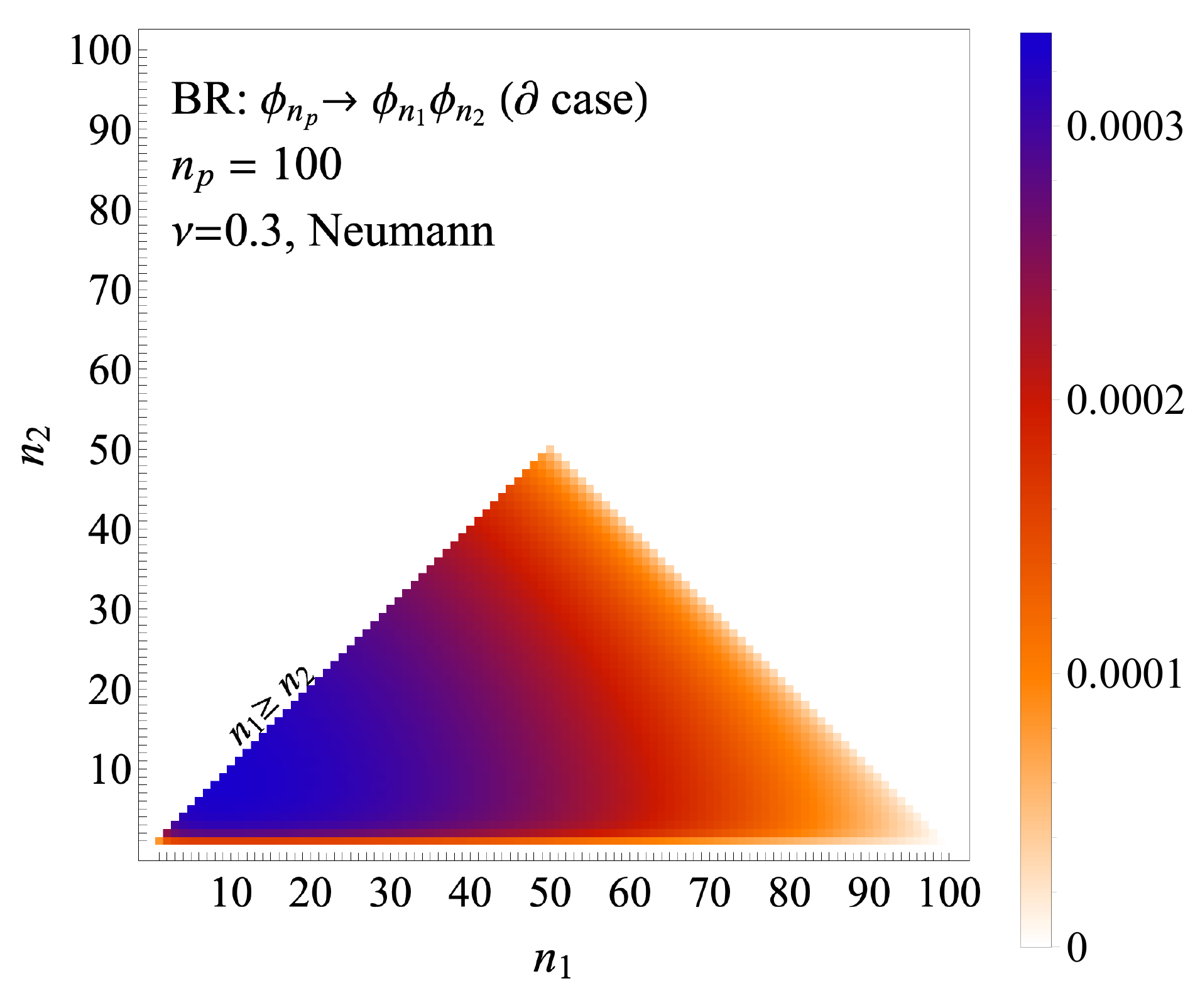}}\\\subfloat[]{\includegraphics[width=0.47\textwidth]{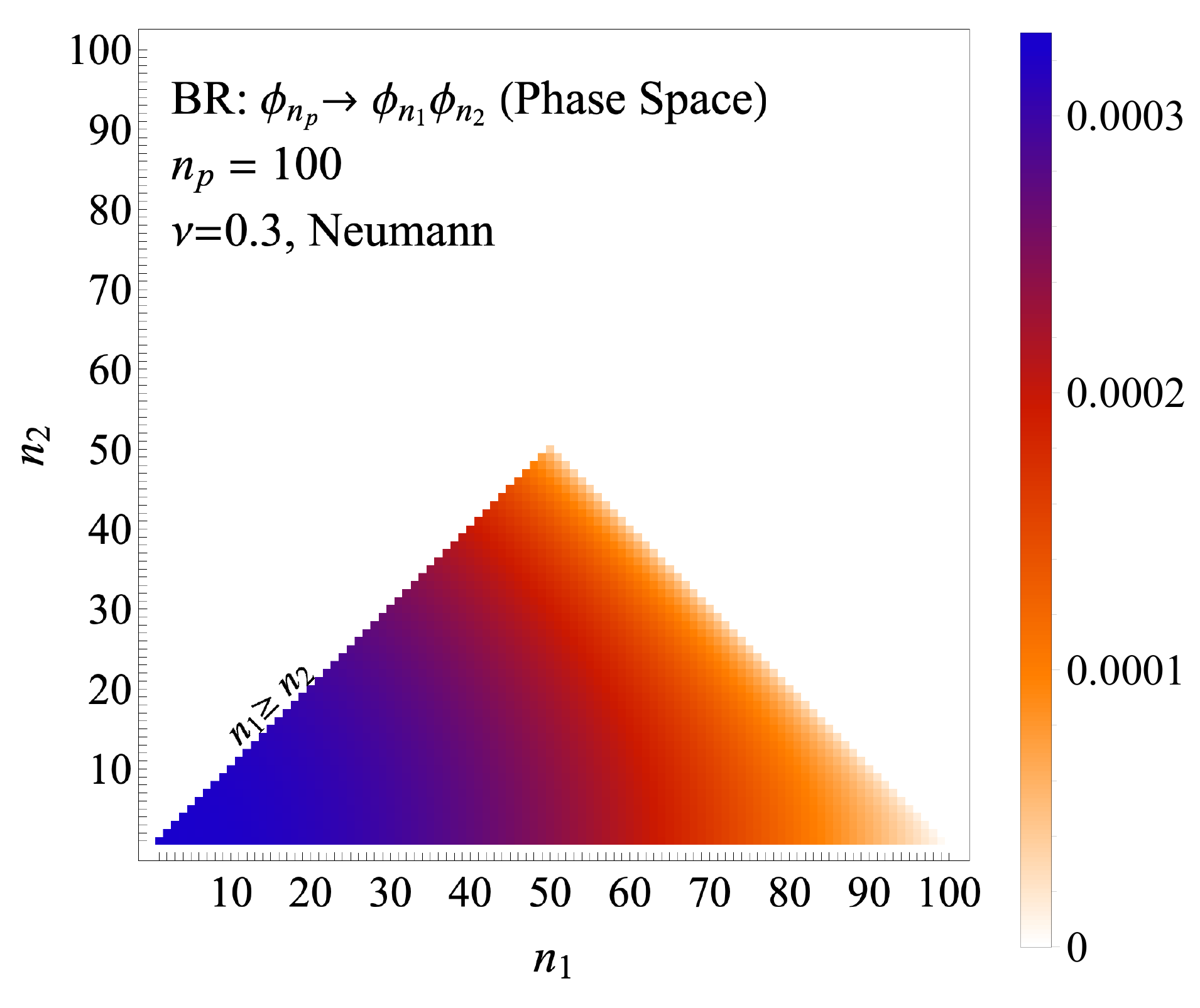}}\qquad\subfloat[]{\includegraphics[width=0.41\textwidth]{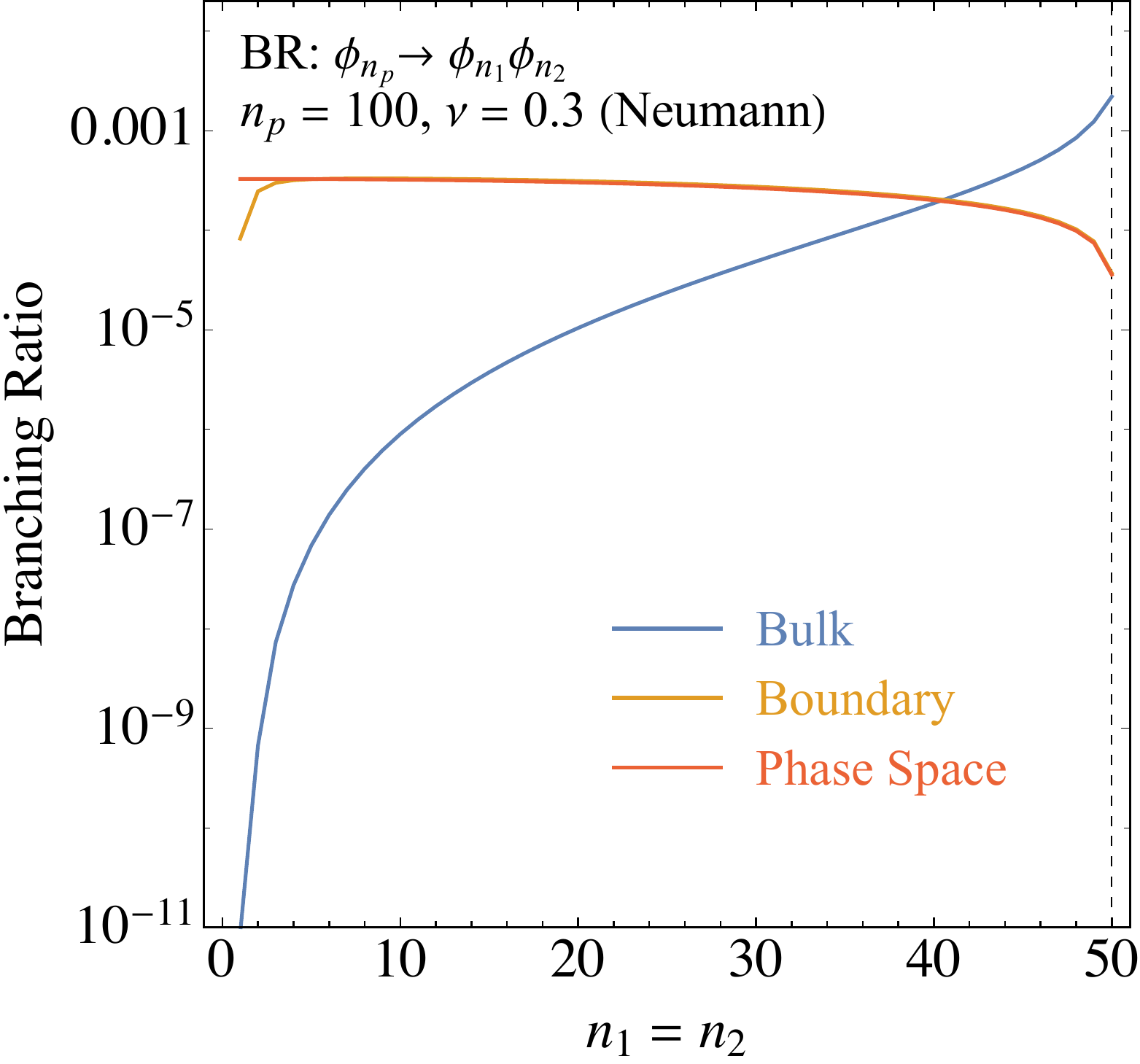}}
\caption{The branching ratios for $\nu=0.3$ of the $100$th KK mode into all kinematically allowed two-body final states, in scenarios with Neumann boundary conditions. The cases are (a) bulk couplings; (b) boundary-localized coupling; and (c) pure phase space. In (d), we show a 1d plot of the branching ratios along the $n_1 = n_2$ line. For the boundary-localized case (b), in contrast to Fig.~\ref{fig:brratios} and the bulk case, the decays significantly populate the full triangle, and favor daughter particles with substantial momentum. The distribution is similar to that of pure phase space (c). 
}
\label{fig:BRnu0p3bdry}
\end{figure}

In the scenarios discussed above, we have assumed that fields interact in the bulk of the extra dimension, as in \eqref{eq:interactions}, and we have taken the boundary conditions to be Dirichlet. A very different phenomenology arises if we assume that the  fields have Neumann boundary conditions, and  add the interaction term \eqref{eq:bdryinteractions} localized on the IR boundary of the space. As indicated in \eqref{eq:neumannIRvalue}, the boundary values of the different KK modes  are approximately equal, and so  decays are governed approximately by the phase space of  the final state. The branching ratios in this case are illustrated in Fig.~\ref{fig:BRnu0p3bdry}, which also shows the (similar) branching ratios that would result from pure phase space factors. Of course, one can also consider a model containing both the bulk coupling $c$ \eqref{eq:interactions} and the boundary coupling $\tilde c$ \eqref{eq:bdryinteractions}, obtaining physics that interpolates between the two. Below we will show some results from such a case with ${\tilde c} = 0.015 c$, which  turns out to produce  event-shape  distributions roughly midway between those of the pure bulk and pure boundary cases.

\begin{figure}[t!]
\centering
\subfloat[]{\includegraphics[width=0.47\textwidth]{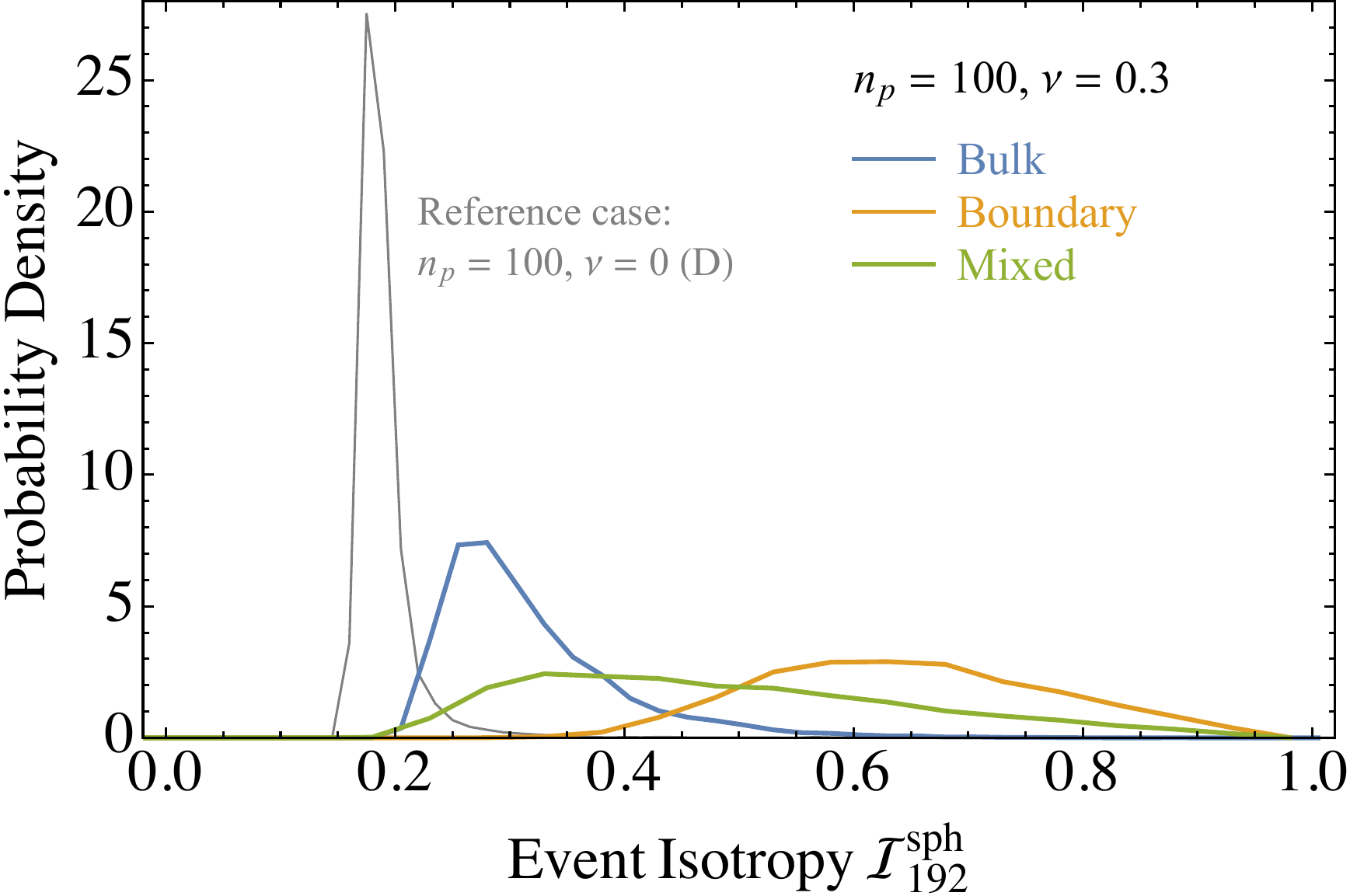}}
\quad \subfloat[]{\includegraphics[width=0.47\textwidth]{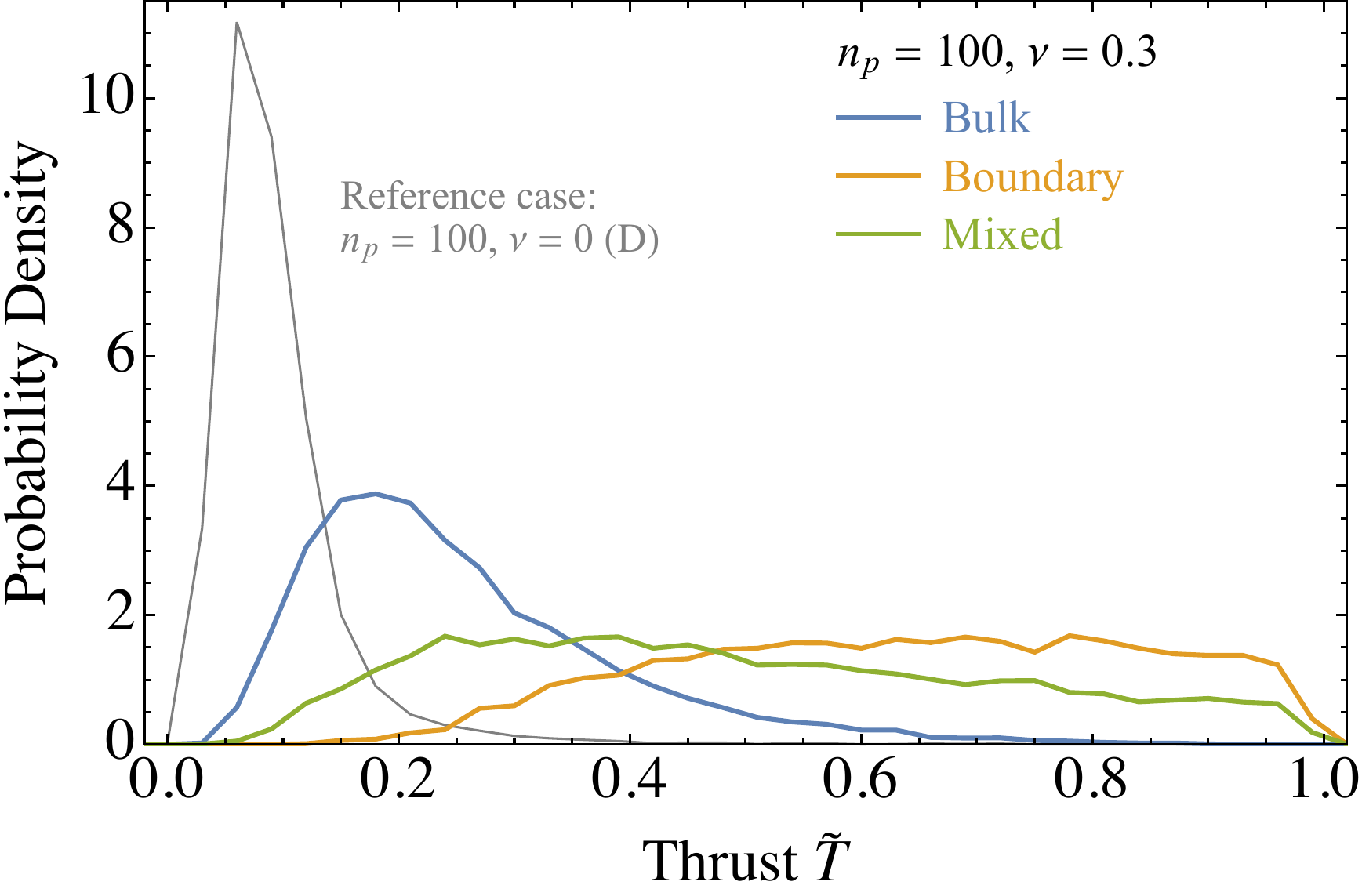}}\\
\subfloat[]{\includegraphics[width=0.47\textwidth]{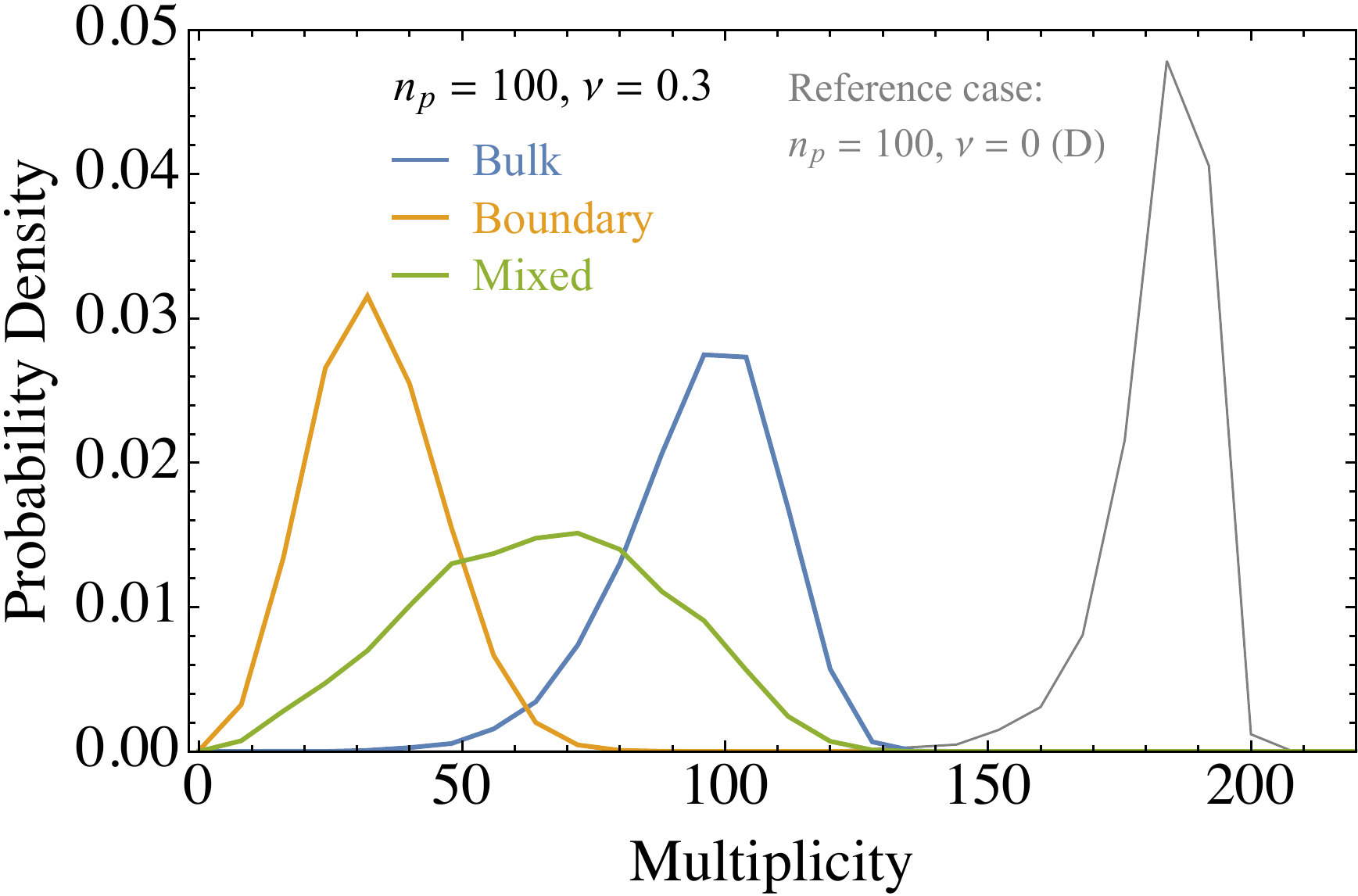}}
\quad
\subfloat[]{\includegraphics[width=0.47\textwidth]{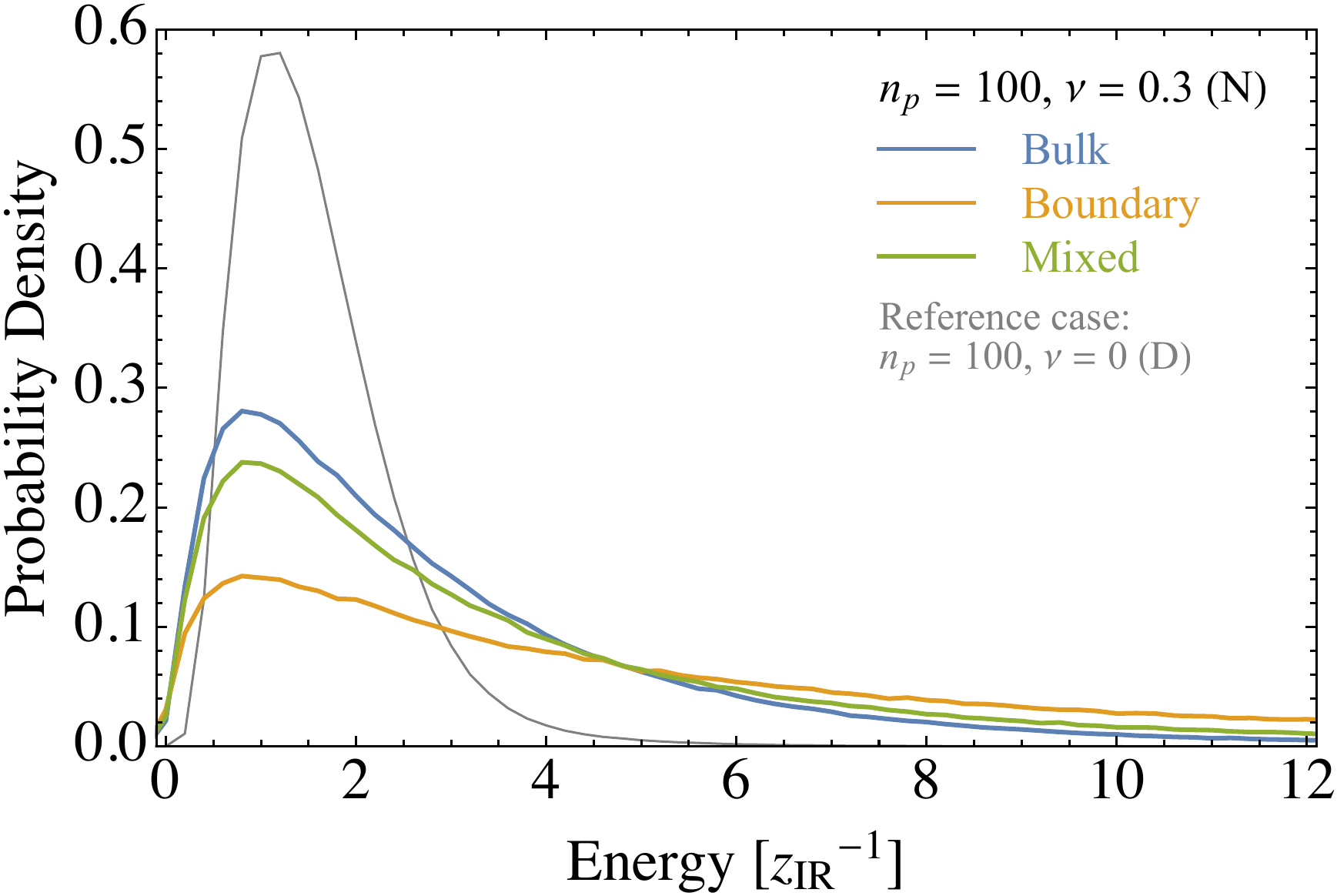}}
\caption{The distributions of (a) event isotropy $\iso{sph}{192}$; (b) rescaled thrust $\tilde T$; (c) particle multiplicity; and (d) energy of final-state particles for three different scenarios, all with $\nu = 0.3$, Neumann boundary conditions, and a decay cascade beginning with the $100$th KK mode. The couplings are: in blue, pure bulk ($\tilde c = 0$); in orange, pure boundary ($c = 0$); in green, a mixed case ($\tilde c = 0.015 c$). For ease of comparison with other figures, we also show the $\nu = 0$ Dirichlet case (originally displayed in Fig.~\ref{fig:toyModels}). We see that the cascade with a boundary coupling  produces less isotropic events (larger $\iso{sph}{192}$). Turning on both couplings interpolates between the bulk and boundary case, with a broad distribution of $\iso{sph}{192}$.
}
\label{fig:evisobulkvbdry}
\end{figure}

\begin{figure}[!h]
\centering
\subfloat[]{
\includegraphics[width=0.31\textwidth]{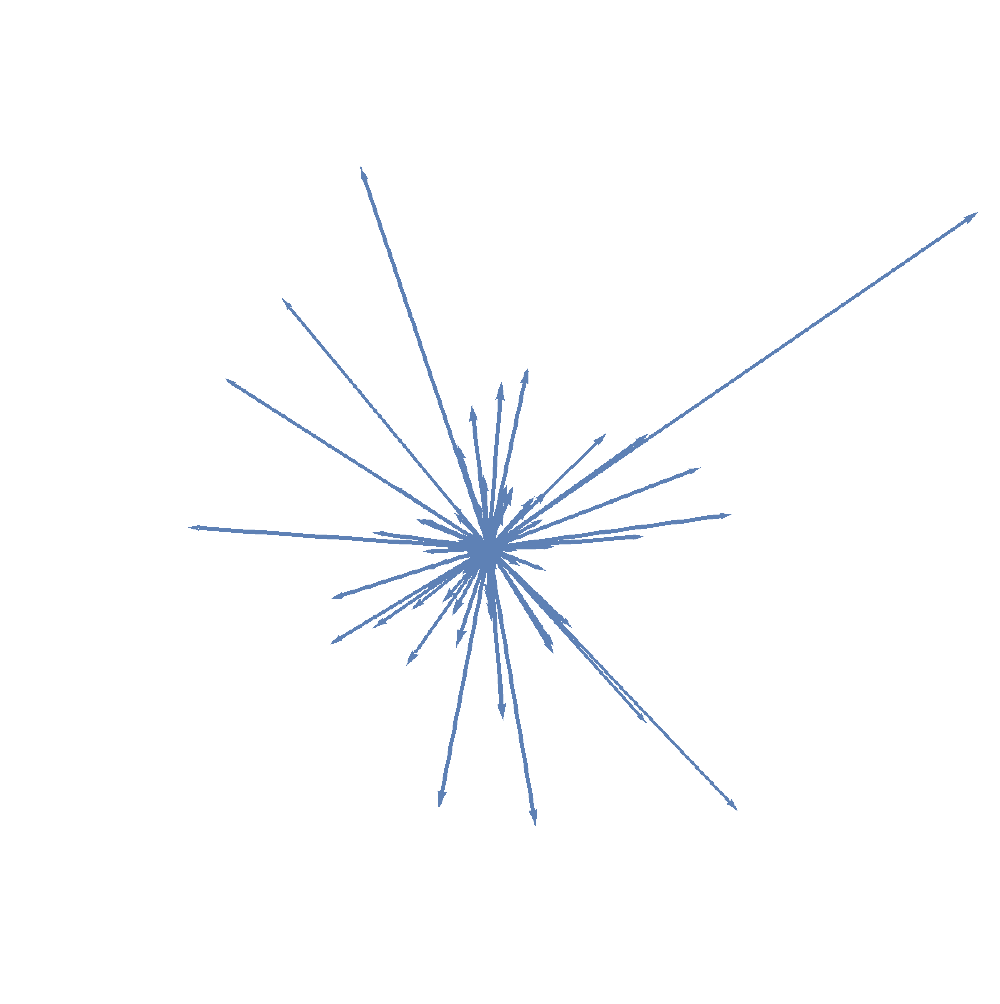}
}
\hfill
\subfloat[]{
\includegraphics[width=0.31\textwidth]{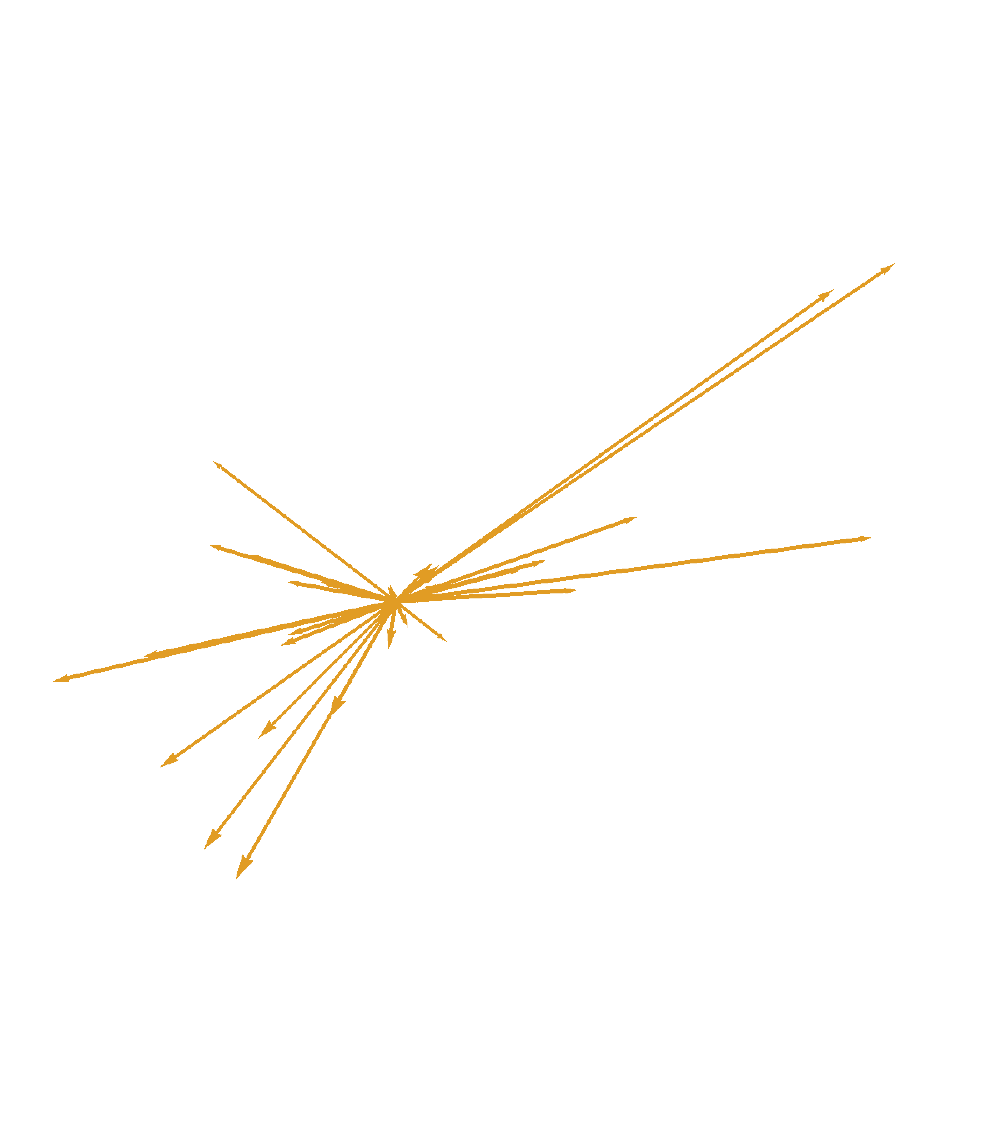}
}
\hfill
\subfloat[]{
\includegraphics[width=0.31\textwidth]{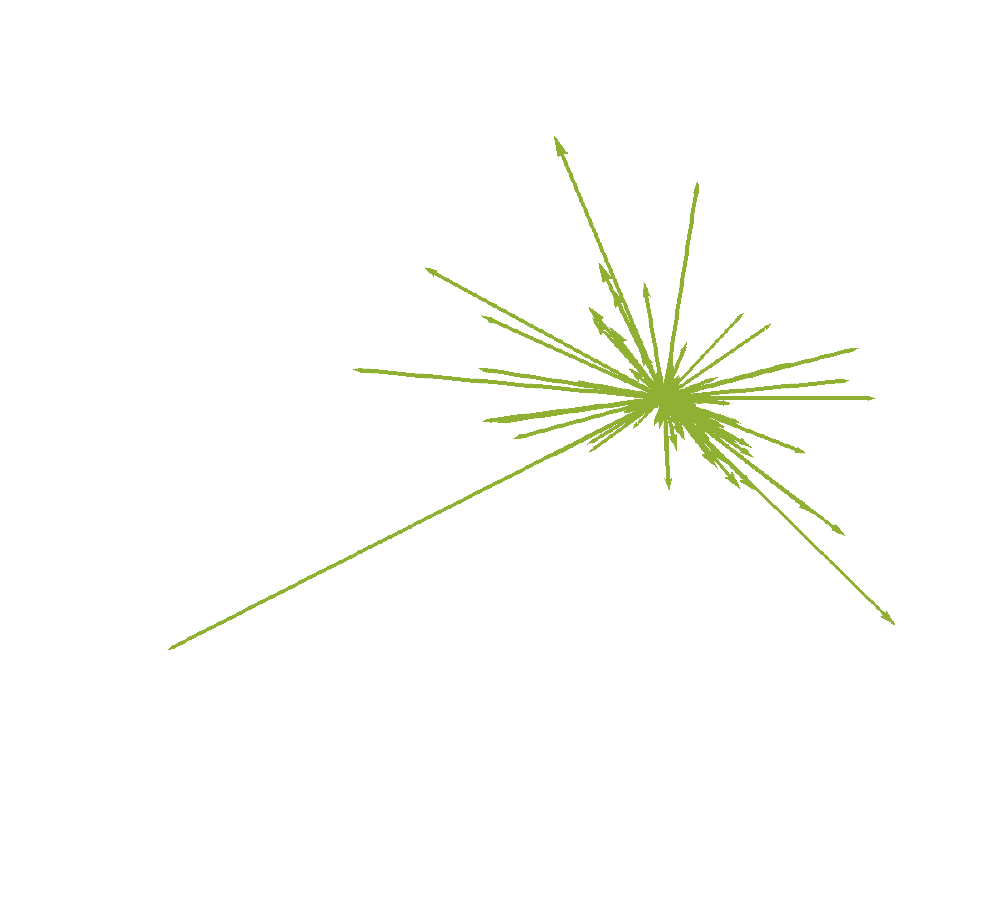}
}
\caption{ Visualizations of a characteristic final state radiation pattern for the (a) bulk coupling, (b) boundary  coupling, (c) mixed ($\tilde c = 0.015 c$) samples. In each case, $\nu  = 0.3$ and we assume Neumann boundary conditions. 
Each event has event isotropy equal to the mean value for the corresponding event sample: (a) $\langle \iso{sph}{192} \rangle = 0.33$; (b) $\langle \iso{sph}{192} \rangle = 0.67$; and (c) $\langle \iso{sph}{192} \rangle = 0.55$.
The plots show that events with boundary couplings are visibly less isotropic than the sample with only a bulk coupling.
}
\label{fig:bdryVis}
\end{figure}

In Fig.~\ref{fig:evisobulkvbdry}, we illustrate how boundary couplings affect the event shape. The blue curve shows the case of pure bulk couplings with Neumann boundary conditions,  for a single self-interacting field with $\nu=0.3$; as in Fig.~\ref{fig:singleFieldDist}, this leads to approximately spherical events with small values of $\iso{sph}{192}$ and thrust.\footnote{In the Neumann case, $I_-$ is suppressed and $I_+$ dominates; consequently the checkerboard pattern seen in \Fig{fig:brratios} is absent.} The orange curve shows events with boundary decays, which are substantially less isotropic. They also have a broader distribution of both event isotropy and thrust, although the event isotropy distribution is more peaked than the thrust distribution. Comparing to Fig.~\ref{fig:SMbenchmark}, we see that the typical event with boundary decays is more isotropic than a QCD dijet event, but has similar isotropy to a near-threshold $t \bar{t}$ event. Finally, we show the case with a mixed bulk/boundary coupling, $\tilde c = 0.015 c$, in green. As expected, this interpolates between the bulk and boundary cases. It does so by broadening the  distribution, rather than producing a narrow peak in  between the two cases. Visualizations of typical events are shown in Fig.~\ref{fig:bdryVis}.

Fig.~\ref{fig:evisobulkvbdry} also shows the typical multiplicity and energy of the individual final-state massless particles in the events. We see that the boundary cascades have much lower multiplicity, because the decays more often go directly to lighter daughters, so it takes fewer steps to reach the HSHs at the bottom of the cascade. Consequently, the individual particles also have more energy than they would for bulk couplings. This figure suggests that the event isotropy may be highly correlated with the particle multiplicity. Although this is true when we start all cascades with the same initial KK-number, it is not the case in general: boundary cascades remain much less isotropic even if they begin with a much larger choice of $n_p$. Thus event isotropy captures {\em different} information from particle multiplicity, or even from pairs of observables like particle multiplicity and thrust.  We will discuss this more in the companion paper \cite{paper2}.

\subsection{Evolution of Multiplicity, Event Isotropy in Cascade}
\label{subsec:evolmult}
\begin{figure}[t!]
\centering
\includegraphics[width=0.46\textwidth]{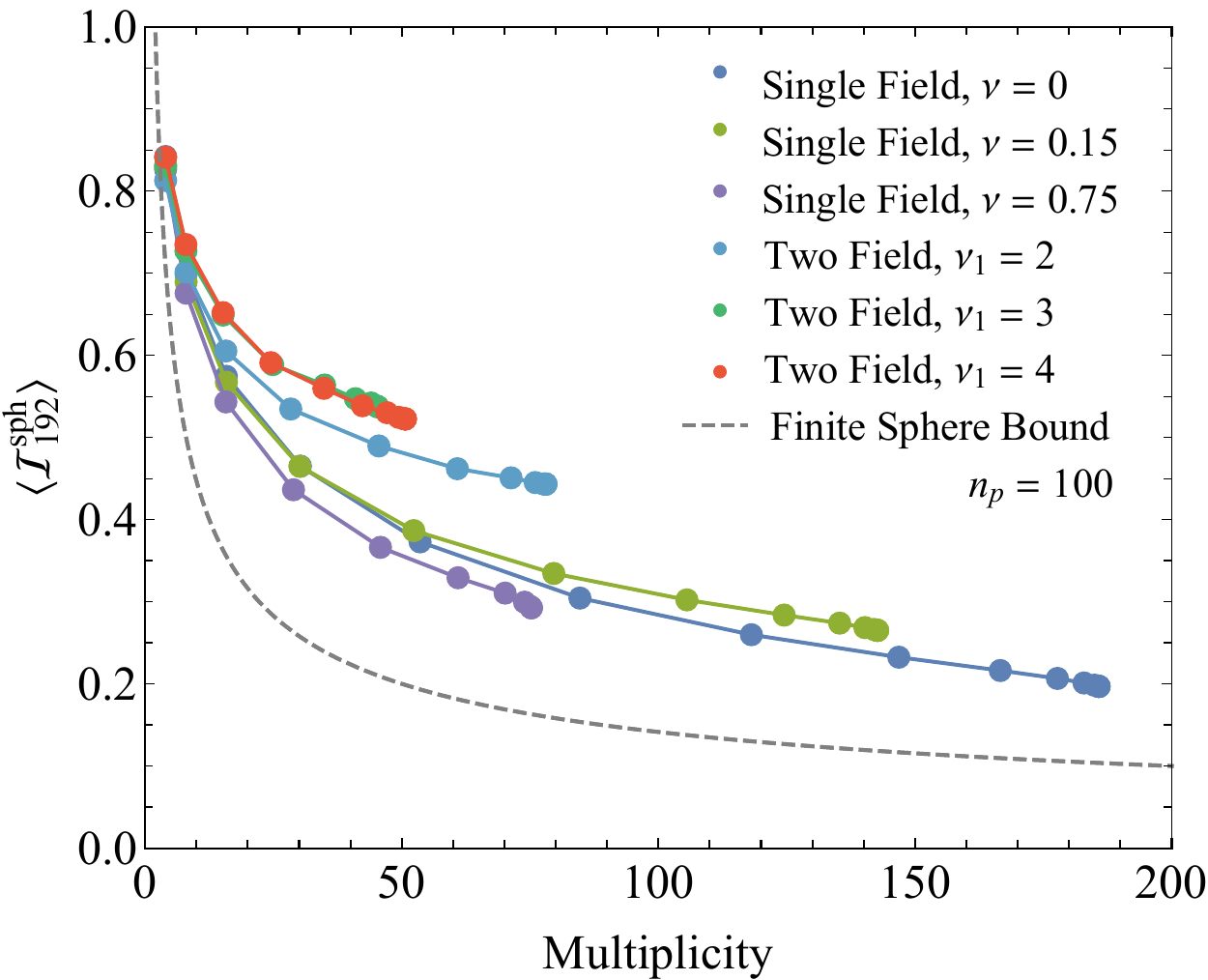}
\caption{The average multiplicity vs. average event isotropy at each step in the cascade decays for the single field samples and two field samples. }
\label{fig:multEMD}
\end{figure}
To investigate further the correlations between multiplicity and isotropy, it is interesting to see how these evolve through the decay cascade.
We do the following exercise. 
At each step in the cascade, we take all the hadrons present at that step (independent of whether they are stable against hadronic decays in following steps) and artificially force them to decay to massless particles. 
At the initial step of the cascade, with just the $n_p$ mode at rest, every event has a pencil-dijet with $\iso{sph}{192}=\tilde T=1$, while at the end of the cascade we obtain the samples studied above.
 In between, the multiplicity in each event gradually increases and $\iso{sph}{192}$ decreases.  

In \Fig{fig:multEMD}, we plot the average multiplicity and the average isotropy (averaged over $10^4$ events) at each step in the cascade. 
We also show, as a dashed line, the theoretical lower limit on $\iso{sph}{192}$ for the corresponding multiplicity; see \Eq{eq:isoest}. 
If multiplicity and event isotropy were perfectly correlated, then all of the cascades would lie on the same curve, even though at any given step, and at the end of the cascade, they would sit at different values. 
Instead, we see that the cascades for different choices of $\nu_i$ can give different curves. 
 This is additional evidence that isotropy measures more than multiplicity.  
 We will explore the independence and correlation of multiplicity, isotropy, thrust and other event shapes in \cite{paper2}.
%
%

\section{Analytical Estimates for Couplings}
\label{sec:analytic}

In \Sec{sec:simResults}, we saw that  different parameter choices lead to qualitatively different patterns of couplings $c_{ijk}$ and branching fractions, and from there to qualitatively different event shapes.  
In this section, we provide analytic calculations of the overlap integrals that determine the couplings, in order to substantiate and explain the results presented above. In particular, an {\em inverse} dependence on the phase-space function $\lambda_\text{PS}$ will be manifest in our results, explaining the cases in which we have observed a preference for near-threshold decays. We will also  understand the oscillatory behavior that gives  rise to the observed plateaus in the two-field case.

\subsection{Basic ingredients and general strategy}

Our goal is to gain an analytic understanding of integrals of the form
\begin{equation}
I(\nu_i, m_i) \equiv \int_0^1 {\rm d}z\, z\, \J{\nu_1}{m_1 z} \, \J{\nu_2}{m_2 z} \, \J{\nu_3}{m_3 z}.
\label{def:Iintegral}
\end{equation}
We choose a convention where particle \#1 is the heaviest particle, i.e., we assume without loss of generality that $m_1 \geq m_2 + m_3$. We can separate our integral into two pieces, 
\begin{align}
I(\nu_i, m_i) &= I_+(\nu_i, m_i) - I_-(\nu_i, m_i), \quad \text{where:} \nonumber \\
I_+(\nu_i, m_i) &\equiv \int_0^{\infty} {\rm d}z\, z\, \J{\nu_1}{m_1 z} \, \J{\nu_2}{m_2 z} \, \J{\nu_3}{m_3 z}, \nonumber \\
I_-(\nu_i, m_i) &\equiv \int_1^{\infty} {\rm d}z\, z\, \J{\nu_1}{m_1 z} \, \J{\nu_2}{m_2 z} \, \J{\nu_3}{m_3 z}. \label{eq:IpImdef}
\end{align}
The reason for doing this is that when all of the masses (in units of the IR brane scale) are sufficiently large, i.e., when $m_i \gg \nu_i^2$, we can approximate the Bessel functions in the integrand of $I_-(\nu_i, m_i)$ by their large-argument asymptotic expansions. This makes approximating $I_-(\nu_i, m_i)$ into an analytically tractable problem. On the other hand, the integral $I_+(\nu_i, m_i)$ over the whole positive real axis is known analytically. By combining the exact analytic answer for $I_+$ and the approximate answer for $I_-$, we obtain an analytic approximation to the $I(\nu_i, m_i)$ and hence to the couplings among Kaluza-Klein modes.

\begin{figure}[!h]
\centering
\includegraphics[width=1.0\textwidth]{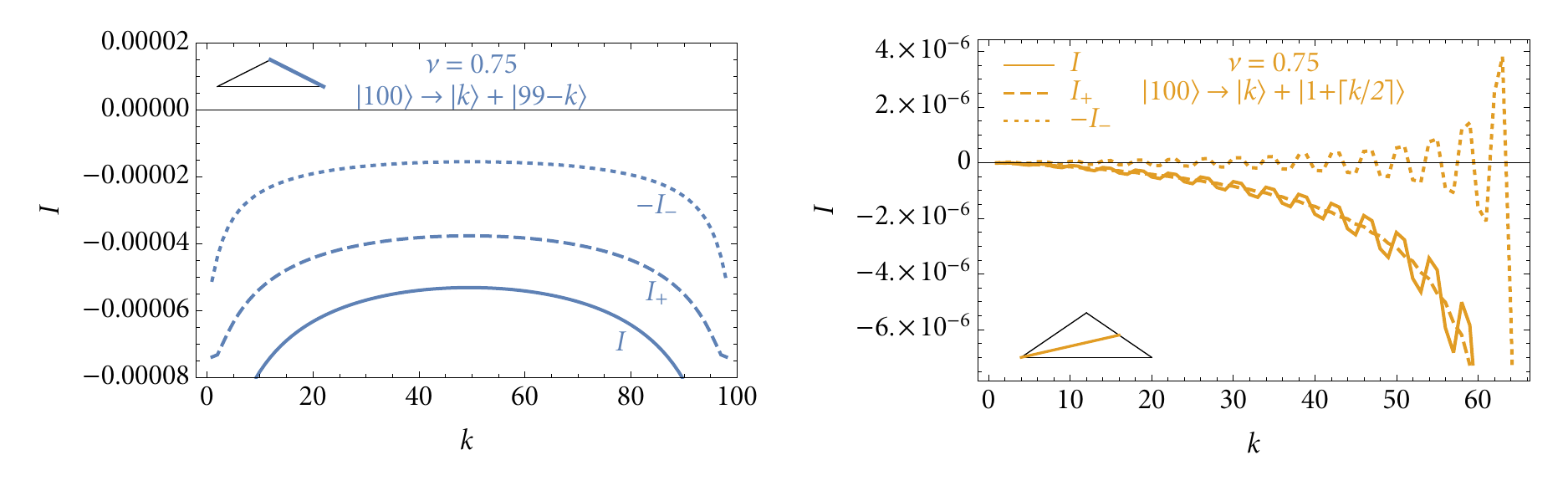}
\caption{Examples of the decomposition of the integral $I$ (solid) into a sum of $I_+$ (dashed) and $-I_-$ (dotted), in the single-field case where $\nu_1 =  \nu_2 = \nu_3 \equiv \nu = 0.75$. In both plots, the inset triangle shows the slice of the branching ratio triangle (as in Fig.~\ref{fig:brratios}) along which we have done the calculation. At left, we take near threshold decays (right-hand edge of the triangle); at right, we take decays along a slice through the middle of the triangle. For near-threshold decays, we see that $I_+$ and $I_-$ are comparable. Away from threshold, $I_+$ dominates, with $I_-$ contributing a small oscillatory pattern.}
\label{fig:Ipmdecomp}
\end{figure} 

In Fig.~\ref{fig:Ipmdecomp}, we show examples of how $I$ breaks down into contributions from $I_+$ and $-I_-$, along two slices of the branching ratio triangles in the single-field case with $\nu = 0.75$. Typically, $I_+$ dominates, and the integral is suppressed for decays to light modes. Near threshold, $I_-$ gives a contribution comparable to that of $I_+$. As we will see below, in certain special cases (like $\nu = 0$), the integral $I_+$ is identically zero, but the behavior seen in the plot is representative of more general $\nu$ values.

Our goal in the remainder of this section is to provide some analytic insight into the behavior of the integrals $I_+$ and $I_-$. We will first discuss an analytic approximation to $I_-(\nu_i, m_i)$, and compare it to numerical results. Then we will present an exact analytic formula for $I_+(\nu_i, m_i)$, and comment on a special case to elucidate the ``plateau'' structure observed in Fig.~\ref{fig:brTwoField}.

\subsection{Approximating the integral $I_-(\nu_i, m_i)$}

We apply the large-argument asymptotic approximation of the Bessel function, \eqref{eq:besselasymptotic}, to estimate $I_-(\nu_i, m_i)$:
\begin{equation}
I_-(\nu_i, m_i) \approx I^{(0)}_-(\nu_i, m_i) \equiv \int_1^\infty {\rm d}z\, z\prod_{i=1}^3\sqrt{\frac{2}{\cpi m_i z}} \cos\left(m_i z - \frac{1}{2} \cpi \nu_i - \frac{1}{4} \cpi\right).
\label{eq:Iminus0def}
\end{equation}
We can rewrite the product of three cosines as a sum of four cosines by repeated use of the identity $2\cos a\cos b = \cos(a+b)+\cos(a-b)$. Given signs $\sigma, \sigma' \in \{+1,-1\}$, we define
\begin{equation}
m_{\sigma  \sigma'} \equiv m_1 + \sigma m_2 + \sigma' m_3, \quad \nu_{\sigma \sigma'} \equiv \nu_1 + \sigma \nu_2 + \sigma' \nu_3.
\end{equation}
In labels, we will suppress the ``1'' and write the $\sigma$'s as $+$ or $-$; for example, we denote $m_1 + m_2 - m_3$ by $m_{+-}$. Then \eqref{eq:Iminus0def} is equivalent to
\begin{equation}
I^{(0)}_-(\nu_i, m_i) = \frac{1}{\cpi^{3/2} \sqrt{2m_1 m_2 m_3}} \int_1^\infty \frac{{\rm d}z}{\sqrt{z}} \sum_{\sigma, \sigma' \in \{+,-\}} \cos\left(m_{\sigma \sigma'}z - \frac{\cpi}{2} \nu_{\sigma \sigma'} - \frac{\cpi}{4}(1 + \sigma + \sigma')\right).
\label{eq:Iminusfourterms}
\end{equation}
This integral can be performed analytically in terms of the Fresnel cosine and sine integrals. Our conventions for these are specified in Appendix \ref{sec:fresnel}. We obtain:
\begin{align}
I^{(0)}_-(\nu_i, m_i) = \frac{1}{\cpi \sqrt{m_1 m_2 m_3}} \sum_{\sigma, \sigma' \in \{+,-\}} \sqrt{\frac{1}{m_{\sigma \sigma'}}} \Bigg[ & \cos\left(\frac{\cpi}{2} \nu_{\sigma\sigma'} + \frac{\cpi}{4}(1+\sigma+\sigma')\right)\left(\frac{1}{2} - \FC{\sqrt{\frac{2m_{\sigma \sigma'}}{\cpi}}}\right) + \nonumber \\
&  \sin\left(\frac{\cpi}{2} \nu_{\sigma\sigma'} + \frac{\cpi}{4}(1+\sigma+\sigma')\right)\left(\frac{1}{2} - \FS{\sqrt{\frac{2m_{\sigma \sigma'}}{\cpi}}}\right)\Bigg].
\label{eq:IminusFresnel}
\end{align}
This is already a useful approximation to $I_-(\nu_i, m_i)$. We can go further by noting that, provided the masses are all large and that we do not consider decays too close to threshold, we can exploit the large-argument asymptotics of the Fresnel integrals, \eqref{eq:Fresnelasymptotics}. In this case, we find that the answer depends on $\sin$ and $\cos$ of $m_{\sigma \sigma'}$, so we make use of the mass eigenvalue estimates in \eqref{eq:KKmassestimate}. We will provide the estimate in the case of Dirichlet boundary conditions.

\subsubsection{$I_-(\nu_i,m_i)$ estimate for Dirichlet boundary conditions}

Keeping the first subleading, ${\cal O}(1/z)$, term in the Fresnel integral asymptotics, the expression in brackets in \eqref{eq:IminusFresnel} becomes
\begin{align}
\frac{1}{\sqrt{2\cpi \,m_{\sigma \sigma'}}} \Bigg[ &- \cos\left(\frac{\cpi}{2} \nu_{\sigma\sigma'} + \frac{\cpi}{4}(1+\sigma+\sigma')\right) \sin(m_{\sigma \sigma'}) + \sin\left(\frac{\cpi}{2} \nu_{\sigma\sigma'} + \frac{\cpi}{4}(1+\sigma+\sigma')\right) \cos(m_{\sigma \sigma'})\Bigg],
\end{align}
Using the sine sum-of-angles identity, then using \eqref{eq:KKmassestimate} to replace the masses in the argument of the sine by an approximate expression in terms of the KK mode numbers $n_i$, we obtain a simple approximate formula for the Dirichlet case:
\begin{align}
I^{(1)}_-(\nu_i, m_i) &\equiv -\frac{1}{\cpi^{3/2} \sqrt{2m_1 m_2 m_3}} \sum_{\sigma, \sigma' \in \{+,-\}} \frac{1}{m_{\sigma \sigma'}} \sin\left(\cpi (n_1 + \sigma n_2 + \sigma' n_3) - \frac{\cpi}{2} (1 + \sigma + \sigma')\right) \nonumber \\
&= \frac{(-1)^{n_1 + n_2 + n_3 + 1}}{\cpi^{3/2} \sqrt{2m_1 m_2 m_3}} \sum_{\sigma, \sigma' \in \{+,-\}} \frac{\sigma \sigma'}{m_{\sigma \sigma'}},
\label{eq:IminusDirichlet}
\end{align}
where $n_i$ is the integer mode number of the corresponding KK mode. In particular, there is a term that scales as $1/(m_1 - m_2 - m_3)$ that dominates when the phase space is relatively small (but not so small that our approximations break down). We can write this term using the phase space factor \eqref{eq:lambdaPS}
\begin{equation}
I^{(1)}_-(\nu_i, m_i) \approx \frac{(-1)^{n_1 + n_2 + n_3 + 1} 4 \sqrt{2}}{\cpi^{3/2}} \frac{\sqrt{m_1 m_2 m_3}}{\lambda_\text{\rm PS}(m_1^2,m_2^2,m_3^2)},
\label{eq:IminusDirichletSmallPS}
\end{equation}
which we used in \eqref{eq:cijk_nu_even}.

The large-argument expansion of the Fresnel functions breaks down at $m_{\sigma \sigma'} \lesssim 1$, very close to threshold. This case generally only arises if we tune the values of $\nu$ so that the offsets in the Bessel function zeros align to allow for $m_3 \approx m_1 + m_2$, so we do not expect that it is generally relevant, but we discuss it for completeness. In this case, the sum is dominated by the {\em small}-argument expansion of the Fresnel functions for the $\sigma, \sigma' = -1$ case,
\begin{align}
I^{(\mathrm{threshold})}_-(\nu_i, m_i) &\approx \frac{1}{\cpi \sqrt{2 m_1 m_2 m_3}}  \frac{1}{\sqrt{m_{--}}} \sin\left(\frac{\cpi}{2} (\nu_1 - \nu_2 - \nu_3)\right) \nonumber \\
&\approx \frac{2}{\cpi} \frac{ \sin\left[\frac{\cpi}{2} (\nu_1 - \nu_2 - \nu_3)\right] }{\lambda_\mathrm{PS}^{1/2}(m_1^2,m_2^2,m_3^2)}.
\label{eq:IminusDirichletVerySmallPS}
\end{align}
Thus, very close to threshold, we expect that the divergence is ameliorated to $\lambda_\mathrm{PS}^{-1/2}$, except in cases where $\nu_1 - \nu_2 - \nu_3$ is an even integer, when this term has coefficient zero and subleading terms dominate. As we will see below, this is also the near-threshold behavior of the integral $I_+$, so we predict that $I_-$ does not parametrically dominate over $I_+$ in the small phase-space region.   In fact the original integral is finite at threshold, so the singular behavior of $I_+$ and $I_-$ must cancel there. 

Unlike the Dirichlet case, note that the large-argument approximation for the Fresnel function would have given zero for the Neumann case, because the different constant term in \eqref{eq:KKmassestimate} removes the $\frac{\cpi}{2} (1 + \sigma + \sigma')$ term in the argument of the sine and leads to zero. To obtain a similar approximation in the Neumann case, we must keep subleading terms in the various approximations we have made. We will not do so here.

\begin{figure}[!h]
\centering
\includegraphics[width=0.6\textwidth]{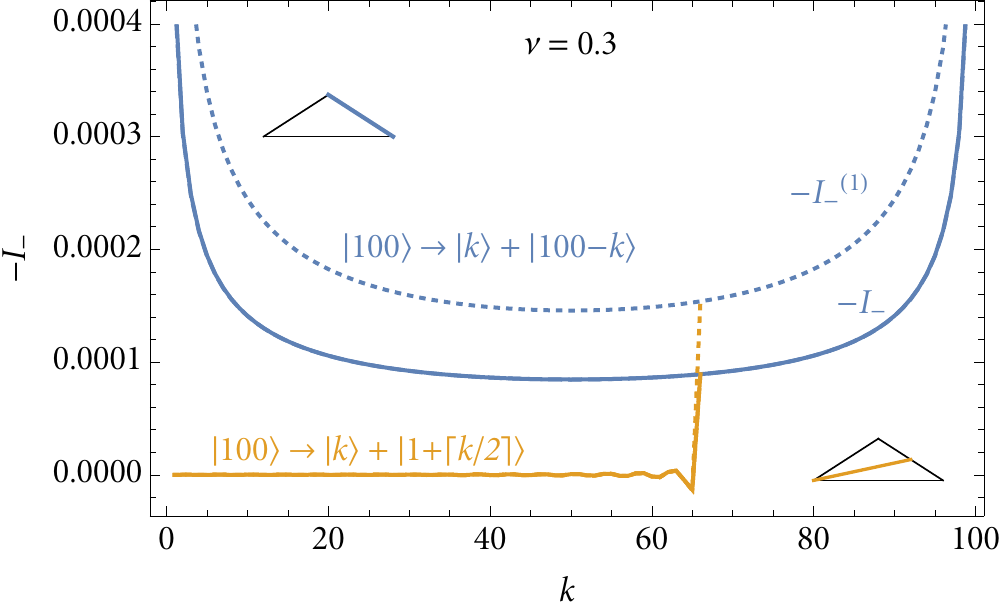}
\caption{Comparison of a numerical computation of the integral $I_-$ (solid curves) with  the two analytic approximations $I_-^{(0)}$, \eqref{eq:IminusFresnel} (dashed curves) and $I_-^{(1)}$, \eqref{eq:IminusDirichlet} (dotted curves). The dashed curves always fall so close to the solid curve that they are indistinguishable. The simpler analytic formulation, $I_-^{(1)}$, works very well away from threshold but deviates close to the threshold, leading to the visible dotted curves. As in Fig.~\ref{fig:Ipmdecomp}, the inset triangles illustrate the slice through the branching ratio triangle that is plotted. We show the case $\nu = 0.3$ because it shows a larger discrepancy, and thus a more visible dotted curve, than the case $\nu = 0.75$.}
\label{fig:Iminusanalytic}
\end{figure} 

We compare a numerical computation of the integral $I_-$ with the two approximations $I_-^{(0)}$ \eqref{eq:IminusFresnel} and $I_-^{(1)}$ \eqref{eq:IminusDirichlet} in Fig.~\ref{fig:Iminusanalytic}. The first approximation, based on the large-argument expansion of the Bessel functions, works extremely well. The subsequent approximations  made in the Dirichlet  case lead to an imprecise estimate, but one which is useful since the formula \eqref{eq:IminusDirichlet} makes the enhancement of the integral in the small phase-space region obvious.

\subsection{Computing the integral $I_+(\nu_i, m_i)$}

\subsubsection{General formula}

In the case of $I_-$, we used the asymptotic expansion of the Bessel function to obtain an analytic approximation. We can do better with $I_+$: the integral is analytically known (eq.~(7.1) of \cite{bailey1936some}, in the special case $\lambda = 2$) to be:
\begin{align}
I_+(\nu_i, m_i) &= \frac{2 m_2^{\nu_2} m_3^{\nu_3} m_1^{-\nu_2 - \nu_3 - 2} \ \Gamma\left(\frac{\nu_1 + \nu_2 + \nu_3}{2}+1\right)}{\Gamma \left( \nu_2 + 1 \right) \Gamma \left(\nu_3 + 1 \right) \Gamma \left( \frac{\nu_1 - \nu_2 - \nu_3}{2}\right)} \nonumber \\
&\times \mathrm{F}_4 \left( 1+\frac{\nu_2+\nu_3-\nu_1}{2},1+\frac{\nu_1+\nu_2+\nu_3}{2}; \nu_2+1, \nu_3 +1;  \frac{m_2^2}{m_1^2}, \frac{m_3^2}{m_1^2}\right).
\label{eq:GandR}
 \end{align}
The function $\mathrm{F}_4$ is known as an Appell function; it is a two-variable generalization of a hypergeometric function. This integral has previously appeared in the physics literature on 3-point correlators in momentum space in conformal field theories \cite{Coriano:2013jba, Bzowski:2013sza, Gillioz:2019lgs} and de Sitter space \cite{Antoniadis:2011ib, Sleight:2019mgd}. For convenience, we include its definition in Appendix~\ref{sec:appell}.

The behavior of the Appell function leads to one of the important qualitative features we have observed in our numerical results: the existence of plateaus of large branching fractions separated by valleys of suppressed branching fractions. It is manifest from the series definition of the Appell function that it becomes a polynomial when its first argument is a non-positive integer. When $\nu_1=\nu_2+\nu_3+2k$, for $k$ a positive integer, the polynomial has degree $k-1$, and has $k-1$ curves of zeroes. Numerical results, supported by incomplete analytic arguments, indicate that  these zeros always lie in the physical region $m_2+m_3<m_1$.\footnote{Along the line $m_3 = 0$, this follows from the fact that $\mathrm{F}_4$ is simply $\Hyp$, discussed in Appendix~\ref{sec:appell}, together with Theorem 3.2.i of Ref.~\cite{driver2008zeros}. In the special case $\nu_2 = \nu_3$, it can also be proven along the line $m_2 = m_3$, using the same theorem together with \Eq{eq:Iplus2F1} below. However, we lack a completely general proof.}  Two special cases of interest to our two field case are $k=1$, for which the Appell function is 1, and $k=2,\nu_2=\nu_3$, for which it is 
\be
\mathrm{F}_4 \left( -1,2 \nu_2+3; \nu_2+1, \nu_2 +1;  \frac{m_2^2}{m_1^2}, \frac{m_3^2}{m_1^2}\right) = 1 - \frac{2\nu_2+3}{\nu_2+1}\left(\frac{m_2^2+m_3^2}{m_1^2}\right).
\label{eq:F4k2}
\ee
The zeros of the Appell functions form the valleys between plateaus observed in Fig.~\ref{fig:brTwoField}. Even in cases where the Appell function is not a polynomial, we expect that $\floor*{\frac{\nu_1 - \nu_2 - \nu_3}{2}}$, when positive, approximately counts the number of plateaus.

The expression \Eq{eq:GandR} can always be reduced to an expression in terms of ordinary hypergeometric functions \cite{bailey1933reducible, gervois1985integrals, gervois1986some, GervoisNaveletUnpublished}
{\small
\begin{align}
I_+(\nu_i, m_i) &= \left(\frac{m_2}{m_1}\right)^{\nu_2}\left(\frac{m_3}{m_1}\right)^{\nu_3} \frac{2}{\lambda_{\rm PS}^{1/2}(m_1^2,m_2^2,m_3^2)} \frac{\Gamma\left(\frac{\nu_1 + \nu_2 + \nu_3}{2}+1\right)}{\Gamma \left( \nu_3 + 1 \right) \Gamma \left(\nu_2 + 1 \right) \Gamma \left( \frac{\nu_1 - \nu_2 - \nu_3}{2}\right)} \nonumber \\
\times &\Bigg[\frac{\nu_1 + \nu_3 - \nu_2}{2\nu_1} \Hyp\left(1 + \frac{\nu_2+\nu_3 - \nu_1}{2},\frac{\sum_i \nu_i}{2},\nu_2 + 1; X\right) \Hyp\left(\frac{\nu_2+\nu_3 - \nu_1}{2},1+\frac{\sum_i \nu_i}{2},\nu_3 + 1; Y\right) \nonumber \\
+ & ~ \frac{\nu_1 + \nu_2 - \nu_3}{2\nu_1} \Hyp\left(1 + \frac{\nu_2+\nu_3 - \nu_1}{2},\frac{\sum_i \nu_i}{2},\nu_3 + 1; Y\right) \Hyp\left(\frac{\nu_2+\nu_3 - \nu_1}{2},1+\frac{\sum_i \nu_i}{2},\nu_2 + 1; X\right) \Bigg],  \nonumber \\
\text{where } X &\equiv \frac{2m_2^2}{m_1^2+m_2^2 - m_3^2 + \lambda_{\rm PS}^{1/2}(m_1^2,m_2^2,m_3^2)}, ~Y \equiv \frac{2m_3^2}{m_1^2-m_2^2 + m_3^2 + \lambda_{\rm PS}^{1/2}(m_1^2,m_2^2,m_3^2)}.
   \label{eq:Iplus2F1}
\end{align}
}
The variables $X$ and $Y$ are defined so that $(m_2/m_1)^2 = X(1-Y)$ and $(m_3/m_1)^2 = Y(1-X)$. 
We comment on some details of the reduction from $\mathrm{F}_4$ to $\Hyp$ in Appendix \ref{sec:hypergeom}.

 We can extract from \eqref{eq:Iplus2F1}  a few simple, general points:
\begin{itemize}
\item The integral $I_+$ has a phase-space factor $\lambda_{\rm PS}^{1/2}(m_1^2,m_2^2,m_3^2)$ in the {\em denominator}. Thus, precisely when the general decay width formula \eqref{eq:width} would lead one to naively expect a decay to be rare, the coupling squared enhances it via an {\em inverse} dependence of the decay width on the phase space. This accounts for the general tendency of the 5d models to feature many near-threshold decays.
\item The hypergeometric functions have the property $\lim_{z \to 0} \Hyp(a,b,c; z) = 1$, so they are unsuppressed when $X$ or $Y$ is small. However, the prefactors $(m_2/m_1)^{\nu_2} (m_3/m_1)^{\nu_3}$ indicate that, in general, the decay rates to light daughters are suppressed. Even in the special case $\nu_2 = \nu_3 = 0$ when this suppression is absent, the normalization factor $N^{(\nu)}_n$ {\em is} smaller for light daughters than heavy ones, as indicated in \eqref{eq:dirNorm}. Thus, in general, we expect that decays to light daughters are rare, as confirmed by the numerical results in Fig.~\ref{fig:brTwoField}, for example. 
\item  Recall that $\Gamma(x)$ has poles whenever $x$ is a nonpositive integer. As a result, whenever $\nu_2 + \nu_3 - \nu_1$ is a positive even integer, the integral $I_+(\nu_i, m_i)$ vanishes. In the special case $\nu_1 = 0$, there are also zeros in the denominator, but taking the limit from nonzero $\nu_1$'s shows that $I_+$ vanishes in this case as well (as we saw in Fig.~\ref{fig:Ipmdecomp}, where all $\nu$'s were zero).
\item We have $X, Y \in (0,1)$, so that the hypergeometric functions are nonsingular in the physical region except perhaps near threshold. In the near threshold region $m_2 + m_3 \to m_1$, we have $X \to m_2/m_1$ and $Y \to m_3/m_1$. The hypergeometric functions may be singular when $X \to 1$ or $Y \to 1$. For concreteness, consider the case $X \to 1$. This requires that $m_3 \to 0, m_2 \to m_1$. In this case, the $\Hyp$ functions of $X$ diverge as $(1-X)^{-\nu_3}$. However, this is compensated by the prefactor $(m_3/m_1)^{\nu_3}$. As a result, the prefactor of $\lambda_{\rm PS}^{-1/2}$ is the only source of singularities at the boundary of the physical region.
\end{itemize}

\subsubsection{Plateau structure in the special case $\nu_2 = \nu_3 = 1/2, \nu_1 \in \mathbb{Z}$}
\label{sec:Cheby}

We already noted that ${\rm F}_4$ becomes a polynomial when $\nu_1 - \nu_2 -  \nu_3$ is a positive even integer; in this case, the hypergeometric functions in \eqref{eq:Iplus2F1} are also simply polynomials in $X$ and  $Y$. We will now present a special case for which we can give a straightforward derivation of the integral $I_+$ in terms of Chebyshev polynomials. This is one of the simplest cases in which the existence of plateaus and valleys can be deduced.

Spherical Bessel functions have simple expressions in terms of trigonometric functions. In particular, for $\nu = 1/2$, we have the identity
\begin{equation}
\J{1/2}{x} = \sqrt{\frac{2}{\cpi x}} \sin x.
\end{equation}
In the special case $\nu_2 = \nu_3 = 1/2$, this reduces our overlap integral to
\begin{equation}
\frac{2}{\cpi} \int_0^{z_{\rm IR}} {\rm d}z\, \J{\nu_1}{m_1 z} \sin(m_2 z) \sin(m_3 z) = \frac{1}{\cpi} \int_0^{z_{\rm IR}} {\rm d}z\, \J{\nu_1}{m_1 z} \left[\cos\left((m_2 - m_3) z\right) - \cos\left((m_2 + m_3) z \right)\right].
\end{equation}
The infinite integral $\int_0^\infty {\rm d}t\, \J{\nu}{a t}\cos(b t)$ is known  \cite[\S 13.42]{WatsonTreatise}. Here, we will provide a  clear derivation in the special case when $\nu$ is an integer. The result is 0 if $\nu$ is odd, whereas if $\nu$ is even and $b < a$, it  is given by $(-1)^{\nu/2} T_{\nu}(b/a) / \sqrt{a^2 - b^2}$ \cite{bateman1954tables,gradshteyn2014table}.

The link between Chebyshev polynomials, defined by $T_n(\cos \phi) = \cos(n \phi)$, and Bessel functions of integer index arises from the decomposition of plane waves in cylindrical coordinates ($\rho, z, \phi$): 
\begin{equation}
\E^{\iu x} = \E^{\iu \rho \cos \phi} = \sum_{n = -\infty}^\infty \iu^n \J{n}{\rho} \E^{\iu n \phi}. \label{eq:besselgenerating}
\end{equation}
One could take this to be a {\em definition} of the integer-index Bessel functions via a generating function. Sending $\iu \mapsto -\iu$, we learn that $\J{-m}{\rho} = (-1)^m \J{m}{\rho}$. By integrating both sides of \eqref{eq:besselgenerating} against $\cos(n \phi)$ from $-\cpi$ to $\cpi$ and exploiting orthogonality, we obtain the integral representation
\begin{equation}
\J{n}{\rho} = \frac{\iu^{-n}}{\cpi} \int_0^\cpi {\rm d}\phi\, \E^{\iu \rho \cos \phi} \cos(n \phi), \label{eq:Jnintegral}
\end{equation}
where we have used that the cosine is even to rewrite the integral from $0$ to $\cpi$ instead of $-\cpi$ to $\cpi$. In this expression we see that taking a derivative with respect to $\rho$ brings down a factor of $\iu \cos \phi$. Using the fact that $\cos(n\phi) = T_n(\cos \phi)$, we then obtain the identity
\begin{equation}
\J{n}{\rho} = \iu^{-n} T_n\left(-\iu \frac{\rm d}{{\rm d}\rho}\right) \J{0}{\rho}. \label{eq:besselchebyshev}
\end{equation}
This will allow us to easily compute the integral $\int_0^\infty {\rm d}t\, \J{n}{a t}\cos(b t)$ for arbitrary integer $n$ once we know the integral for the special case $n = 0$.

For $n = 0$, we directly use the representation \eqref{eq:Jnintegral} of $J_0$:
\begin{equation}
\int_0^\infty {\rm d}t\, \J{0}{a t}\cos(b t) = \frac{1}{\cpi} \int_0^\cpi {\rm d}\phi\, \int_0^\infty {\rm d}t\, \E^{\iu a t \cos \phi} \cos(b t) = \begin{cases}
                                        \frac{1}{\sqrt{a^2 - b^2}}, & 0 < b < a \\
                                        0, & 0 < a < b
                                 \end{cases} .
\end{equation}
One way to see this is to view the integral over $\phi$ as a contour integral, which picks up poles where $\cos \phi = \pm b/a$ which are in the domain of integration ($-1 \leq \cos \phi \leq 1$) when $b < a$ but not otherwise. 
Next, we can obtain the result for general $J_n$ by using \eqref{eq:besselchebyshev} and then integrating by parts to move the derivatives onto the $\cos(b t)$ factor. If $n$ is even, $T_n(x)$ contains only even terms and the derivatives produce a $\cos(b t)$ factor; if $n$ is odd, a similar argument leads to a $\sin(b t)$ factor. Every two derivatives acting on a $\cos(bt)$ factor will multiply by $-b^2 = (\iu b)^2$, effectively absorbing an extra factor of $\iu b$ into the argument of the polynomial. Hence:
\begin{equation}
\begin{aligned}
\int_0^\infty {\rm d}t\, \J{2k}{a t}\cos(b t) &= \iu^{-2k} \int_0^\infty {\rm d}t\, \cos(b t) T_{2k}\left(-\iu \frac{1}{a} \frac{{\rm d}}{{\rm d}t}\right) \J{0}{a t}  \\
&= (-1)^k \int_0^\infty {\rm d}t\, \J{0}{a t}  T_{2k}\left(-\iu \frac{1}{a} \frac{{\rm d}}{{\rm d}t}\right) \cos(b t) \\
&= (-1)^k \,T_{2k}(b/a) \int_0^\infty {\rm d}t\, \J{0}{a t} \cos(b t) = \frac{(-1)^k \,T_{2k}(b/a)}{\sqrt{a^2 - b^2}}, 
\end{aligned}
\end{equation}
agreeing with the results in the literature \cite{bateman1954tables,gradshteyn2014table}.

The polynomial behavior of $T_{2k}(b/a)$ leads to several zeros as a function of $b/a$, and hence to ``plateau'' structure in the Bessel overlap integrals like that we have previously observed (for different choices of $\nu$) in Fig.~\ref{fig:brTwoField}. More generally, the plateau structure arises due to similar oscillatory  behavior in the hypergeometric functions in \eqref{eq:Iplus2F1}.

\section{Conclusions}
\label{sec:conclusions}

 One of the challenges facing LHC studies is that new physics with unusual signatures might be able to escape trigger strategies or hide in the large data sets.  It is important therefore to consider these unusual signatures carefully, especially those that rarely appear in classic BSM models and are difficult or impossible to calculate with confidence.  Among these signatures are complex high-multiplicity final states.  Events with a small number of QCD-like jets are well-studied, and various approximately spherical high-multiplicity signals have also been considered, but little is known about signatures that, in some sense, lie between these extremes. 

It is well-known that large 't Hooft coupling gauge theories can produce spherical events \cite{Strassler:2008bv,Hofman:2008ar,Hatta:2008tx}, and that RS-like 5d models with cascade decays of Kaluza-Klein modes can produce {\em approximately} spherical events \cite{csaki2009ads}. Here we have  shown  that, in fact, 5d simplified models with tunable parameters (including a small number of bulk  fields with bulk  or boundary interactions) can produce a wide range of event shapes  in cascade decays of their heavy states.  These 5d simplified models are well-suited to serve as templates when designing collider searches for unusual events that might be hiding in samples of events with high jet multiplicity.

 Specifically, we saw that a key determinant in the cascade decays was the degree to which KK-number is violated, which in turn determines how close to threshold are the majority of decays. With some choices of bulk parameters, KK-number is approximately conserved, leading to quasi-spherical event samples; see \Fig{fig:singleFieldDist}.  For other choices of bulk parameters, or with the addition of boundary interactions, KK-number can be strongly violated, making events with a few hard jets commonplace.  In the latter cases, one can find samples as jetty as threshold $t\bar t$ events, with many individual events as jetty as SM $q\bar q$ events; compare \Fig{fig:emdMultTwoField} or \Fig{fig:evisobulkvbdry} to \Fig{fig:SMbenchmark}.  In demonstrating this, we have relied not only on thrust, a classic event shape variable, but also on event isotropy, a newly-introduced variable which appears well-suited to this purpose.  
 
One of the main results of this paper is an approximate analytic understanding of the Bessel function overlap integrals \Eq{def:Iintegral} that determine the couplings among different KK modes in the 5d simplified model,  which are the most important source of the KK-number violation. We are not aware of any previous detailed studies of this definite integral. In particular, we have seen that, aside from decays very near threshold, the integral over the finite fifth dimension is generically well approximated by the integral $I_+$ over an infinite interval; see \Eq{eq:IpImdef}. This integral has an analytic expression, \eqref{eq:Iplus2F1}, in terms of hypergeometric functions. This integral has a factor of the phase space function $\lambda_{\rm  PS}^{1/2}$,  defined in \eqref{eq:lambdaPS}, in the {\em denominator}. As a result, the naive expectation that decays far from threshold are favored due to the larger available phase space is precisely inverted within the context of these extra-dimensional models. Decays with small phase space are  often  favored. Similar results hold for the complementary integral $I_-$, which dominates in special cases, and for which we have provided an approximate analytic understanding.   However, there are also cases where  $I_+$  exhibits quasi-polynomial behavior with multiple zeroes, and is consequently enhanced far from the threshold region and/or suppressed at the threshold region. Then decays near threshold are no longer dominant and the cascades are more likely to produce kinematic jets.

In a companion paper \cite{paper2}, we will explore the event shapes that arise from our 5d simplified models in more  depth. In particular, we will illustrate, using both simulation and analytic estimates, that event isotropy provides an important complementary probe, capturing aspects of event shapes that are distinct from those captured by thrust, the eigenvalues of the sphericity tensor, or jet multiplicities. 

A remaining task is to connect our 5d simplified models with the Standard Model. In this paper, we have started our events with a single heavy KK mode which then cascades into many daughter particles. In order to use event shape observables based on massless momenta, we have assumed that all of the final-state daughter particles decay into two massless particles. However, we have stopped short of a full model for the interaction with the SM, which would allow for the production of many different modes and for a more general set of decays.
More complete models could be an interesting topic of further investigation, and could allow our 5d simplified models to be used in full event generators for experimental studies.   In this regard, it would be interesting to determine the effect on event shape variables of replacing our massless particles, which stand in for SM particles in the current study, with QCD jets from quarks and gluons, or with relatively soft photons. In the LHC context, one would need to consider carefully the impact  of initial state radiation, the underlying event and pileup.  It is also not clear what event-shape variables would be most effective in reducing backgrounds at a hadronic collider.  All of these issues must be addressed before optimized searches for phenomena of this type can be designed. 

\section*{Acknowledgments}

We thank Marat Freytsis, Gavin Salam, Jesse Thaler, David Pinner, and Andr{\'e} Frankenthal for useful discussions. MR and CC are supported in part by the DOE Grant DE-SC0013607. CC is supported in part by an NSF Graduate Research Fellowship Grant DGE1745303. We have made use of the Python Optimal Transport \cite{flamary2017pot} package when computing event isotropy.

\appendix

\section{Definitions and useful properties of special functions}
\label{sec:specialfunc}

\subsection{Fresnel integrals}
\label{sec:fresnel}

We define the Fresnel cosine and sine integrals by
\begin{align}
\FC{x} &= \int^x_0 {\rm d}t \, \cos\left(\frac{\cpi}{2} t^2\right), \nonumber \\
\FS{x} &= \int^x_0 {\rm d}t \, \sin\left(\frac{\cpi}{2} t^2\right).
\end{align}
This convention follows \cite{abramowitz1948handbook} and Mathematica, and differs from another common convention in which the argument is simply $t^2$. With this normalization, both of these functions tend to $1/2$ when their argument tends to infinity. In more detail, the large-argument asymptotics are
\begin{align}
\FC{z\rightarrow \infty} \rightarrow \frac{1}{2} + \frac{1}{\cpi z }\sin\left( \frac{\cpi z^2}{2} \right) - \frac{1}{\cpi^2 z^3}\cos\left( \frac{\cpi z^2}{2}\right) + \cdots, \nonumber \\
\FS{z\rightarrow \infty} \rightarrow \frac{1}{2} - \frac{1}{\cpi z }\cos\left( \frac{\cpi z^2}{2} \right) - \frac{1}{\cpi^2 z^3}\sin\left( \frac{\cpi z^2}{2}\right) + \cdots.
\label{eq:Fresnelasymptotics}
\end{align}

\subsection{Appell function}
\label{sec:appell}

The fourth Appell hypergeometric function $\mathrm{F}_4$ is defined by
\begin{equation}
\mathrm{F}_4(a,b; c,d; x,y) \equiv \sum_{m,n = 0}^\infty \frac{(a)_{m+n} (b)_{m+n}}{(c)_m (d)_n\, m!\, n!} x^m y^n,
\end{equation}
where the Pochhammer symbol $(a)_n$ denotes the rising factorial $a (a+1) \cdots (a+n-1)$, and $(a)_0 \equiv 1$.

In particular, notice that if $y = 0$, only terms with $n = 0$ contribute, and the formula becomes independent of $d$ and reduces to an ordinary hypergeometric function $\Hyp$:
\begin{align}
\mathrm{F}_4(a,b; c,d; x, 0) &= \sum_{m = 0}^\infty \frac{(a)_m (b)_m}{(c)_m m!} x^m \nonumber \\
&= \Hyp(a, b; c; x).
\end{align}
Similarly, $\mathrm{F}_4(a,b; c,d; 0, y) = \Hyp(a, b; d; y)$.

\subsection{Hypergeometric function reduction}
\label{sec:hypergeom}

There is a well-known identity that relates the fourth Appell hypergeometric function of two variables $z$ and $w$, $\mathrm{F}_4(\alpha, \beta; \gamma, \delta; z, w)$, to a product of two single-variable hypergeometric functions $\Hyp$ in the special case $\alpha + \beta = \gamma + \delta - 1$. Specifically \cite{bailey1933reducible}:
\begin{equation}
\mathrm{F}_4(\alpha, \beta; \gamma, 1 + \alpha + \beta - \gamma; z, w) = \Hyp(\alpha, \beta; \gamma; X)\, \Hyp(\alpha, \beta; 1 + \alpha + \beta - \gamma; Y), 
  \label{eq:F4reduction}
\end{equation}
where $X$ and $Y$ are chosen so that $z = X(1-Y)$ and $w = Y(1-X)$.

There is a general result that relates an infinite integral of products of three Bessel functions times a power to the function $\mathrm{F}_4$ \cite{bailey1936some}:
\begin{align}
\int_0^\infty {\rm d}z\, z^{\lambda-1} \J{\nu_1}{m_1 z} &\J{\nu_2}{m_2 z}\J{\nu_3}{m_3 z} = \frac{2^{\lambda - 1} m_2^{\nu_2} m_3^{\nu_3}  \Gamma\left(\frac{\nu_1 + \nu_2 + \nu_3 + \lambda}{2}\right)}{m_1^{\nu_2 + \nu_3 + \lambda} \Gamma \left( \nu_2 + 1 \right) \Gamma \left(\nu_3 + 1 \right) \Gamma \left( 1-\frac{\nu_1 - \nu_2 - \nu_3 - \lambda}{2}\right)} \nonumber \\
&\times \mathrm{F}_4 \left(\frac{\lambda + \nu_2+\nu_3-\nu_1}{2},\frac{\lambda + \nu_1+\nu_2+\nu_3}{2}; \nu_2+1, \nu_3 +1;  \frac{m_2^2}{m_1^2}, \frac{m_3^2}{m_1^2}\right),
\end{align}
valid (at least) when the $\lambda$, $\nu$ and $m$ parameters are real, $m_1 > m_2 + m_3$, $\lambda + \sum_i \nu_i > 0$, and $\lambda < 5/2$. 

Notice that the couplings we wish to calculate, \eqref{def:Iintegral}, have the form of this integral in the special case $\lambda = 2$. On the other hand, identifying the first four arguments of the $\mathrm{F}_4$ function as $\alpha, \beta; \gamma, \delta$, we see that they obey
\begin{equation}
\alpha + \beta = \gamma + \delta + \lambda - 2.
\end{equation} 
Thus, the reduction \eqref{eq:F4reduction} to ordinary hypergeometric functions applies when $\lambda = 1$, which is not the case of our interest.

However, integrals with different values of $\lambda$ may be related to each other using the identity
\begin{equation}
z J_\nu(m z) = \left(\frac{\partial}{\partial m} + \frac{\nu+1}{m}\right) J_{\nu + 1}(m z).
\end{equation}
This allows us to obtain the $\lambda = 2$ case of our interest from the $\lambda = 1$ case with a known reduction by taking a derivative with respect to a mass parameter. This method of relating different integrals was discussed in \cite{gervois1985integrals, gervois1986some}, and a specific formula relevant for the case of our interest was given in \cite{GervoisNaveletUnpublished}. This approach leads to the equation \eqref{eq:Iplus2F1} that we have given in the main text.

\section{Dependence on $n_p$}
\label{app:npdependence}

In this section we explore the event shape dependence on the initial mode $n_p$ with Dirichlet boundary conditions. 
As we increase $n_p$, this can affect both the coupling structure of the initial decays (where the degree of KK-number violation affects the event shape) and the kinematics of the final state.

\begin{figure}[t!]
\centering
\includegraphics[width=0.65\textwidth]{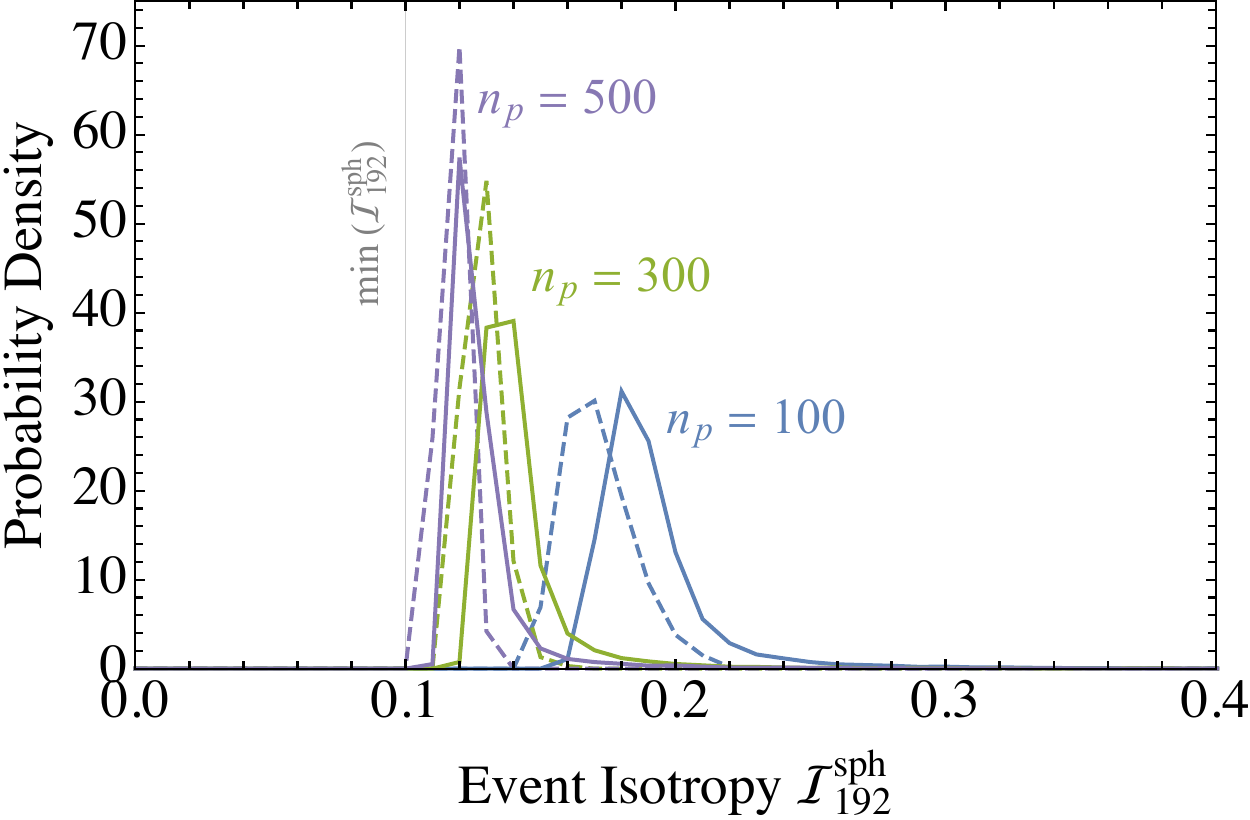}
\caption{The event isotropy $\iso{sph}{192}$ distributions for $n_p = 100$, 300, and 500 of the spherical toy model (dashed lines) and $\nu=0$ (solid line) samples. 
As $n_p$ increases, the $\nu=0$ sample approaches the spherical toy distribution, and all distributions move towards the theoretical bound set by \Eq{eq:isoest}.}
\label{fig:npInc}
\end{figure}

To understand how the kinematics of the final state depends on $n_p$, we first reconsider the spherical toy model described in \Sec{sec:sphereNearsphere}, i.e., events with $n_p$ identical particles at rest, which split into $n_p$ pairs of massless particles with equal energy and random orientation.
For sufficiently large $n_p$, the randomly distributed particles will appear isotropic. 
In \Fig{fig:npInc}, we see that as $n_p$ increases, the distribution of event isotropy $\iso{sph}{192}$ converges toward the theoretical limit of approximately 0.1,  by \Eq{eq:isoest}.

Similar behavior is seen in \Fig{fig:npInc} for the single field example with $\nu=0$, which has slightly larger event isotropy than the spherical toy model due to non-zero KK-number violation and energy anisotropy.
More generally, we expect near-spherical examples to converge to the theoretical bound at large $n_p$.

To explore the highly non-spherical regime, we consider another toy model that produces boosted particles early in the cascade. 
We generate a back-to-back pair of initial particles of equal mass $m = \frac{n_p}{2} m_0$, where $m_0$ is the mass of the lightest HSH, and with boost $\gamma$.
Each particle decays to $n_p/2$ HSHs which are at rest in the particle's decay frame.
The HSHs then decay to two massless particles each. 
Thus the two initial particles produce two quasi-spheres, made of $n_p$ massless particles with equal energy, that are boosted by $\gamma$ in opposite directions. 

\begin{figure}[t!]
\centering
\subfloat[]{
\includegraphics[width=0.45\textwidth]{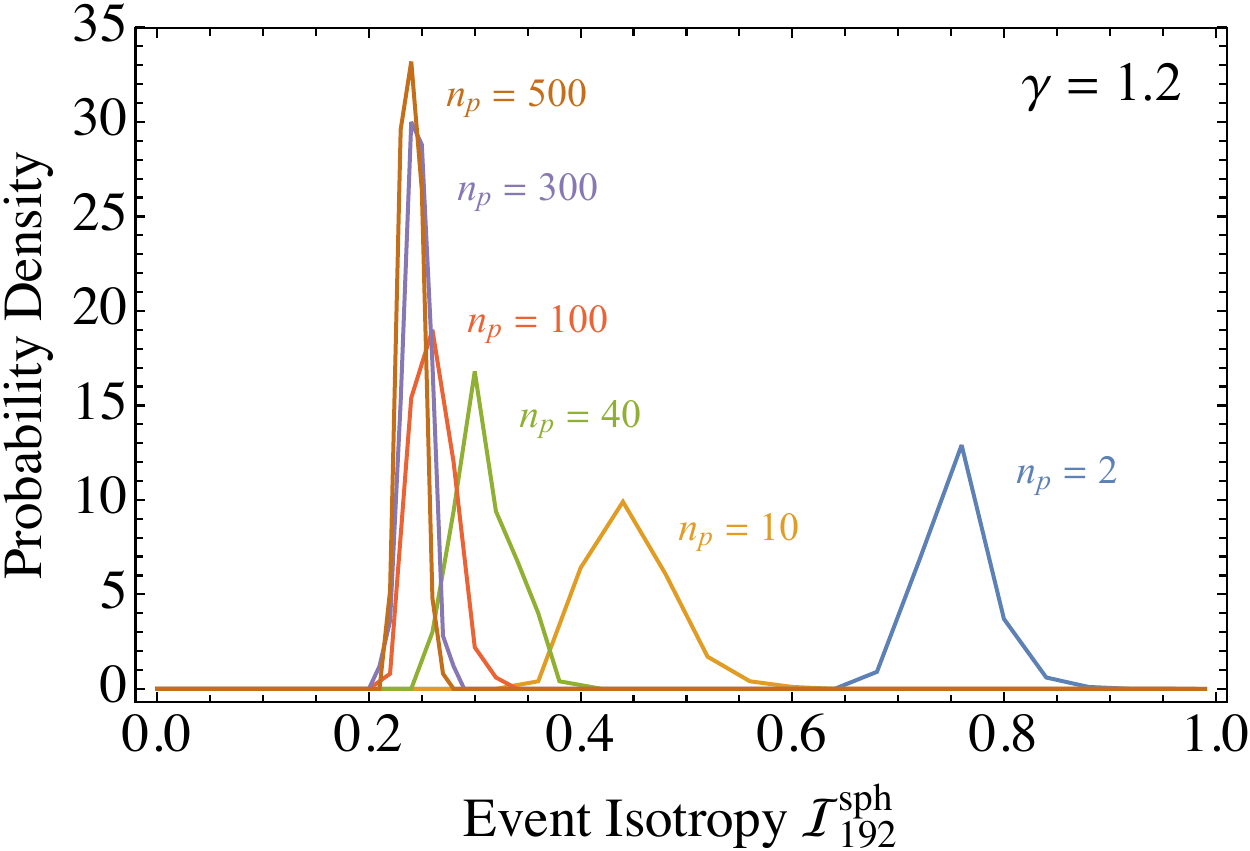} 
}
\hfill
\subfloat[]{
\includegraphics[width=0.45\textwidth]{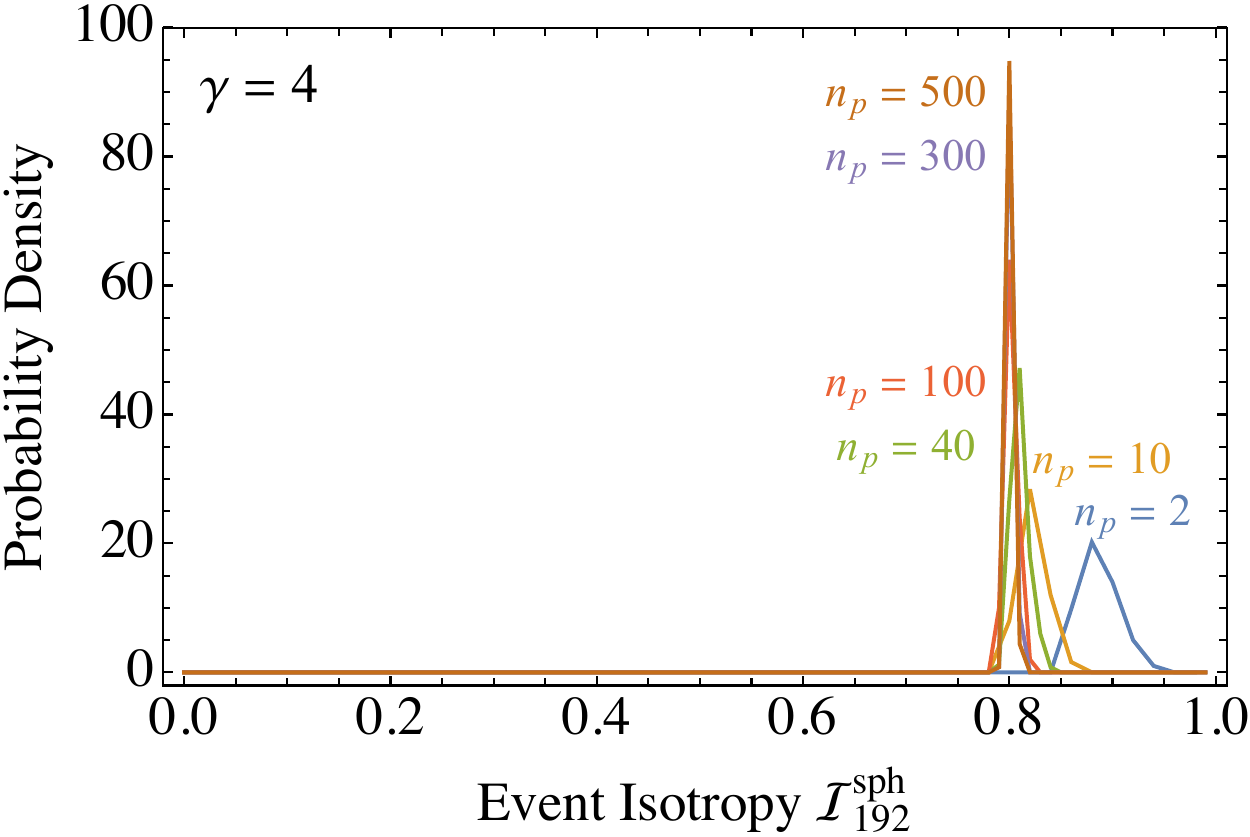} }
\hfill
\subfloat[]{
\includegraphics[width=0.45\textwidth]{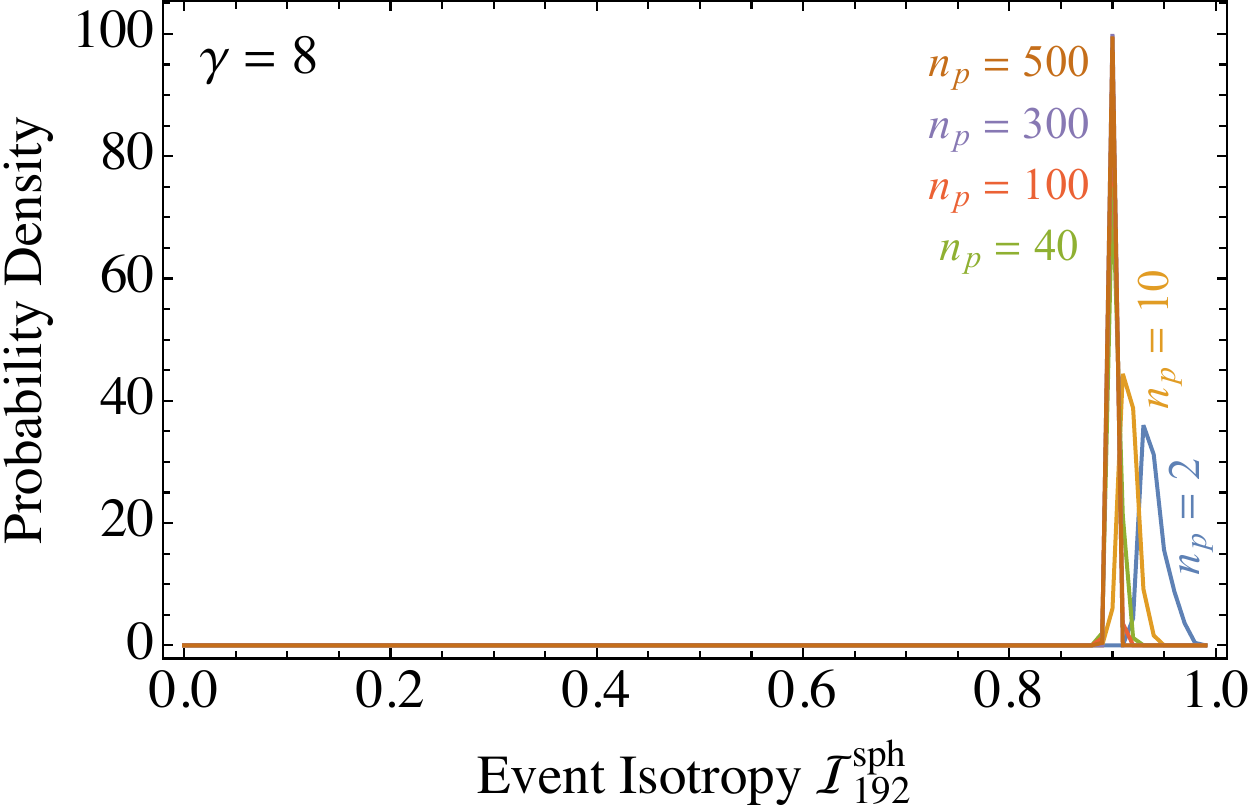}
}
\caption{Distributions in event isotropy $\iso{sph}{192}$ of the boosted toy model for $n_p =$ 2, 10, 40, 100, 300, 500 at boost (a) $\gamma = 1.2$, (b) $\gamma = 4$, and (c) $\gamma = 8$.
As the boost increases, all of the distributions are shifted towards jettier values of event isotropy ($\iso{sph}{192} \sim 1$) and overlap substantially.}
\label{fig:boostApp}
\end{figure}

We study the event shape dependence on $n_p$ as we increase the boost $\gamma$. 
When the initial particles are produced at rest ($\gamma = 1$), we recover the spherical toy model, as in \Fig{fig:npInc}, where the isotropy gradually decreases with $n_p$. 
In the opposite limit of large boost ($\gamma \rightarrow \infty$), each initial particle decays to a pencil-like jets, giving unit event isotropy for any $n_p$. 
Thus we expect the typical event isotropy to increase with $\gamma$, and to decrease with $n_p$, albeit with reduced $n_p$ dependence at large $\gamma$.  
These expectations\footnote{The value of event isotropy at large $n_p$ seems to increase as $1-1/\gamma$, though we have no analytic proof of this.}
 are borne out  in \Fig{fig:boostApp}, which shows event isotropy distributions for $n_p = 2, 10, 40, 100, 300$ and $500$ and $\gamma = 1.2$, $4$, and $8$.

An independent source of $n_p$ dependence is through the pattern of partial widths as seen in the branching fraction triangles shown in Figs.~\ref{fig:brratios}, \ref{fig:brTwoField}, and \ref{fig:brTwoField2}.
It turns out that for large $n_p$ these patterns become constant, though for two different reasons, as we will now show.

It is useful to divide the branching fraction triangles into three regions.  
The first is the ``immediate threshold'' (IT) region, the upper-right edge of the triangle, where the parent particle of mass $m_1$  decays to particles with mass $m_2$, $m_3$, where all $m_i \gg \cpi$. Using \Eq{eq:KKmassestimate}, we define the mass splitting as
\begin{equation}
m_{--}\equiv  m_1-m_2-m_3 = \left[(n_1-n_2-n_3)+\frac12(\nu_1-\nu_2-\nu_3)+\frac14 \right ]\cpi  .
\end{equation}
Note that the mass splitting is much less that the typical mass spacing in the KK tower: $m_{--} \ll \cpi$. 
The rest of the triangle may be divided roughly into an ``unboosted'' region where neither decay product is boosted (the center-right of the triangle) and a ``boosted'' region where at least one decay product is boosted.  

Each decay in the IT region has a partial width of order $n_p^{-3/2}$.
This follows from taking the integral \eqref{def:Iintegral}, using the asymptotic expansion of the Bessel functions for large argument (which is increasingly accurate across the integration region as the masses become large), and subjecting it to the same approximations that lead to 
\eqref{eq:Iminusfourterms}.
The integrand is a sum of four terms, of which one is constant at the extreme threshold $m_{--}\to 0$ and nearly so for $m_{--}\lesssim \cpi \ll m_i$.
Its integral, combined with the overall coefficient $N_i^{(\nu)}$ defined in \Eq{eq:dirNorm}, is then of order $(m_1m_2m_3)^{-1/2}\sim n_p^{-3/2}$. 
The couplings $c_{ijk}$ of the IT region, written explicitly in \Eq{eq:4dcoup}, are therefore independent of $n_p$.
Since the phase space $\lambda_\text{PS}\sim n_p^3 m_{--}$ in this regime, partial widths for individual decays given by \Eq{eq:width} are proportional to $n_p^{-3/2}$.
Summing over the $\sim n_p/2$ decay modes in the IT region, we find the partial width $\Gamma_\text{IT}\sim (n_p)^{-1/2}$ for the region.  
This applies unless $\nu_1-\nu_2-\nu_3$ an odd integer, in which case the partial widths fall even faster with $n_p$.

Away from the IT region, $I_+/I_-\sim \sqrt{\lambda_\text{PS}}/\sqrt{m_1 m_2 m_3}\sim \sqrt{n_p}$, so unless $I_+$ vanishes, it dominates at large $n_p$.
A partial width of a typical decay dominated by $I_+\sim\lambda_\text{PS}^{-1/2}$ scales as $n_p^{-2}$.  
Since there are of order $n_p$ decay modes, the total width $\Gamma_\text{tot}$ is independent of $n_p$.  
The IT region is therefore subleading and scales away as $n_p\to\infty$, while the rest of the branching fraction triangle becomes constant at sufficiently large $n_p$.
Thus if the events are far from spherical at $n_p=100$, they remain so at larger $n_p$.
Alternatively, for events that are already quasi-spherical at $n_p = 100$, we expect the distribution in event isotropy to approach the theoretical bound as we increase $n_p$.

If $I_+$ vanishes, which occurs when $\nu_2+\nu_3-\nu_1$ is a positive even integer, then there is a subtlety.  
The partial width of a typical decay dominated by $I_-\sim\sqrt{m_1m_2m_3}\lambda_{PS}^{-3/2}$ scales as $n_p^{-3}$.
This would suggest that IT region dominates and that the total width scales as $n_p^{-1/2}$.
While the latter conclusion is correct, near-threshold decays, with small kinematic boosts but larger KK-number violation, can be as important as the minimally-KK-violating decays at the IT.
There is therefore a band in the branching fraction triangle, including but extending beyond the immediate threshold, which remains important as $n_p\to\infty$.
The boosted region scales away, so the events are far from jetty.
However, because KK-number violation is non-minimal, the events may not become  spherical in the $n_p\to\infty$ limit.  

\bibliography{ref}
\bibliographystyle{utphys}

\end{document}